\documentclass[a4paper,10pt]{article}
\usepackage{fullpage}
\usepackage{rotating}
\usepackage{setspace}
\usepackage{amsmath}
\usepackage{amsthm}
\usepackage{natbib}
\usepackage{enumerate}
\usepackage{algorithmic, algorithm}
\usepackage{graphicx, psfrag}
\usepackage[font=footnotesize,labelfont=bf]{caption}
\usepackage{subfig}
\usepackage{txfonts}
\usepackage{color}
\renewcommand{\tabcolsep}{3pt}

\title{Markov chain Monte Carlo for exact inference for diffusions}

\author{Sermaidis, G.\thanks{Lancaster University}, Papaspiliopoulos,
O.\thanks{corresponding author,Universitat Pompeu
Fabra,omiros.papaspiliopoulos@upf.edu}, Roberts, G.O.\thanks{Warwick
University}, Beskos, A.\thanks{UCL} and Fearnhead, P.$^*$}
\date{}

\def\myh{3.8cm}                 
\def\myw{5.2cm}                 

\newcommand{\lp}{\left(}
\newcommand{\rp}{\right)}
\newcommand{\lsb}{\left[}
\newcommand{\rsb}{\right]}
\newcommand{\lcb}{\left\{}
\newcommand{\rcb}{\right\}}
\newcommand{\tl}[2]{{\tilde #1}^{\{#2\}}}
\newcommand{\bl}[2]{{\bar #1}^{\{#2\}}}
\newcommand{\Gir}[2]{G_{#1}\lp #2;\theta \rp}
\newcommand{\MSTATE}[1]{\STATE\hspace{#1}}

\def\isi{HFA}
\def\ipi{PA}
\def\eda{EDA}
\def\amcmc{AMCMC}
\def\emcmc{EMCMC}
\def\LAM{{\small \bf [LAM]}}
\def\SMO{{\small \bf [SMOOTH]}}
\def\GRA{{\small \bf [GRAD]}}
\def\LBO{{\small \bf [LBOUND]}}

\def\lh{\epsilon}
\def\ci{k}                
\def\dd{d}
\def\db{d}
\def\Det{{\rm det}}
\def\idm{{\it I}}
\def\rr{{\{\ci\}}}
\def\ri{{\{i\}}}
\def\mbb{\mu}

\def\pint{\lambda}
\def\obd{q}
\def\obs{Y}
\def\nobs{Z}

\newcommand{\DD}[1]{D(#1;\spar)}

\newtheorem{lemma}{\bf Lemma}
\newtheorem{theorem}{\bf Theorem}

\newtheorem{condition}{\it Condition}

\newcommand{\ind}[1]{\ensuremath{{\mathbb I}#1}}
\newcommand{\cor}[2]{#1^{\{#2\}}}
\def\xb{\bar{x}}
\def\yb{\bar{y}}
\def\bxt{\bar{x}^\rr(\spar)}
\def\byt{\bar{y}^\rr(\spar)}
\def\tX{{\tilde X}}
\def\sX{X^{*}}

\def\tE{{\tilde E}}
\newcommand{\xt}[1]{x_{#1}(\spar)}
\newcommand{\yt}[1]{y_{#1}(\spar)}
\newcommand{\xtr}[1]{x^{\{#1\}}(\spar)}
\newcommand{\ytr}[1]{y^{\{#1\}}(\spar)}

\def\Leb{{\rm Leb}}

\def\pPois{\Phi}
\def\pPsi{\Psi}
\def\ppsi{\psi}
\def\pkappa{\kappa}
\def\pUps{\Upsilon}
\def\pups{u}

\def\apath{\tilde X}
\def\aPois{\Phi}
\def\aPsi{\Psi}
\def\apsi{\psi}
\def\akappa{\kappa}
\def\aUps{\Upsilon}

\def\pPoisnc{\tilde\Psi}
\def\pPsinc{\tilde\Psi}
\def\ppsinc{\tilde\psi}
\def\pkappanc{\tilde\kappa}
\def\pUpsnc{\tilde\Upsilon}
\def\pupsnc{\tilde u}

\def\pxinc{\tilde\xi}

\def\aPoisnc{\tilde\Psi}
\def\aPsinc{\tilde\Psi}
\def\apsinc{\tilde\psi}
\def\akappanc{\tilde\kappa}

\def\axinc{\tilde\xi}

\newcommand{\ncdensf}{\pi_{nc}}

\def\ptheta{\theta^{*}}

\newcommand{\skel}[1]{\ensuremath{S(#1)}}

\newcommand{\tinc}[1]{\ensuremath{\Delta t_{#1}}}

\def\player{\tilde L}
\def\plpath{\tX}

\def\alayer{\tilde L}
\def\alpath{\tilde X}
\def\aPois{\Phi}
\def\aPsi{\Psi}
\def\apsi{\psi}
\def\akappa{\kappa}
\def\aUps{\Upsilon}

\newcommand{\minbb}{\ensuremath{m}}
\newcommand{\locbb}{\ensuremath{{\tau}}}

\def\gx{X}
\newcommand{\maxint}[1]{\ensuremath{r(#1;\theta)}}
\newcommand{\maxinttheta}{\ensuremath{r(\theta)}}

\def\d{\mathrm{d}}
\def\mR{\mathbb{R}}
\def\sp{v}
\def\ep{w}
\def\pd{p}
\def\Po{\mbox{Po}}
\def\Un{\mathrm{Un}}
\def\mE{\mathbb{E}}

\newcommand{\gdens}[2]{\ensuremath{{\cal N}_{#2}\lp #1 \rp}}
\newcommand{\gdenscb}[2]{\ensuremath{{\cal N}_{#2}\lcb #1 \rcb}}

\newcommand{\lf}[2]{\eta_{#1}(#2;\spar)}
\newcommand{\lfrcb}[2]{\eta^{\{#1\}}\{#2;\spar\}}

\newcommand{\ilf}[1]{\eta^{-1}(#1;\spar)}

\newcommand{\ilfe}{\eta^{-1}}

\newcommand{\trans}[3]{\ensuremath{p_{#1}(#2,#3;\theta)}}
\newcommand{\ttrans}[3]{{\tilde p}_{#1}( #2,#3;\theta )}
\newcommand{\ttranscb}[3]{{\tilde p}_{#1}\lcb #2,#3;\theta \rcb}

\def\dpar{\theta_1}
\def\spar{\theta_2}
\def\dparstate{\Theta_1}
\def\sparstate{\Theta_2}
\newcommand{\odrift}[2]{\beta_{#1}(#2; \dpar)}
\newcommand{\odriftr}[2]{\beta^{\{#1\}}(#2; \dpar)}

\newcommand{\drift}[2]{\alpha_{#1}(#2; \theta)}
\newcommand{\driftTr}[2]{\alpha^{T}_{#1}(#2; \theta)}
\newcommand{\driftr}[2]{\alpha^{\{#1\}}(#2; \theta)}
\newcommand{\driftnp}{\alpha}

\newcommand{\driftsqd}[1]{\ensuremath{||\drift{}{#1}||^2 + \Delta_x
\adrift{#1}}}
\newcommand{\driftsqdcb}[1]{\ensuremath{\lcb ||\drift{}{#1}||^2 + \Delta_x
\adrift{#1}
\rcb}}

\newcommand{\asqapf}[1]{\ensuremath{f\left( #1;\theta\right)} }

\newcommand{\phif}[1]{\ensuremath{\phi\lp#1;\theta\rp} }

\newcommand{\phifcb}[1]{\ensuremath{\phi\lcb#1;\theta\rcb} }

\newcommand{\adrift}[1]{\ensuremath{{H}(#1;\theta)}}
\newcommand{\adriftcb}[1]{H\{#1;\theta\}}

\newcommand{\difcoef}[1]{\sigma(#1; \spar)}
\newcommand{\difcoefnp}{\sigma}
\newcommand{\difcoefinv}[1]{\sigma^{-1}(#1; \spar)}

\newcommand{\volr}[2]{\gamma^{\{#1\}}(#2;\spar)}
\newcommand{\volnp}{\gamma}

\newcommand{\mP}[1]{\mathbb P^{(#1)}}
\newcommand{\tmL}[1]{\tilde{\mathbb L}^{(#1)}}

\newcommand{\cmWn}[1]{\ensuremath{\mathbb{W}^{(#1)}}}
\newcommand{\cmMn}[1]{\ensuremath{\mathbb{M}^{(#1)}}}
\newcommand{\cmW}[1]{\ensuremath{\mathbb{W}_{\theta}^{(#1)}}}
\newcommand{\cmQ}[1]{\ensuremath{\mathbb{Q}_{\theta}^{(#1)}}}

\newcommand{\cmQt}[2]{\ensuremath{{\tilde {\mathbb Q}}_{#1}^{(#2)}}}

\def\mN{{\mathbb N}^{*}}

\begin{document}

\maketitle
\begin{abstract}
  We develop exact Markov chain Monte Carlo methods for
  discretely-sampled, directly and indirectly observed diffusions. The
  qualification "exact" refers to the fact that the invariant and
  limiting distribution of the Markov chains is the posterior
  distribution of the parameters free of any
  discretisation error. The class of processes to which our methods
  directly apply are those which can be simulated using the most
  general to date exact simulation algorithm. The article introduces
  various methods to boost the performance of the basic scheme,
  including reparametrisations and auxiliary Poisson sampling. We
  contrast both theoretically and empirically how this new approach
  compares to irreducible high frequency imputation, which is the
  state-of-the-art alternative for the class of processes we consider,
  and we uncover intriguing connections. All methods discussed in the
  article are tested on typical examples.
\end{abstract}


\section{Introduction}
\label{sec:intro}

Diffusion processes provide a flexible framework for modelling
phenomena which evolve randomly and continuously in time and are
extensively used throughout Science, e.g. in finance
\citep{saha:real}, biology \citep{gol:wil:2006}, molecular kinetics
\citep{hore:hmmsde}, pharma\-co\-ki\-netics/pharma\-co\-dy\-na\-mics \citep{ditl} and
spatio-temporal modelling \citep{tone}.

A time-homogeneous diffusion process $V\in\mR^\dd$ is a Markov process
defined as the solution to a stochastic differential equation (SDE):
\begin{align}\label{eq:model} \d V_s = \odrift{}{V_s}\d s +
\difcoef{V_s}\d W_s, ~~~~~~V_0=\sp,\,s\geq 0,
\end{align}
where $W$ is a $\db$-dimensional standard Brownian motion. The
functions $\beta: \mR^\dd\times \dparstate\rightarrow \mR^\dd$ and
$\sigma: \mR^\dd\times \sparstate\rightarrow \mR^{\dd\times\db}$ are
known as the {\it drift} and {\it diffusion coefficient} respectively,
and are allowed to depend on an unknown parameter
$\theta=(\dpar,\spar)\in\Theta\subset \mR^\pd$. The assumption of
distinct parameters for each functional is by no means restrictive and
is adopted here for ease of presentation. We assume that $\sigma$ is
invertible and make the usual set of assumptions on $\beta$ and $\sigma$ to
ensure that (\ref{eq:model}) has a unique weak non-explosive solution,
see for example Theorem 5.2.1 of \cite{oksendal}; see also Section
\ref{sec:pre}.

Even though the process is defined in continuous time, the available data
consist of observations recorded at a set of discrete time points, 
\begin{align*}
 \obs := \lcb V_{t_0}, V_{t_1}, \ldots, V_{t_n} \rcb, ~~~~~~0\leq t_0<
t_1<\ldots<t_n.
\end{align*}
Statistical inference is pursued in a Bayesian framework where 
prior beliefs about the parameters, encoded via a prior density  $\pi(\theta)$, are updated on the
basis of the available data through the discrete-time likelihood to
yield the posterior beliefs, encoded via the posterior density: 
\begin{equation}
  \label{eq:post}
  \pi(\theta \mid \obs) \propto \pi(\theta) \prod_{i=1}^n \trans{\Delta t_i}{V_{t_{i-1}}}{V_{t_i}}\,,
\end{equation}
where $\tinc{i}=t_i-t_{i-1}$ is the time increment between consecutive
observations and 
\begin{align*}
  \trans{t}{\sp}{\ep} = P(V_t \in \d\ep \mid V_0 = \sp) /\d w,
  ~~~~~~t>0, \sp,\ep\in\mR^\dd,
\end{align*}
is the transition density of the process.
 
Bayesian and generally likelihood-based inference in this context is
hindered by the unavailability of the transition density and
sufficiently accurate approximations to the density exist only when
$t$ is sufficiently small. One strand of the literature approaches the
inference problem by resorting to Monte Carlo data augmentation
(DA), according to the following principle. First, a
  DA scheme is constructed by identifying auxiliary variables such
  that the joint density of those and the observations, known as complete
  likelihood, is analytically available. Subsequently, inference is
  performed by employing a Markov chain Monte Carlo (MCMC) algorithm
  which targets the posterior density of parameters and auxiliary
  variables. Early DA schemes were based on
imputation of a finite number of points, say $M$, of the latent
diffusion bridges $\lcb V_s, s\in(t_{i-1},t_i)\rcb$, see for example
\cite{eraker:2001, eler:chib:shep:2001}. The complete likelihood is
still intractable but can be reasonably approximated using an Euler
scheme which now operates on smaller time increments. The bias
introduced in this approximation is eliminated by increasing $M$.
There are three serious challenges with this approach. First, the
simulation of the latent bridges conditionally on the parameters; this
simulation is required in the "Imputation'' step of a DA
algorithm. This problem has been intensively studied, see for example
\cite{paprobsemstat} for a recent account. Second, the choice of $M$,
at least in practice. A good approximation usually requires a large
value of $M$ which is typically found by repeated runs of the
algorithm until the estimated posterior distributions show no
change. This adds a substantial computational burden. Third,
\cite{robe:stra:2001} showed that when $\theta_2$ is unknown the
mixing time of the MCMC algorithm is ${\cal O}(M)$. This is due to the
quadratic variation identity, according to which any continuous-time
path contains infinite information about $\theta_2$.  Thus, in these
early DA schemes reduction in bias comes with unbounded increase in
the Monte Carlo variance. \cite{robe:stra:2001} constructed
appropriate path-parameter transformations in order to yield a working
DA scheme which is valid even in the limit $M\to \infty$, unlike those
of the previous generation. We refer to the limiting case $M\to\infty$
as path augmentation ({\ipi}). An MCMC
algorithm based on {\ipi} is not implementable in practice, since
it involves infinite amount of computation. For finite $M$, we
refer to the DA scheme as high frequency augmentation ({\isi}) and to
the MCMC algorithm which targets it as approximate MCMC (\amcmc), the
term reflecting the fact that
bias is introduced due to the discretisation of the paths. More details on
these schemes are given in Section \ref{sec:ihfi}.

A new generation of Monte Carlo schemes for diffusions was initiated
with the introduction of the exact algorithm (EA) for the exact
simulation of non-linear diffusions.  The potential of using the EA to
build an MCMC algorithm for parameter estimation was sketched in
\cite{besk:papa:robe:fear:2006} for a restrictive class of univariate
diffusions.

In this paper we present a novel augmentation scheme, called exact
data augmentation ({\eda}), and develop MCMC algorithms for all
diffusions which can be simulated under the broader framework of the
so-called EA3 \citep{besk:papa:robe:2008}. This generation of MCMC
algorithms based on EDA is referred to as exact MCMC ({\emcmc}), and
is such that their equilibrium distribution is the exact posterior
distribution of the parameters, i.e., free of any discretisation
error. We enhance algorithmic performance by designing noncentred
reparametrisations and extend our methods to the case of indirect
observations where interest lies in estimating both parameters and the
latent diffusion process. A further contribution of this work is
  a theoretical investigation of the connection between {\eda} and
  {\ipi}. First, it is shown that {\eda} augments more information
  than {\ipi}. This is rather surprising since the former appears to
augment only a small finite-dimensional distribution of the missing
paths whereas the latter in principle augments continuous paths and in
practice high-frequency approximations thereof. The key is that the
extra augmentation in {\eda} creates conditional independence
relationships which are exploited to apply an algorithm which targets
an infinite-dimensional state using finite computation. This
  result also suggests that {\amcmc} for the same amount of
  computation is expected to mix faster than {\emcmc}; this is
effectively another instance of the bias-variance tradeoff. This
connection motivates a further observation which links the two
approaches and suggests a way to improve the convergence rate of
{\emcmc} using auxiliary Poisson sampling.

A comment on the applicability of the methods proposed here is
due. The methods rely on a variance-stabilising transformation after
which the diffusion has constant diffusion
matrix and drift which is of gradient form, see Section
\ref{sec:trans} for details. The transformation poses little
limitations for univariate processes, but might not even
exist for general multivariate SDEs with coupling in the diffusion. On
the other hand, multivariate processes with gradient drift structure
and no coupling in the diffusion are rather standard in the framework
of physical systems. Note that the variance-stabilising transformation
is necessary for the {\isi} approach as well. In summary, the
technology we develop here is not directly applicable to general
stochastic volatility models, say, although exploiting particular
structures can push considerably these limitations, see for example
\cite{kalog}. Additionally, advances in the exact simulation of
diffusions, as for example in \cite{ea:local,ea:jump} would \emph{eo
  ipso} lead to exact MCMC methods following the framework of this
article. Irreducible DA schemes avoiding this transformation are also
currently under investigation, see for example \cite{gol:wilk:2008}.

The article is structured as follows. Section \ref{sec:pre} contains
the background on assumptions, notations and recalls the EA. Section
\ref{sec:eda} presents formally the {\eda} and contrasts it to
{\ipi}. Section \ref{sec:repar} describes noncentred
reparametrisations of the {\eda} and auxiliary Poisson sampling from
improving algorithmic performance. Section \ref{sec:error}
discusses extensions to indirect observations. Section
\ref{sec:numerics} carries out a careful and extensive numerical
comparison of several schemes. Section \ref{sec:discuss} closes with a
discussion and the Appendix contains the proofs of main results.


\section{Preliminaries}\label{sec:pre}

In this section we collect some necessary background.  In terms of
notation, $x^{\{i\}}$ or $x^{\{ij\}}$ denote the $i$th or $(i,j)$th
element of a vector or matrix $x$, $\Det[x]$, $x^T$ and $x^{-1}$
denote the determinant, transpose and inverse of a matrix $x$ where
appropriate, and $\idm_d$ denotes the $d\times d$ identity matrix. For two
vectors $x$ and $y$, we define vectors $\xb$ and $\yb$ such that
$\xb^\ri=x^\ri\wedge y^\ri$ and $\yb^\ri=x^\ri\vee y^\ri$. The
Euclidean norm is denoted by $||.||$. $\nabla_x$ and $\Delta_x$ denote
the Jacobian matrix and Laplacian operators respectively, that is if
$x \in \mR^d$, then for functions $f_1:\mR^\dd\rightarrow\mR^{m}$ and
$f_2:\mR^\dd\rightarrow\mR$
\begin{align*}
  \nabla_x f_1(x) = \lsb \frac{f_1^{\{j\}}(x)}{\partial x^{\{i\}}}
  \rsb_{i=1,\ldots,d;j=1,\ldots,m} \quad \Delta_x f_2(x) =
  \sum_{i=1}^d \frac{\partial^2 f_2(x)}{\partial (x^{\{i\}})^2}\,.
\end{align*}
We define $D:\mR^\dd\times\Theta_2\rightarrow\mR$ as
$D=|\det[\difcoefnp]|^{-1}$ and
$\volnp:\mR^\dd\times\Theta_2\rightarrow\mR^{\dd\times\dd}$ as
$\volnp=\difcoefnp\difcoefnp^T$, where $\difcoefnp$ is the diffusion
coefficient. For function $f:\mR^\dd\rightarrow\mR$ twice continuously
differentiable on its domain, we denote the generator of
\eqref{eq:model} by
\begin{align*}
 A_{\theta} f(v) =
\sum_{i=1}^{\dd}\odriftr{i}{v}\frac{\partial{f(v)}}{\partial
v^{\{i\}}} +
\frac{1}{2}\sum_{i,j=1}^{\dd}\volr{ij}{v}\frac{\partial^2 f(v) } { \partial v^{\{i\}}
\partial v^{\{j\}}}.
\end{align*}
Finally, $\gdens{u}{t}$ denotes the density of a Gaussian random
variable with mean vector $0$ and covariance matrix $t\idm_{\dd}$
evaluated at $u\in\mR^\dd$.

\subsection{Reducible diffusions of gradient type}
\label{sec:trans}

The methods in this paper rely on the existence of a transformation
$\eta$, known as Lamperti transformation, such that
$\eta(V_s;\theta_2)$ solves an SDE with constant diffusion
matrix. This transformation can be obtained for univariate diffusions
rather trivially. For multivariate processes its existence is a subtle
matter. In the elliptic case a sufficient condition is
\begin{description}
\item \LAM: $\lsb\nabla_v \eta(v;\theta_2)\rsb^{T} = \difcoefinv{v},$
\end{description}
which can be simplified to yield explicit conditions on the elements
of $\sigma^{-1}$; see for example \cite{sah:2008}. In the rest of the
article we will assume the existence of this transformation and denote
its inverse by $\ilfe$. If $X_s:= \lf{}{V_s}$
is the transformed diffusion, then by It\^o's formula $X$ solves
\begin{align}\label{eq:unit_SDE1}
 \d X_s = \drift{}{X_s}\d s + \d W_s,~~~~~~X_0 = x =\lf{}{\sp},\, s\geq 0,
\end{align}
where $\driftnp:\mR^{\dd}
\times\Theta\rightarrow\mR^{\dd}$ , with
\begin{align*}
\driftr{\ci}{u} = A_{\theta} \lfrcb{\ci}{\ilf{u}},~~~~~~\ci=1,\ldots,\dd.
\end{align*}
If $\ttrans{t}{x}{z},\,x,z\in \mR^\dd$ is the transition density
of $X$, then the transition density of $V$ can be
expressed as
\begin{align}\label{eq:Jacobian}
  \trans{t}{\sp}{\ep} =
\DD{\ep}\ttranscb{t}{\lf{}{\sp}}{\lf{}{\ep}}.
\end{align}

The methodology also requires certain conditions on the drift $\alpha$
of the transformed process. 
The following set of assumptions should hold for any $\theta \in
\Theta$:
\begin{description}
 \item[\SMO:] $\driftr{\ci}{\cdot}$ is continuously differentiable for $\ci=1,\ldots,\dd$.\label{as:as1}

\item[\GRA:] There exists $H:\mR^d\times\Theta\rightarrow\mR$ such that $\nabla_x \adrift{x}
=
\drift{}{x}$.\label{as:as2}

\item[\LBO:] There exists $l(\theta)>-\infty$, such that $l(\theta) \leq
\inf_{u\in\mR^\dd}\frac{1}{2}\driftsqdcb{u}$.\label{cond:lower_bound}\label{as:as3}
\end{description}
The first is a very weak condition and the third is rather mild
too. The second identifies $X$ as a diffusion of gradient-type, where
$H$ is called the potential function. When the diffusion is ergodic,
its invariant log-density can be expressed directly in terms of
$H$. This condition is trivially satisfied for univariate processes but
is more restrictive for multivariate ones. Finally, we require that
$\alpha$ is such that the probability law generated by the solution of
(\ref{eq:unit_SDE1}) is absolutely continuous with respect to the
Wiener measure. A particularly useful and weak set of conditions are
given in \cite{Rydberg:1997}; in the case of (\ref{eq:unit_SDE1}) if
$\alpha$ is locally bounded the conditions simply require that the SDE
be not explosive.

\subsection{Exact simulation of diffusions}\label{sec:ea}

The EA is a rejection sampling algorithm on the space of
diffusion paths, which uses Brownian path proposals and delivers the
diffusion path revealed at a finite collection of random points. The
path can be filled in later with no further
reference to the target process. The main attraction of the algorithm
is that the draws are from the exact finite-dimensional distribution.
Here we focus on exact {\it diffusion bridge} simulation, i.e.,\
obtain samples from (\ref{eq:model}) conditionally on the origin
$V_0=\sp$ and terminal point $V_t=\ep$.  It turns out that this
conditional simulation is really the key to DA methods for parameter
estimation.

The target process \eqref{eq:model} is transformed into one of unit
diffusion matrix as described in Section \ref{sec:trans}. The problem
is therefore reduced to the simulation of (\ref{eq:unit_SDE1})
conditionally on the origin $x=\xt{} := \lf{}{\sp}$ and terminal point
$y = \yt{} := \lf{}{\ep}$. An $X$-bridge yields a $V$-bridge by
applying the inverse transformation.  Let $\cmQ{t,x,y}$ denote the law
of the $X$-bridge starting at $x$ and terminating at $y$ at time $t$, and
$\cmW{t,x,y}$ the law of a Brownian bridge conditioned on the same
endpoints. The following lemma, which is a
restatement of Lemma 1 in \cite{besk:papa:robe:fear:2006}, derives the
density of the target law with respect to the Brownian bridge law.
%
\begin{lemma}
 The law $\cmQ{t,x,y}$ is absolutely continuous with respect to $\cmW{t,x,y}$
with density 
\begin{align}
 \frac{\d \cmQ{t,x,y}}{\d \cmW{t,x,y}}(\gx) &=
\frac{\gdens{y-x}{t}}{\ttrans{t}{x}{y}} \exp\left\{
\adrift{y} - \adrift{x} - \frac{1}{2}\int_0^t\driftsqdcb{X_s}\d s
\right\}\label{eq:girs}
\\
& \propto \exp\left\{-\int_0^t\phif{\gx_s}\d s \right\} \leq 1,
\label{eq:ea_ratio}
\end{align}
where $\phi:\mR^\dd\times\Theta\rightarrow \mR_{+}$ is defined by
\begin{align*}
 \phif{u} = \frac{1}{2}\driftsqdcb{u} - l(\theta).
\end{align*}
\end{lemma}
%


The EA is based on recognising (\ref{eq:ea_ratio}) as the probability
of a specific event from an inhomogeneous Poisson process of intensity
$\phif{X_s}$ on $[0,t]$. Such processes can
be simulated by constructing an upper bound for the variable intensity
and using Poisson thinning. Assume
that there exists a finite-dimensional random variable $L:=L(X)$ and a
positive function $r$ such that
\begin{align*}
\maxint{L} \geq \sup_{s\in[0,t]}\phif{\gx_s},
\end{align*}
and let $\Phi=\lcb \Psi,\Upsilon \rcb$ be a homogeneous Poisson process of
intensity $\maxint{L}$ on $[0,t]\times[0,1]$, with uniformly
distributed points $\Psi=\lcb \psi_1, \ldots,\psi_\kappa \rcb$ on
$[0,t]$ and marks $\Upsilon=\lcb
u_1,\ldots,u_\kappa\rcb$ on $[0,1]$, where
$\kappa\sim\Po[\maxint{L}t]$. If $N$ is the number of points of $\Phi$
below the graph $s\rightarrow\phif{X_s}/\maxint{L}$, then
\begin{align*}
 P\lp N=0\mid \gx \rp = \exp\left\{-\int_0^t\phif{\gx_s}\d s
\right\}.
\end{align*}
This implies a
rejection sampler where a proposed path $X\sim\cmW{t,x,y}$ is accepted as a path
from $\cmQ{t,x,y}$ according to
the indicator
\begin{align}\label{eq:ind_ea}
 I (L, \gx, \Phi, \sp, \ep,\theta):=
\prod_{j=1}^{\kappa}\ind{\lsb\phif{\gx_{\psi_j}}/\maxint{L}
<u_j\rsb }.
\end{align}
The EA output is the collection $\lcb
L(\gx), \Phi,
S(X)
\rcb$, where 
\begin{equation*}
  \skel{X}:=\{
(0,\gx_0),(\psi_1,\gx_{\psi_1}),
\ldots,(\psi_{\kappa},\gx_{\psi_{\kappa}}),
(t,\gx_t) \}
\end{equation*}
is a skeleton of the accepted path. 
The algorithm is presented in Algorithm
\ref{alg:gen_CEA}.
The technical difficulty that underlies the implementation of the EA
is the simulation of $L(X)$, and primarily the conditional simulation
of a Brownian bridge given $L(X)$ for the evaluation of
\eqref{eq:ind_ea}. This has led to the construction of three EAs
that share the rejection sampling principle, but have a different
range of applicability.
\begin{algorithm}
\caption{\textbf{EA for diffusion bridges}}
\label{alg:gen_CEA}

\begin{spacing}{1.4}

\begin{algorithmic}[1]{\tt\footnotesize

\STATE simulate $L = L(X)$, where $\gx\sim\cmW{t,x,y}$,

\STATE simulate $\kappa\sim\Po\lsb
\maxint{L}t\rsb$ and $\Phi=\{(\psi_j,\upsilon_j)\}_j,\,1\leq j\leq \kappa$
uniformly on $[0,t]\times[0,1]$,

\STATE conditionally on $L$ sample Brownian bridge $X_{\psi_j}$,\label{step:LBB}
\vspace{-1ex}
\STATE evaluate $I=I(L, \gx, \Phi, \sp, \ep,\theta)$ as in
  \eqref{eq:ind_ea}; if $I=1$ then return $\lcb L, \Phi, S(X) \rcb$,
  otherwise go to 1.}\vspace{-0.2cm}

\end{algorithmic}
  
\end{spacing}

\end{algorithm}
\vspace{-0.3cm}

\subsubsection{The family of EAs}\label{sec:ea_family}

Each EA exploits the specific structure of the drift to construct
$L(X)$. The EA1 \citep{besk:papa:robe:2004} is the simplest EA type
and its framework is restricted by
\begin{condition}\label{cond:ea1_bound} $\phif{\cdot}$ is bounded
above.
\end{condition}
\noindent This condition ensures that
$\maxint{L}\equiv\maxinttheta{}$, implying there is no need for
constructing $L(X)$. As a consequence, step \ref{step:LBB} of
Algorithm \ref{alg:gen_CEA} merely requires simulation of a Brownian
bridge at time instances $\psi_1,\ldots,\psi_{\kappa}$.  The EA2
\citep{besk:papa:robe:2004} is applicable only when $\dd=1$ and
relaxes Condition \ref{cond:ea1_bound} to a more mild one:
\begin{condition}\label{cond:ea2_bound} Either
$\lim\sup_{u\rightarrow\infty}\phif{u}<\infty$ or
$\lim\sup_{u\rightarrow-\infty}\phif{u}<\infty$.
\end{condition}
\noindent For simplicity consider only the first case. The algorithm
constructs the proposed path by first simulating its minimum, say
$\minbb$, and subsequently the remainder of the path conditioned on
$\minbb$. In this setting $L(X)$ is defined as the two-dimensional
random variable $L(X)=\{\minbb,\locbb\}$, where $\locbb$ is the time
instance the minimum is attained. Then, the required upper bound is
found as $\maxint{L} = \sup_{u}\{ \phif{u}; u\geq \minbb\}.$
Simulating a Brownian bridge conditionally on its minimum is based on
a path transformation of two independent Bessel bridges, see
\cite{besk:papa:robe:fear:2006} for more details.

The EA3 \citep{besk:papa:robe:2008} poses no upper boundedness
conditions.  For the moment assume $\dd=1$ and consider a path $X$
with initial point $x$ and terminal point $y$ at time $t$.
The algorithm is based on creating a partition on the path space using
a series of lower and upper bounds. For a
user-specified constant $\delta > \sqrt{t/3}$ the partition consists
of the sets $C(\lh,x,y)$, $\lh\in\mN$, defined by
%
%
%
\begin{align*}
 &A(\lh,x,y) = \lcb \sup_{0\leq s \leq t}
\gx_s\in\lsb\yb+(\lh-1)\delta,
\yb+\lh\delta\rp
\rcb\cap\lcb \inf_{0\leq s \leq t} \gx_s > \xb-\lh\delta \rcb,\notag\\
&B(\lh,x,y) = \lcb \inf_{0\leq s \leq t} \gx_s\in\lp\xb-\lh\delta,
\xb-(\lh-1)\delta\rsb
\rcb\cap\lcb
\sup_{0\leq s \leq t} \gx_s < \yb+\lh\delta \rcb,\notag\\
&C(\lh,x,y)=A(\lh,x,y)\cup B(\lh,x,y),
\end{align*}
where $\xb, \yb$ as defined in the beginning of  Section
\ref{sec:pre}.

In the multidimensional case, sets $C_{\ci}(\lh):=C(\lh,x^\rr,y^\rr)$ are
constructed for each coordinate $\gx_s^\rr,\,\ci=1,\ldots,\dd$, and
$L(\gx)$ is defined as the $\dd$-dimensional discrete random variable
$L(\gx)=(L^{\{1\}},\ldots,L^{\{d\}})$ where $L^\rr=\lh_{\ci}$,
$\lh_{\ci}\in\mN$, if $X \in \cap_{\ci=1}^{\dd} C_{\ci}(
\lh_{\ci})$. Hence, $\{L^\rr\leq
\lh_{\ci},\,\ci=1,\ldots,\dd\}\equiv\{ \xb^\rr-\lh_{\ci}\delta <
\gx_s^\rr < \yb^\rr + \lh_{\ci}\delta,\,0\leq s\leq
t,\,\ci=1,\ldots,\dd\}$. Figure \ref{fig:layers_t_a} illustrates the
construction for an arbitrary coordinate.

The random variable $L$ is referred to as the \textit{Brownian bridge
  layer} and a Brownian bridge path conditioned on this layer as the
\textit{layered Brownian bridge}. Conditioned on $L$, and using the
continuity of $\phif{\cdot}$, the Poisson rate can be found as
\begin{align*}
 \maxint{L} = \sup\lcb \phif{u};u^\rr\in(\xb^\rr-L^\rr\delta,
\yb^\rr+L^\rr\delta),\,\ci=1,\ldots,\dd \rcb.
\end{align*}
The exact mathematical and implementation details of sampling a
layered Brownian bridge can be found in the original paper.


\subsubsection{Computational considerations}\label{sec:ea_cost}

The computational performance of EA depends on its
acceptance probability. For any two fixed points $x$ and $y$,
expression (\ref{eq:ea_ratio}) implies that the probability of
accepting a proposed path is
\begin{align}\label{eq:ea_prob}
 a(x,y,t,\theta) := \mE_{\cmW{t,x,y}}\lsb \exp\lcb-\int_0^t\phif{\gx_s}\d s
\rcb \rsb,
\end{align}
%
thus suggesting that the
acceptance probability decays exponentially to $0$ as
$\dd$ or $t$ increase. 

Finally, we note that although EA3 has the widest framework of
applicability, EA1 and EA2 should still be preferred whenever
possible; simulating a layered Brownian bridge is a non-trivial task
and is achieved by means of rejection sampling, thus adding to EA3 an
extra level of computational complexity. In particular, an extensive
empirical study by \cite{pel:robe:ea} suggests as a rule of thumb that
EA3 is approximately $10$ times slower than EA1.

\section{Data augmentation for discretely observed
  diffusions}
\label{sec:eda}

In this section we describe irreducible DA approaches for parameter
estimation and present formally our EDA scheme. The auxiliary
variables involved in {\eda} are intimately related to the EA output
for diffusion bridges and lead to MCMC algorithms which involve no
approximation to the statistical model of interest. At first, {\eda}
appears totally different from the {\ipi} paradigm of
\cite{robe:stra:2001}. However, there are close but subtle links and
the presentation in this section has been structured to naturally
bring those out. Effectively, the {\ipi} scheme, which is recalled
below, leads to two possibilities. One is its approximation by an
{\isi} with some finite $M$, which leads to bias. The other is to
consider an EA for the simulation of the auxiliary process identified
by \cite{robe:stra:2001} and identify finite-dimensional auxiliary
variables which then lead to the {\eda} scheme. Section
\ref{sec:emcmc} identifies those variables and Section
\ref{sec:emcmc_integrated} gives a theorem which establishes the
connection between {\ipi} and {\eda}.

\subsection{Irreducible path imputation and finite-dimensional approximations}
\label{sec:ihfi}

We first outline the {\ipi} approach of \cite{robe:stra:2001}. This
scheme corresponds to the limiting case $M=\infty$, and leads to an
idealised, yet impossible to implement, MCMC algorithm which involves
imputing continuous path trajectories. The auxiliary processes are
obtained after two path-parameter transformations of the original
latent bridges. We then review the {\isi} scheme which is constructed
by approximating the {\ipi} scheme with some finite $M$. The {\isi}
scheme requires only condition {\LAM}, whereas {\eda} additionally
requires {\SMO}, {\GRA} and {\LBO} for being able to employ
EA. However, {\isi} can be considerably improved when {\SMO} and
{\GRA} are also satisfied.

Consider for the moment only two observations from the diffusion
process, $V_0 = \sp$ and $V_t = \ep$. The process is first transformed
as $V_s\rightarrow X_s = \lf{}{V_s}$ as in Section \ref{sec:trans}.
$X$-paths over bounded time increments only contain finite information
for $\theta$, since all parameters now relate with the drift, see
\eqref{eq:unit_SDE1}. The transformed path starts at $\xt{}$ and
terminates at $\yt{}$, which are both deterministic functions of
$\spar$ and the observations. This suggests that a DA scheme based on
$X$ will not work when $\spar$ is unknown, since a realisation of $X$
determines $\spar$ through its endpoints. An alternative way to see
the problem is to note that the collection of dominating measures $\{
\cmW{t,x,y}, \theta\in\Theta \}$ are mutually singular, and therefore
a Gibbs algorithm based on this augmentation would be trapped in the
support of one of these measures. This necessitates a further
reparametrisation from $X_s\rightarrow\tX_s$, where
\begin{align}\label{eq:path_rep}
 \tX_s := X_s - \lp 1-\frac{s}{t}\rp x(\spar) -
\frac{s}{t}y(\spar),\quad s\in[0,t],
\end{align}
which forces the path to start and finish at $0$ and essentially
transforms the
distribution $\cmQ{t,x,y}$ so that the dominating measure, now given
by $\cmWn{t,0,0}$, is
independent of $\theta$.  

The {\ipi} is based on imputing $\tX$. Accounting for all
observations, let $x_i(\spar)
:=\lf{}{V_{t_i}}$ and denote by $\tX_i=\{ \tX_{i,s},
s\in[0,\tinc{i}]\},\,i=1,\ldots,n$ the imputed paths. We introduce
\begin{align*}
  \Gir{t}{X} = \exp\lp \int_0^t \driftTr{}{X_s}\d X_s -
  \frac{1}{2}\int_0^t||\drift{}{X_s}||^2\d s\rp,\quad
  \mbb_{i,s}(\theta) = \lp 1-\frac{s}{\tinc{i}}\rp x_{i-1}(\spar) + \frac{s}{\tinc{i}}x_i(\spar),
\end{align*}
and note that the inverse transformation of \eqref{eq:path_rep},
$\tX_i\rightarrow X_i$ is given by $\tX_{i,s} + \mbb_{i,s}(\theta)$.
Then the joint posterior density of $\theta$ and imputed paths,
$\pi_{\ipi}(\theta, \{\tX_i, 1\leq i \leq n \}\mid \obs)$, is
proportional to
\begin{align}\label{eq:cd_lik}
&\pi(\theta)\prod_{i=1}^n\DD{V_{t_i}}\gdenscb { x_i(\spar) -
x_{i-1}(\spar)}{\tinc{i}}\Gir{\tinc{i}}{g_i(\tX_i;\theta)},
\end{align}
with respect to $\Leb^p\otimes_{i=1}^n\cmWn{\tinc{i},0,0}$, where
$g_i(\tX_i;\theta):=\{\tX_{i,s} +
\mbb_{i,s}(\theta),\,s\in[0,\tinc{i}]\}$. For a detailed
derivation, the reader is referred to Section 3 of
\cite{robe:stra:2001}.

The joint posterior density is not be computable since the augmented
paths cannot be represented by a finite number of variables, hence the
integrals cannot be computed. Instead, the paths are approximated by
vectors of size $M+2$, $\{ \tX_{i,j \Delta t_i/(M+1)},
j=0,\ldots,M+1\}$, and the integrals are approximated numerically,
typically by Riemann sums, to yield $\pi_{{\isi},M}(\theta, \{\tX_i,
1\leq i \leq n \}\mid \obs)$, where by an abuse of notation we let
$\tX_i$ denote the path and its discretisation. This introduces a bias
in the inference for $\theta$. The approximated posterior is targeted
by an MCMC algorithm which updates in turns $\theta$ and
$\tX_i,\,i=1,\ldots,n$ according to their conditional
densities. Crucially for the efficiency of the algorithm, the
auxiliary processes $\tX_i$ are independent over $i$ conditionally on
$\theta$ and $\obs$, and thus can be updated sequentially. Each update
is typically performed by proposing Brownian bridge skeletons and
accepting them according to \eqref{eq:cd_lik}.  Algorithm
  \ref{alg:hfi_mcmc} is a typical {\amcmc} implementation, where
updates of $\theta$ are obtained from a Metropolis-Hastings step with
proposal kernel $q$.

\begin{algorithm}
\caption{{\bf {\amcmc}}}
\label{alg:hfi_mcmc}

\begin{spacing}{1.4}

\begin{algorithmic}[1]{\tt\footnotesize

\STATE Choose $\theta^0$, set $\delta_i=\tinc{i}/(M+1)$ and $\tX_i^0 =
  \lcb \tX^0_{i,j\delta_i}\rcb_j$, $0\leq j \leq M+1$, $1\leq i\leq n$. Set
  $t=0$.

\STATE For $1\leq i \leq n$, $0\leq j \leq M+1$


\MSTATE{4ex} simulate $\sX_i = \lcb \sX_{i,j\delta_i}\rcb_j$ from
$\cmWn{\tinc{i},0,0}$, and sample $U\sim\Un(0,1)$,

\MSTATE{4ex} if
\begin{align*}
  U < \frac{\pi_{\isi,M}(\sX_i\mid
    \theta^t,\obs)}{\pi_{\isi,M}(\tX^t_i\mid \theta^t,\obs)}
\end{align*}
\hspace{4ex} then set $\tX_i^{t+1} = \sX_i$, else set $\tX_i^{t+1} = \tX_i^{t}$.

\STATE Sample $\ptheta\sim q(\theta^t,\cdot)$ and $U\sim\Un(0,1)$,

\STATE if 
\begin{align*}
  U < \frac{\pi_{\isi,M}(\ptheta, \{\tX^{t+1}_i, 1\leq i
    \leq n \}\mid \obs)~q(\ptheta,\theta^t)}{\pi_{\isi,M}(\theta, \{\tX^{t+1}_i,
1\leq i
\leq n \}\mid \obs)~q(\theta^t,\ptheta)}
\end{align*}
then set $\theta^{t+1}=\ptheta$, else set
$\theta^{t+1}=\theta^t$.

\STATE Set $t=t+1$ and go to 2.\vspace{-0.2cm}
}

\end{algorithmic}
\end{spacing}
\end{algorithm}

An alternative approximation to $\pi_{\ipi}$ exists if conditions
{\SMO}, {\GRA} and {\LBO}
are satisfied. In particular, by using
integration by parts we can transform the stochastic integrals in
$G_{\tinc{i}}$ into time integrals, and rewrite \eqref{eq:cd_lik}
as
\begin{align}\label{eq:cd_lik_ibp} &\pi(\theta)\exp\lsb
\adriftcb{x_n(\spar)} - \adriftcb{x_0(\spar)} - l(\theta)(t_n -
t_0)\rsb\notag\\
&\times\prod_{i=1}^n\DD{V_{t_i}}\gdenscb { x_i(\spar) -
x_{i-1}(\spar)}{\tinc{i}}\exp\lcb
-\int_0^{\tinc{i}}\phif{\tX_{i,s}+\mbb_{i,s}(\theta)} \d s \rcb.
\end{align}
The finite-dimensional approximation to \eqref{eq:cd_lik_ibp}, denoted
by $\bar{\pi}_{{\isi},M}$, will typically be less biased
than $\pi_{{\isi},M}$, as illustrated in Section \ref{sec:numerics}.
The relative {\amcmc} algorithm follows along the lines of Algorithm
\ref{alg:hfi_mcmc} by replacing $\pi_{{\isi},M}$ with
$\bar{\pi}_{{\isi},M}$.

\subsection{Exact Data Augmentation and MCMC}
\label{sec:emcmc}

The main contribution of this section is to demonstrate that an exact
rejection sampling algorithm for simulating diffusion bridges implies
appropriate auxiliary variables which can be used to design
  EMCMC algorithms for exact inference for diffusions.  We describe
the data augmentation and the EMCMC algorithm which corresponds to the
EA3 case, and later comment on the simplifications that arise when a
more basic EA is applicable to the model of interest.

The augmentation is first identified for a pair of consecutive
observations, $V_0=\sp$ and $V_t=\ep$, and is then extended to an
arbitrary number of observations by using the Markov property. This is
achieved using the two main tools developed so far. First, the two
path-parameter transformations for irreducible DA; recall that these
are $V \to X$ and $X \to \tX$, which can be written in one step as
\begin{align}
  \label{eq:path_tr}
  \tX_s = \lf{}{V_s} - \lp 1-\frac{s}{t}\rp \lf{}{\sp} -
\frac{s}{t}\lf{}{\ep},\quad s\in[0,t].
\end{align}

 Second, the EA for simulating an
$X$-bridge according to $\cmQ{t,x,y}$, with $x=x(\spar)$ and
$y=y(\spar)$ as in Section \ref{sec:ea}.  Let $\cmQt{\theta}{t}$
denote the measure induced by the linearly transformed bridge $\tX$,
when $X$ is drawn from $\cmQ{t,x,y}$. In passing, recall that when $X$
is drawn from $\cmW{t,x,y}$, $\tX$ is distributed according to
$\cmWn{t,0,0}$.

\begin{figure}[!t]
\captionsetup[subfigure]{margin=17pt, position=top, singlelinecheck=false,
captionskip=0pt}
\setlength{\abovecaptionskip}{7pt}
\scriptsize
\subfloat[]{\psfrag{x_0}{\hspace{-0.5ex}$0$}
\psfrag{x_1}[r]{$-\delta$}
\psfrag{x_2}[r]{$-2\delta$}
\psfrag{x_3}[r]{}
\psfrag{x_4}[r]{}
\psfrag{y_0}[l]{}
\psfrag{y_1}[l]{$\delta$}
\psfrag{y_2}[l]{$2\delta$}
\psfrag{y_3}[l]{}
\psfrag{y_4}[l]{}
\includegraphics[width=7.5cm, height=4.5cm]{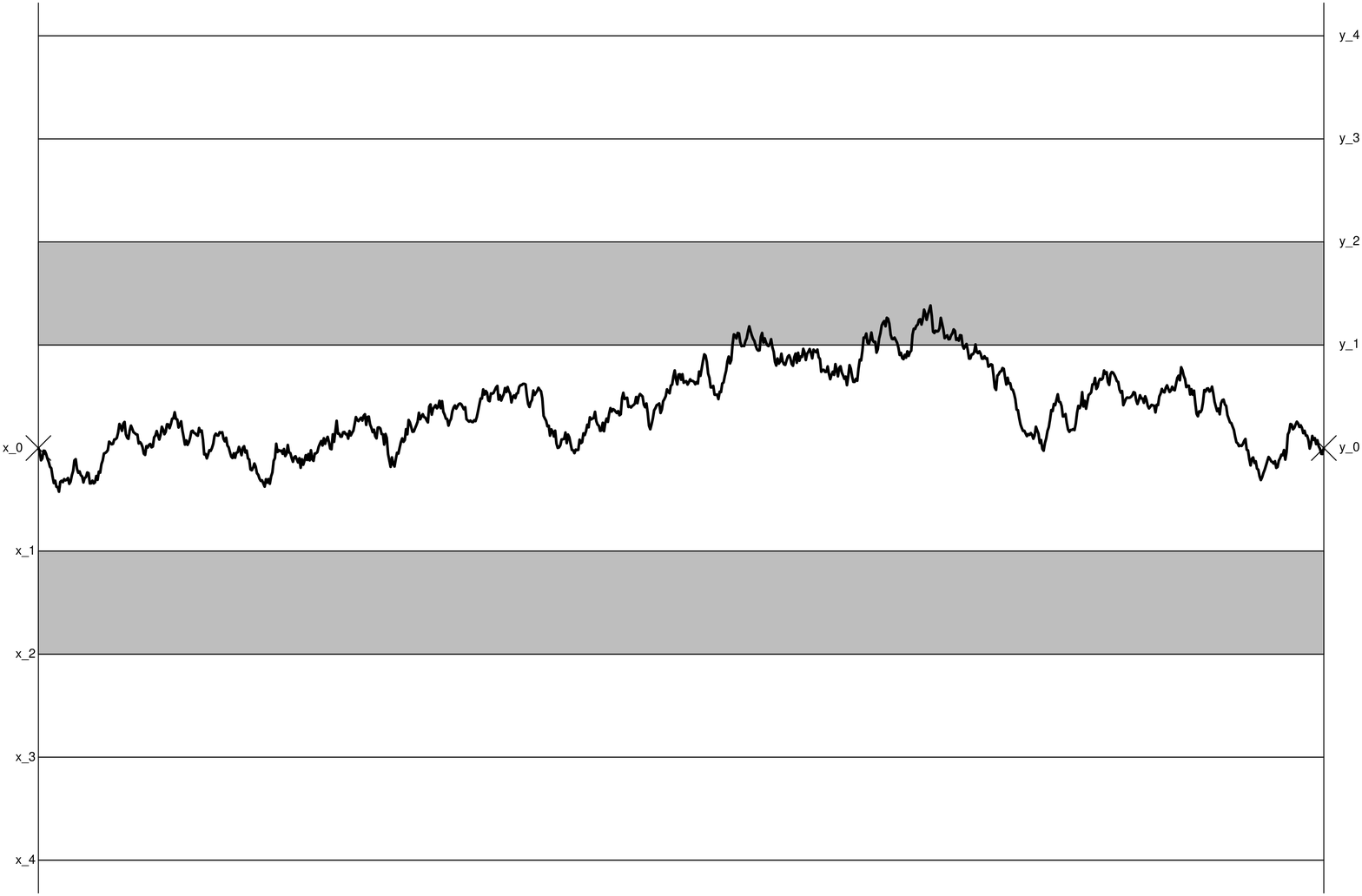}
\label{fig:layers_t_a}}\hspace{1ex}
 \subfloat[]{\psfrag{x_0}[r]{\hspace{-0.5ex}$\xtr{\ci}$}
\psfrag{x_1}[r]{\hspace{-0.5ex}$\xtr{\ci}-\delta$}
\psfrag{x_2}[r]{}
\psfrag{x_3}[r]{\hspace{-0.5ex}$\xtr{\ci}+\delta$}
\psfrag{x_4}[r]{}
\psfrag{y_0}[l]{\hspace{0.5ex}$\ytr{\ci}$}
\psfrag{y_1}[l]{\hspace{0.5ex}$\ytr{\ci}-\delta$}
\psfrag{y_2}[l]{}
\psfrag{y_3}[l]{\hspace{0.5ex}$\ytr{\ci}+\delta$}
\psfrag{y_4}[l]{}
\includegraphics[width=7.5cm, height=4.5cm]{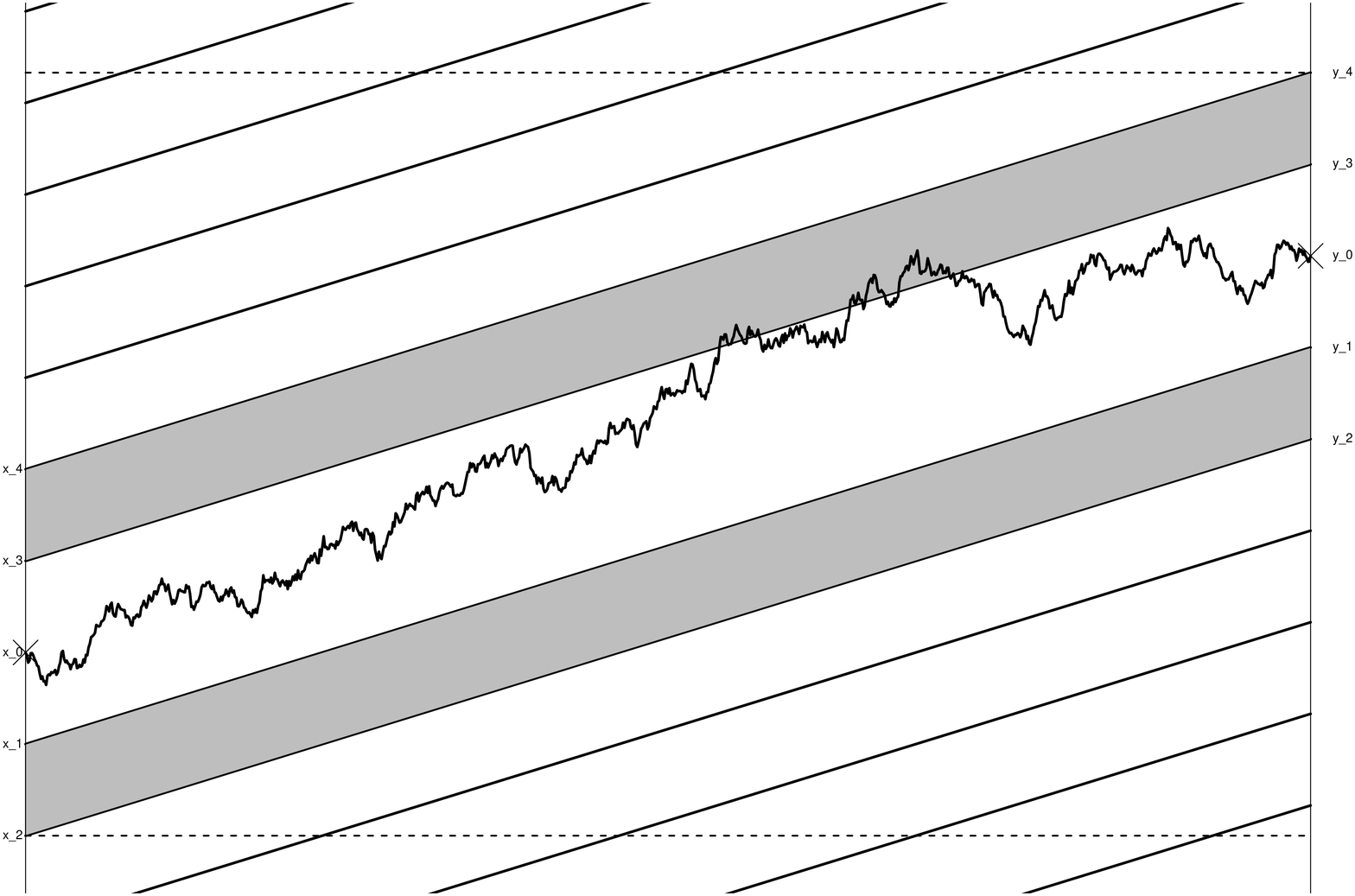}
\label{fig:layers_t_b}}
\caption[]{\subref{fig:layers_t_a} The $\ci$th coordinate of a Brownian
  bridge path $\plpath$ starting and terminating at $0$. In this
  example the event $C_{\ci}(2)$ has occurred. \subref{fig:layers_t_b}
  The transformed process $\plpath_s^\rr + \lp 1-s/t\rp \xtr{\ci} +
  s\ytr{\ci}/t$ starts at $\xtr{\ci}$ and terminates at $\ytr{\ci}$,
  with $\xtr{\ci}<\ytr{\ci}$. The dashed lines provide a lower and
  upper bound for the coordinate of the transformed path.}
  \label{fig:layer_BB}
\end{figure}

We first sketch an EA for simulating from $\cmQt{\theta}{t}$ using
proposals from $\cmWn{t,0,0}$. This is a minor modification of the EA
for $X$, since a proposed path $\tX\sim\cmWn{t,0,0}$ is accepted as a
path from $\cmQt{\theta}{t}$ if and only if $\tX_s + ( 1-s/t)
x + s y/t,\,s\in[0,t]$, is accepted as a path from
$\cmQ{t,x,y}$.  Let $\player$ denote the layer of a Brownian bridge
that starts and terminates at $0$, and denote by $\cmMn{t}$ the joint
law of $(\alayer,\alpath)$; hence if $(\alayer, \alpath)
\sim\cmMn{t}$, marginally $\alpath \sim \cmWn{t,0,0}$. The pair
$\player,\plpath$ imply realisations for the corresponding variables
in the $X$-space. These are easy to obtain, since conditionally on a
given $\player$, $\{\plpath_s^\rr,s\in[0,t]\}$ moves within
$(-\player^\rr\delta, \player^\rr\delta)$, and thus
\begin{align*} \cor{\xb}{\ci}(\spar) - \player^\rr\delta <
\plpath_s^\rr + \lp 1-\frac{s}{t}\rp \cor{x}{\ci}(\spar) +
\frac{s}{t}\cor{y}{\ci}(\spar) < \cor{\yb}{\ci}(\spar) +
\player^\rr\delta,
\end{align*}
where we recall that $\cor{\xb}{\ci}(\spar)=\cor{x}{\ci}(\spar)\wedge
\cor{y}{\ci}(\spar)$ and
$\cor{\yb}{\ci}(\spar)=\cor{x}{\ci}(\spar)\vee
\cor{y}{\ci}(\spar)$. The above construction and the derivation of the
bounds are illustrated in Figure \ref{fig:layer_BB}.

\graphicspath{{Figures/}}

An EA which samples from $\cmQt{\theta}{t}$ follows easily. If
$\aPois=\{\aPsi,\aUps\}$ is a homogeneous Poisson process of intensity
$\maxint{\player}$ on $[0,t]\times[0,1]$, where $\aPsi$ is the
projection of the points on $[0,t]$ and $\aUps$ the projection on
$[0,1]$, and
\begin{align}\label{eq:poisson_rate_ea3}
 \maxint{\player} = \sup\lcb \phif{u};u^\rr\in(\bxt-\player^\rr\delta,
\byt+\player^\rr\delta),\ci=1,\ldots,\dd \rcb,
\end{align}
then the algorithm accepts the proposed  $(\player,
\plpath, \pPois)$ according to
\begin{align}\label{eq:ind_ea3}
 I(\player,\plpath, \pPois,\sp,\ep,\theta):=
\prod_{j=1}^{\pkappa}\ind{\lsb\frac{1}{\maxint{\player}}\phifcb{\plpath_{\ppsi_j
}
+\lp 1-\frac{\psi_j}{t}\rp\xt{} + \frac{\psi_j}{t}\yt{}}
<\pups_j\rsb },
\end{align}
which is the familiar indicator function (\ref{eq:ind_ea}), reformulated in
terms of $(\player,\plpath,\pPois)$. 

The {\eda} scheme for a pair of observations $V_0, V_t$ is now defined
by the random variables $(\player,\plpath,\pPsi)$. Note that we can
avoid augmenting $\aUps$ and still obtain a tractable density. The
density of the auxiliary variables is derived in the following lemma
proved in the Appendix.

\begin{lemma}
\label{lem:aux-den}
Let $(\player,\plpath)\sim\cmMn{t}$ and
$\pPsi$ be a homogeneous
Poisson process of intensity $\maxint{\player}$ on $[0,t]$. If $I$
is the
acceptance indicator in (\ref{eq:ind_ea3}), then the conditional density
of $(\player, \plpath,\pPsi)$ given $I=1$, $\pi(\player, \plpath, \pPsi\mid
u,v,\theta)$, is
\begin{align}\label{eq:md_density_ea3}
\frac{\maxint{\player}^{\pkappa}}{a(x,y,t,\theta)}\exp\lcb
t\lsb 1
-
\maxint{\player}\rsb\rcb\prod_{j=1}^{\pkappa}\lsb1-\phifcb{\plpath_{\ppsi_j
}
+\lp 1-\frac{\psi_j}{t}\rp\xt{} + \frac{\psi_j}{t}\yt{}}/\maxint{\player}\rsb, 
\end{align}
%
with respect to the product measure $\cmMn{t}\times\mP{t}$, where
$\mP{t}$
is
the measure of a homogeneous Poisson process on $[0,t]$ with unit
intensity, and
$a(x,y,t,\theta)$ is the acceptance probability of the EA.
\end{lemma}
%

Extending the augmentation scheme to account for all observations is
straightforward. Specifically, recall that $x_i(\spar)
=\lf{}{V_{t_i}}$ and let $\alayer_i,\,\aPsi_i,\,\alpath_i=\{
\alpath_{i,s},s\in[0,\tinc{i}] \}$ denote the accepted elements of EA
applied to the interval $[t_{i-1}, t_i]$, for $1\leq i\leq n$. The
Markov property of the diffusion process implies that the bridges
conditionally on the observations are independent, and thus
%
$ \pi (\{\alayer_i, \alpath_i,\aPsi_i,1\leq i \leq n\}\mid \obs,\theta) =
\prod_{i=1}^n
\pi
(\alayer_i, \alpath_i,\aPsi_i\mid V_{t_{i-1}}, V_{t_i}, \theta)\,.$
%

We complete the development of {\eda} with the following theorem which
specifies the joint density of data, auxiliary variables and
parameters. This has a simple computable form and it admits the target
posterior $\pi(\theta \mid \obs)$ as a marginal with respect to the
auxiliary variables and conditional with respect to the data. The key
observation is that the joint density is only a function of the
finite-dimensional $\{ S(\alpath_i),\alayer_i, 1\leq i \leq n \}$,
which are delivered by the EA. Additionally, the intractable
normalising constants have been cancelled out. The proof of the
theorem is given in the Appendix.

\begin{theorem}\label{th:mcmc_ea3_ca} 
  The joint density of data $\obs$, the $p$-dimensional
  parameters $\theta$ and auxiliary variables $\{\alayer_i,
  \alpath_i,\aPsi_i,1\leq i \leq n\}$, is given below with respect to
  the $\theta$-independent dominating measure
  $\Leb^{n+p}\otimes_{i=1}^n\lp\cmMn{\tinc{i}}\times\mP{\tinc{i}}\rp$:
\begin{align}
\label{eq:eda-den}
& \pi(\obs,\theta,\{\skel{\alpath_i}, \alayer_i, 1\leq i \leq
n\}) = \pi(\theta) \prod_{i=1}^n \trans{\Delta t_i}{V_{t_{i-1}}}{V_{t_i}} \prod_{i=1}^n \pi
(\alayer_i, \alpath_i,\aPsi_i\mid V_{t_{i-1}}, V_{t_i}, \theta) =
\nonumber \\
&\pi(\theta)\exp\lp
\adriftcb{x_n(\spar)} - \adriftcb{x_0(\spar)} -
\lsb l(\theta) - 1\rsb (t_n-t_0) -
\sum_{i=1}^n\maxint{\alayer_i}\tinc{i}\rp\notag\\
&\times
\prod_{i=1}^n\lcb\DD{V_{t_i}}\gdenscb{x_i(\spar) - x_{i-1}(\spar)}{\tinc{i}}
\maxint{\alayer_i}^{\akappa_i}
%
\prod_{j=1}^{\akappa_i}\lsb 1 -
\phifcb{\alpath_{i,\apsi_{i,j}}+\mbb_{i,\apsi_{i,j}}(\theta)
}/\maxint{\alayer_i}
\rsb\rcb,
\end{align}
and it admits (\ref{eq:post}) as a marginal when $\{\alayer_i,
\alpath_i,\aPsi_i,1\leq i \leq n\}$ is integrated out and $\obs$
conditioned upon. 
\end{theorem}


This density can be targeted by MCMC methods; actually at this stage
we are only interested in the conditional density given $\obs$, i.e.,
the joint posterior of parameters and auxiliary variables.  We
advocate a Gibbs sampler variant since conditionally on $\obs$ and
$\theta$ the auxiliary variables $\{\skel{\alpath_i}, \alayer_i, 1\leq
i \leq n\}$ are independent over $i$ and can be generated using the
EA. The conditional density of $\theta$ is computable and if it
cannot be directly sampled, a Metropolis-Hastings step can by
employed. Depending on the EA type used to construct the
  augmentation scheme, we distinguish between EMCMC1, 2 and 3. A
typical implementation of EMCMC3 is given in Algorithm
\ref{alg:emcmc3}.

\begin{algorithm}
\caption{\textbf{EMCMC3}}
\label{alg:emcmc3}

\begin{spacing}{1.4}

\begin{algorithmic}[1]\tt\footnotesize
\STATE Choose $\theta^{0}$ and set $t=0$.

\STATE For $1\leq i\leq n$, set $I_i=0$ and repeat the following until $I_i=1$,

\MSTATE{4ex} sample layer $\player_i$ of Brownian bridge path
$\plpath_i\sim\cmWn{\tinc{i},0,0}$,

\MSTATE{4ex} sample $\pkappa_i\sim\Po[r(\player_i;\theta^t)\tinc{i}]$ and $\pPois_i=\{(\ppsi_{i,j},\pups_{i,j})\}_j,\,1\leq
  j\leq \pkappa_i$, uniformly on
$[0,\tinc{i}]\times[0,1]$,

\MSTATE{4ex} conditionally on $\player_i$, sample Brownian bridge
$\plpath_{i,\ppsi_{i,j}}$,

\MSTATE{4ex} set $I_i = I( \player_i, \plpath_i,
  \pPois_i,V_{t_{i-1}},V_{t_i},\theta^t)$ as in (\ref{eq:ind_ea3}); if
  $I_i=1$, then set $\alayer_i^{t+1} =
  \player_i,\,\aPsi_i^{t+1}=\pPsi_i,\,\alpath_i^{t+1}=\plpath_i$.
 
\STATE Sample $\ptheta\sim q(\theta^t,\cdot)$ and $U\sim\Un[0,1]$,

\STATE if 
$$U<\frac{\pi(\obs,\ptheta,
\{\skel{\alpath_i^{t+1}},\alayer_i^{t+1}, 1\leq i \leq
n\})~q(\ptheta, \theta^t)}{\pi(\obs,\theta^t,
\{\skel{\alpath_i^{t+1}}, \alayer_i^{t+1}, 1\leq i \leq
n\})~q(\theta^t, \ptheta)}$$
then set $\theta^{t+1}=\ptheta$, else set $\theta^{t+1}=\theta^t$.

\STATE Set $t=t+1$ and go to 2.\vspace{-0.2cm}
\end{algorithmic}
  
\end{spacing}

\end{algorithm}
\vspace{-0.3cm}

\subsubsection*{Special cases: EMCMC1 and EMCMC2}

Certain simplifications are feasible when a simpler EA can be applied
to the process of interest.  The {\eda} based on EA1 requires less
imputation than that of EA3, since exact simulation from
$\cmQt{\theta}{t}$ no longer requires the variable $\player$. Thus,
the augmentation scheme involves only $\{\alpath_i,\aPsi_i,1\leq i
\leq n\}$. The joint density analogous to (\ref{eq:eda-den}) can be
easily obtained and amounts to simply replacing $\maxint{\alayer}$ by
$r(\theta)$. The conditional of $\theta$ trivially follows; originally
it was given in Theorem 3 of \cite{besk:papa:robe:fear:2006}.

A DA scheme can be built using the auxiliary variables used in
EA2. We do not present the details of this, since it is not a direct
modification of the general scheme, as it is the case with EA1, but
instead involves a different construction of bridges. Details can be
found in Chapter 7 of \cite{sermaidis:thesis}.

\subsection{Interpreting {\eda} in terms of {\ipi}}\label{sec:emcmc_integrated}

The following result provides the connection between {\eda} and
{\ipi}. It effectively shows that the {\ipi} scheme is a
\emph{collapsed} version of {\eda}, i.e., when we integrate out a
subset of the latent variables we obtain the distribution which is
targeted by {\ipi}.
%
\begin{theorem}
  \label{th:connection} Let $\pi(\theta,\{ \tX_i,1\leq i \leq n \}
  \mid \obs)$ be the density obtained from (\ref{eq:eda-den}) by
  conditioning on $\obs$, and marginalising with respect to $\{
  \alayer_i,\aPsi_i,1\leq i \leq n \}$, and $\pi_{\ipi}(\theta,
  \{\tX_i, 1\leq i \leq n \}\mid \obs)$ the density targeted by the
  irreducible path imputation algorithm defined in
  (\ref{eq:cd_lik_ibp}). Then, the two densities are equal a.s.
\end{theorem}

The result is insightful towards the comparison of the computational
efficiency of {\emcmc} to that of {\amcmc}, as it suggests that the
mixing time of the former might generally be larger due to its higher
degree of augmentation; a price one has to pay in order to eliminate
the discretisation error. Nonetheless, Section \ref{sec:repar} shows a
variety of ways with which one can increase the performance of {\emcmc}
and achieve good mixing rates. A numerical comparison of {\emcmc} and
{\amcmc} is investigated in Section \ref{sec:numerics} in concrete
examples.

\subsection{Qualitative characteristics of EMCMC and
  {\amcmc}}
\label{sec:emcmc_eff}

We can make some qualitative remarks about the efficiency of the MCMC
schemes based on {\isi} and {\eda}.  These remarks are based on
general properties of DA methods, as for example discussed in
\cite{papa:robe:2007,meng:yun} but also the particular structure of
the models at hand.  These qualitative statements are
backed up by numerical evidence in Section \ref{sec:numerics}, and
they motivate the three approaches we propose in the next section to
boost the algorithmic efficiency.

To fix terminology, we will identify DA with a Gibbs-type sampler
which updates auxiliary variables and parameters according to their
conditional distributions. In general, DA works well when the
fraction of missing information is not too large relative to the
observed one, i.e., when the augmented dataset is not
considerably more informative than the observed regarding the
parameters. 

In that respect, both {\emcmc} and {\amcmc} become more
efficient when the time increment $t$ between a pair of observations
decreases. In the context of {\amcmc}, this is due to the fact that as
$t$ decreases,  the latent
bridges increasingly look like Brownian 
bridges, hence they do not carry information about the drift over and
above the one contained in the observed endpoints and the augmented information converges to
the observed one.  The efficiency of {\emcmc} improves as
the Poisson rate, $\maxint{\alayer} t$, decreases. 
In the limiting case when
the rate is $0$, the missing data and parameters are
independent, since the skeleton is empty, and {\emcmc} achieves maximal
efficiency. 

On the other hand, as $t$ increases, the augmented paths contain
increasing amount of information about the parameters relative to the
information content in the observed data, therefore any algorithm
which iteratively simulates from the conditional distributions of
parameters and missing data will degrade in this sparse-data limit,
even if the conditional distributions can be simulated exactly and
efficiently. {\eda} has a further weakness over {\isi} because the
number of points in a EA skeleton is informative about
$\maxint{\alayer}$, and thus about $\theta$. However, we can deal with
this dependence, which increases with $t$, by means of a
reparametrisation that is described in Section \ref{sec:repar}.

Additionally, both {\emcmc} and {\amcmc} suffer at the step of
updating missing data given parameters as $t$ increases. In {\emcmc}
the acceptance rate of EA decays to $0$ exponentially with $t$,
implying that the algorithm can spend a large amount of time by
simulating proposed paths until acceptance. Similarly in {\amcmc} the
acceptance rate of the independent Metropolis-Hastings step decays to
$0$ at the same rate (expression \eqref{eq:cd_lik_ibp}), suggesting
that the algorithm will be rarely updating the last accepted path,
thus leading to a slower exploration of the state space. However, this
problem can be improved by resorting to alternative update schemes for
the imputation step. One option for {\amcmc} is to use local
algorithms, see for example \cite{robe:stu:2012}. An other
alternative, which can apply to both algorithms, is to update the
paths in smaller time segments by imputing ${\cal O}(t)$ additional
points between pairs of observations. This can turn the complexity
from exponential to linear in $t$.

\section{Boosting EMCMC efficiency}\label{sec:repar}


\subsection{Noncentred reparametrisation}

One general approach for improving efficiency of data augmentation in
hierarchical models and auxiliary variable models is to adopt a
reparametrisation. Indeed, we have already done so in {\isi} and in
{\eda} by transforming $V \to \tX$ as in
\eqref{eq:path_tr}. Following \cite{papa:robe:2007}, for a generic
random variable $E$ and data $\obs$, a reparametrisation of an
augmentation scheme $E$ is defined by any random pair $(\tE,\theta)$
together with a function $h$ such that
%
 $E = h ( \tE, \theta, \obs )$,
%
where $h$ need not be 1-1. A reparametrisation is called {\it
  noncentred} when the distribution of $\tE$ is independent of
$\theta$. Intuitively, in cases where $\obs$ is not strongly
informative about $E$, a noncentred scheme can perform well due to the
prior independence of $\tE$ and $\theta$. We will attempt to reduce
the dependence between the Poisson process and $\theta$ by resorting
to a noncentred reparametrisation.


Noncentred reparametrisations for decoupling the dependence between
Poisson processes and their intensity were originally proposed in
\cite{robe:papa:della:2004}. Applying their idea in this context, for
two observations $V_0=\sp,\,V_t=\ep$, if $\pPsi$ is a Poisson process
of rate $\maxint{\player}$ on $[0,t]$ and $\pPoisnc$ is a Poisson
process of unit intensity on $[0,t]\times[0,\infty)$ with
point-coordinates $\{(\ppsinc_j, \pxinc_j)\}$ then
\begin{align}\label{eq:rep_pois}
 \pPsi = h (\pPoisnc,\alayer, \theta) = \lcb \ppsinc_j;\,
\pxinc_j<\maxint{\player}\rcb.
\end{align}
Although $\pPoisnc$ includes an infinite number of points, $\pPsi$
only depends on those for which the
second coordinate is below $\maxint{\player}$.
Accounting for all the observations, the noncentred reparametrisation
is $(\theta, \{\alayer_i,\alpath_i,\aPsi_i,1\leq i\leq
n\})\rightarrow(\theta, \{\alayer_i,\alpath_i,\pPoisnc_i,1\leq i\leq
n\})$, where $\pPsi_i = h (\pPoisnc_i,\player_i,\theta)$. The theorem
below derives the conditional density of $\theta$ given the latent
variables and observations, and is proved in the Appendix.

\begin{theorem}\label{th:mcmc_ea3_nca}
The conditional density of $\theta$ given the auxiliary variables $\{\alayer_i,
  \alpath_i,\pPsinc_i,1\leq i \leq n\}$, $\pi_{nc}(\theta\mid \{\alayer_i,
  \alpath_i,\pPsinc_i,1\leq i \leq n\}, \obs)$ is proportional to
\begin{align}\label{eq:mcmc_ea3_nca}
&\pi(\theta)\exp\lsb
\adriftcb{x_n(\theta)} - \adriftcb{x_0(\theta)} -
l(\theta)(t_n-t_0) \rsb\times
\prod_{i=1}^n\DD{V_{t_i}}\gdenscb{x_i(\spar) - x_{i-1}(\spar)}{\tinc{i}}
\notag\\
&\times \prod_{i=1}^n\prod_{j=1}^{\infty}\lsb 1
-\ind{\lsb\axinc_{i,j}<\maxint{\alayer_i}\rsb}
\phifcb{\alpath_{i,\apsinc_{i,j}}+\mbb_{i,\apsinc_{i,j}}(\theta)
}/\maxint{\alayer_i}
\rsb.
\end{align}
\end{theorem}

\noindent Notice that evaluation of \eqref{eq:mcmc_ea3_nca} for any
value of $\theta$ requires only finite computation and therefore
discretisations are avoided.

The MCMC algorithm based on this reparametrisation is practically a
small modification of that based on the original scheme (Algorithm
\ref{alg:emcmc3}). First, we wish to draw from the distribution of
$\{\alayer_i, \alpath_i,\aPoisnc_i,1\leq i \leq n\}$ given $\obs$ and
the current parameter value, say $\theta^t$. It is clear from
(\ref{eq:rep_pois}) that $\pPoisnc_i$ need only be revealed on
$[0,\tinc{i}]\times[0,r(\player_i;\theta^t)]$, which involves a finite
number of points $\pkappanc_i\sim\Po[r(\player_i;\theta^t)\tinc{i}]$
and is sufficient for the implementation of the EA. The $i$th output
consists of $\aPoisnc_i$ partially observed at $\{
(\apsinc_{i,j},\axinc_{i,j}),1\leq j\leq \akappanc_{i}\}$ and the pair
$\{\alayer_i, \alpath_i\}$ discretely observed at times
$\apsinc_{i,j}$. However, sampling from the distribution of $\theta$
given the latent variables and $\obs$ is more tricky. Specifically, if
the proposed value, say $\ptheta$, is such that
$r(\alayer_i;\ptheta)>r(\alayer_i;\theta^t)$, then evaluating
(\ref{eq:mcmc_ea3_nca}) at $\ptheta$ requires revealing $\tX_i$ at
additional time points
$\{\apsinc_{i,j};r(\alayer_i;\theta^t)<\axinc_{i,j}<r(\alayer_i;\ptheta)\}$,
which have not been revealed in the EA output. Notice that this does
not occur when $r(\alayer_i;\ptheta)<r(\alayer_i;\theta^t)$.

We propose two ways to overcome this. The first is based on
prospectively revealing $\aPsinc_i$ at any additional time instances
by simply simulating extra $\tilde\kappa_i^*$ uniform random variates
on $[0,\tinc{i}]\times[r(\alayer_i;\theta^t), r(\alayer_i;\ptheta)]$,
where $\tilde \kappa_i^*\sim\Po\lcb\lsb
r(\alayer_i;\ptheta)-r(\alayer_i;\theta^t)\rsb\tinc{i}\rcb$. The path
$\alpath_i$ is then filled in at the additional points by Brownian
bridge interpolations. The second is closest in spirit to the
retrospective nature of the EA. In particular, $\ptheta$ can be
simulated prior to the application of the EA and therefore
$r(\alayer_i;\theta^t)$ and $r(\alayer_i;\ptheta)$ are known before
the simulation of the latent path. Consequently, we can simulate
$\aPoisnc_i$ directly on $[0, \tinc{i}]\times[0,r(\alayer_i;\ptheta)]$
and reveal $\plpath_i$ at all required time instances during the
implementation of the EA.

In this paper, we adopt the retrospective approach because it can be
applied in a similar fashion to all three EAs. The prospective
approach is simple in the EA1 case due to simple Brownian bridge
interpolations, but becomes more involved in the EA2 and EA3 cases. 

Again, a simplification can be achieved when EA1 is applicable, where
the transformation in that case becomes $\{\apath_i,\aPsi_i,1\leq
i\leq n\}\rightarrow\{\apath_i,\aPoisnc_i,1\leq i\leq n\}$, where
$\aPsi_i = h (\aPoisnc_i,\theta)=\{\ppsinc_{i,j};\,
\pxinc_{i,j}<r(\theta) \}$. The full conditional density of $\theta$
is essentially given by expression (\ref{eq:mcmc_ea3_nca}) replacing
$\maxint{\alayer_i}$ with $r(\theta)$. Noncentred reparametrisations
for EMCMC2 also exist; details can be found in Section 7.4.3 of
\cite{sermaidis:thesis}.


\begin{algorithm}
\caption{\textbf{Noncentred EMCMC3}}
\label{alg:emcmc3_nca}

\begin{spacing}{1.4}
  \begin{algorithmic}[1]\tt \footnotesize
\STATE Choose $\theta^{0}$ and set $t=0$.

\STATE Sample $\ptheta\sim q(\theta^t,\cdot)$ and $U\sim\Un[0,1]$.

\STATE For $1\leq i\leq n$, set $I_i=0$ and repeat the following until $I_i=1$,

\MSTATE{4ex} sample layer $\player_i$ of Brownian bridge path
$\plpath_i\sim\cmWn{\tinc{i},0,0}$, and set $r_{max} = r(\player_i;\theta^t)\vee r(\player_i;\ptheta)$,

\MSTATE{4ex} sample $\pkappanc_i\sim\Po[r_{max}\tinc{i}]$ and
  $\{(\ppsinc_{i,j},\pupsnc_{i,j},\pxinc_{i,j})\}_j,\,1\leq j\leq \pkappanc_i$
  uniformly on $[0,\tinc{i}]\times[0,1]\times[0,r_{max}]$, 
  
\hspace{4ex}  set $\pPsinc_i=\{ (\ppsinc_{i,j},\pxinc_{i,j})\}_j,\,\pUpsnc_i=\{
  (\pupsnc_{i,j},\pxinc_{i,j})\}_j$,

\MSTATE{4ex} set $\pPois_i=\{ \pPsi_i, \pUps_i\}$, where
  $\pPsi_i=h(\pPsinc_i,\player_i,\theta^t)$ and
  $\pUps_i=h(\pUpsnc_i,\player_i,\theta^t)$ as in (\ref{eq:rep_pois}),

\MSTATE{4ex} conditionally on $\player_i$, sample Brownian bridge
$\plpath_{i,\ppsinc_{i,j}}$,

\MSTATE{4ex} set $I_i = I( \player_i, \plpath_i,
\pPois_i,V_{t_{i-1}},V_{t_i},\theta^t)$ as
in (\ref{eq:ind_ea3}); if $I_i=1$, then set $\alayer_i^{t+1} = \player_i$,  
$\aPsinc_i^{t+1}=\pPsinc_i$ and
$\alpath_i^{t+1}=\plpath_i$.
 
\STATE If 
$$U<\frac{\pi_{nc}(\ptheta\mid
\{\alayer_i^{t+1}, \alpath_i^{t+1}, \aPsinc_i^{t+1}, 1\leq i \leq
n\},\obs)~q(\ptheta, \theta^t)}{\pi_{nc}(\theta^t\mid
\{\alayer_i^{t+1}, \alpath_i^{t+1}, \aPsinc_i^{t+1}, 1\leq i \leq
n\},\obs)~q(\theta^t, \ptheta)}$$
then set $\theta^{t+1}=\ptheta$, else set $\theta^{t+1}=\theta^t$.

\STATE Set $t=t+1$ and go to 2.\vspace{-0.2cm}
  \end{algorithmic}
\end{spacing}
\end{algorithm}
\vspace{-0.3cm}

\subsubsection*{An interweaving strategy}

When a noncentred transformation is available, it is not necessary to
choose between that and the original parametrisation.  \cite{meng:yun}
propose instead to \emph{interweave} the two, by creating a single
algorithm which mixes steps of both algorithms. This requires
practically no extra coding work, but as it is demonstrated in the
article, in certain cases the interweaved algorithm outperforms its
parent algorithms even when taking the added computational cost into
account.

In the EMCMC context, if $\theta^t$ and
$\{\alayer_i^t,\alpath_i^t,\aPoisnc_i^t,1\leq i\leq n\}$ denote the
current state of the chain, then the algorithm can effectively be
described in four steps. The first two are identical to sampling from
the noncentred algorithm, i.e., the latent variables are updated to
$\{\alayer_i^{t+1},\alpath_i^{t+1},\aPoisnc_i^{t+1},1\leq i\leq n\}$
using the EA and then the parameter is updated by drawing
$\theta^{t+\frac{1}{2}}$ conditionally on these latent data and
observations.  Subsequently, the latent data are transformed to their
original parametrisation using
$\aPsi_i=h(\aPoisnc_i^{t+1},\theta^{t+\frac{1}{2}})$; notice that this
step does not impose any computational difficulties, since it merely
involves a deterministic transformation. Finally, the parameter is
re-drawn under the original parametrisation,
$\theta^{t+1}\sim\pi(\cdot\mid \{S(\alpath_i^{t+1}),\alayer_i^{t+1},
1\leq i \leq n\}, \obs)$.

\subsection{Auxiliary Poisson sampling}
\label{sec:aux}

An alternative way to improve the mixing time by increasing the computational
cost is to exploit the connection between {\eda} and {\ipi}. For simplicity
we present the result for EMCMC1, and then
discuss extensions to EMCMC3.  

Note that if $r(\theta)$ is the Poisson rate, then it is valid to
apply EA1 with any Poisson sampling rate $R(\theta)>r(\theta)$; the
acceptance probability is invariant to that choice. Increasing the
value of $R(\theta)$ leads to an increase in the computational cost
since the number of points at which the path is evaluated gets
larger. Thus, in terms of computing time it is optimal to implement
the algorithm with the smallest possible $R(\theta)$. This is not the
case though for EMCMC1 that iterates between imputation and
estimation. As it turns out, the dependence between missing data and
parameters decreases with $R(\theta)$ and optimal implementation in
terms of execution time and Monte Carlo error can be achieved for
$R(\theta)>r(\theta)$. Thus, we consider data augmentation where the
auxiliary variables are chosen according to the output of EA1, as
described in Section \ref{sec:emcmc}, but where the Poisson rate is
$R(\theta)$.


A theoretical result (not included here) goes along the following
lines. Let $R(\theta)=r(\theta)+\pint$, where $\pint\geq 0$ is a
  user-specified constant independent of $\theta$. Then, the joint
law of $(\theta,\{\tX_i,\aPsi_i,1\leq i \leq n \})$ after a step of
the EMCMC1 algorithm with parameter $\lambda$, converges to the law of
one step of {\ipi} when $\lambda \to \infty$. An argument for proving
this is based on properties of series expansions for exponential
functionals, as discussed for example in \cite{pap}.

Hence, in a sense the auxiliary variables $\aPsi_i$ are effectively
integrated out by increasing computation and an improvement in the
convergence of EMCMC is expected as $\lambda$ increases. Similar
  arguments are valid for the noncentred algorithm since it merely
  involves a reparametrisation of $\aPsi_i$.
%
A similar property is enjoyed by a generic EMCMC3. In fact, the
limiting algorithm as $\lambda$ increases is a {\ipi} which also
imputes the layer, although the latter is immaterial in the limit
since it does not contain additional information about the parameters.


\section{Diffusion observed with error}\label{sec:error}

The methodology described so far can be easily extended to account for
cases where the diffusion is not directly observed. We
assume that the observations inform only indirectly about the value of
the process  (\ref{eq:model}) at discrete times  $t_i,i=0,1,\ldots,n$,
according to the following 
observation equation
\begin{align*}
 {\nobs}_{t_i} \sim \obd (\cdot\mid V_{t_i},\tau),
\end{align*}
where $\nobs:=\{\nobs_{t_0},\nobs_{t_1},\ldots,\nobs_{t_n}\}$ are conditionally independent
given $\obs=\{V_{t_0},
V_{t_1},\ldots,V_{t_n}\}$, and $\obd$ is a known density function
parametrised by an unknown parameter $\tau$.

The {\eda} described in Section \ref{sec:emcmc} is not appropriate
anymore since the end points $V_{t_i}$ are not directly observed. On
the other hand, Theorem \ref{th:mcmc_ea3_ca} can be used to design an
augmentation scheme where apart from the auxiliary variables involved
in {\eda}, the latent points $\obs$ are imputed as well. A direct
application of Bayes' theorem yields that the joint posterior
$\pi(\obs,\theta,\tau,\{\skel{\alpath_i}, \alayer_i, 1\leq i\leq n\}
\mid \nobs)$ is proportional to
\begin{equation}\label{eq:joint-error} \pi(\theta,\tau) \,
\pi(\obs,\{\skel{\alpath_i}, \alayer_i, 1\leq i \leq n\} \mid \theta)
\, \pi(V_{t_0}\mid\theta) \, \prod_{i=0}^n\obd(Y_{t_i}\mid
V_{t_i},\tau),
\end{equation}
where the second term is obtained directly from Theorem
\ref{th:mcmc_ea3_ca}, $\pi(\cdot\mid\theta)$ is a prior density for
the initial point of the diffusion process, and $\pi(\theta, \tau)$ is
a prior density of the parameters. As before, a direct simplification
is available when the EA1 can be applied to simulate from $\tX_i$.

A simple scheme for simulating from (\ref{eq:joint-error}) is by a
component-wise updating algorithm.  $\{\skel{\alpath_i}, \alayer_i,
1\leq i \leq n\}$ and $\theta$ are simulated conditionally on $\obs$
and $\tau$, using any of the EMCMC schemes we have proposed, and
subsequently $\obs$ and $\tau$ conditionally on $\{\skel{\alpath_i},
\alayer_i, 1\leq i \leq n\} $ and $\theta$ according to the
conditional derived from (\ref{eq:joint-error}). When
$\pi(\theta,\tau)=\pi(\theta) \pi(\tau)$, $\tau$ is conditionally
independent from $\{\skel{\alpath_i}, \alayer_i, 1\leq i \leq n\} $
and $\theta$ given $\obs$, and may have a conditional density which
can be easily simulated. Simulation of the latent points $\obs$ can be
done with various ways. The simplest is to update them one-at-a-time
according to their conditional density, an approach often called
single-site updating. Such schemes for time series are known to be in
general problematic, especially when the latent process exhibits high
persistence and the observations are not very informative about the
latent points, see for instance \cite{pitt1999,papa:robe:2007}. In the
example we consider in the next section we adopt this simplistic
approach since it works quite well. In applications where the process
$\obs$ exhibits very high persistence a joint update of the endpoints
can be done using a Metropolis-adjusted Langevin algorithm, or more
general version of such algorithms, as discussed for example in
\cite{giro:riem}. Other possibility is to resort to an overlapping
block scheme, as for example in \cite{pitt1999,gol:wilk:2008}.

%

\setlength{\abovecaptionskip}{0pt}

 \section{Numerical investigation of MCMC algorithms}
\label{sec:numerics}

We investigate the numerical performance of the several algorithms we
have presented on some standard examples. One is a diffusion which
belongs in the Pearson family, see for example \cite{forman:sorensen},
and it is an example of a process that can be simulated using the
EA1. The second is a univariate double well potential model, typical
of models that are used to describe processes with metastable
behaviour, see for example \cite{Metzner:2006}. This is an example of
a process that can only be simulated using the EA3. The third is a
double well potential model in two dimensions.

We compare several algorithms. The plain-vanilla EMCMC together which
its elaborations: using noncentred reparametrisation, the interweaved
strategy and auxiliary Poisson sampling. Additionally, we compare
against both versions of {\amcmc} described in Section
\ref{sec:ihfi}. The first is the plain one as introduced in
\cite{robe:stra:2001} and is based on \eqref{eq:cd_lik}; we refer to
this basic version simply as {\amcmc}. The second eliminates the
stochastic integrals by integration by parts as in
\eqref{eq:cd_lik_ibp} and will be denoted in the text by ``{\amcmc}
(int-by-parts)''. This is expected to enhance algorithmic performance,
hence we evaluate the effect of this approach. For both versions, we
add the suffix -$M$ to indicate the number of imputed points. In
general, the quality of each MCMC output is assessed with an adjusted
effective sample size, defined as the ratio of the effective sample
size (ESS) to the computational (CPU) time to run each algorithm. The
adjusted ESS is essentially the number of independent draws per second
generated by the Markov chain. The ESS is calculated with the R
\citep{rlang} package {\tt coda} \citep{rcoda}.

The existence of {\emcmc} allows us to have a realistic evaluation of
the performance of the biased approaches. We carry out bootstrap
Kolmogorov-Smirnov tests \citep{rmatching} for checking whether the
marginal posterior distributions on the parameters obtained by
different levels of imputation are significantly different from the
exact samples. These evaluations are based on thinning the original
Markov chain output so that to obtain practically independent draws
from the corresponding distributions.

\subsection{A Pearson diffusion}

Consider  the univariate diffusion
process specified by
\begin{align*}
 \d V_s = -\rho (V_s-\mu)\d s + \sigma\sqrt{1+V_s^2}\d W_s,
\end{align*}
where $\sigma>0$, $\rho>0$ is a mean reverting parameter and
$\mu\in\mR$ is the stationary mean. The parameter vectors are
identified as $\dpar=(\rho,\mu)^{T}$ and $\spar=\sigma$. This model
belongs to a rich class of diffusion processes, known as the Pearson
diffusions, and admits a stationary distribution with skewness
and heavy tails that decay at the same rate as those of a $t$-distribution.
%
%

The unit volatility process is obtained as $X_s := \sinh(V_s)/\sigma$,
with drift given by
\begin{align*} a(x;\theta) = -\lp\frac{\rho}{\sigma} +
\frac{\sigma}{2}\rp\tanh(\sigma x) + \frac{\rho\mu}{\sigma\cosh(\sigma
x)}.
\end{align*}
This is a process which belongs in the EA1 class, and exact inference
can be performed with
\begin{align}\label{eq:tanh_lr} l(\theta) =
-\frac{1}{2}\lp\rho+\frac{\sigma^2}{2}+\frac{\rho\mu}{2}\rp,\quad
r(\theta) = \frac{1}{8}\lcb \rho(6\mu+8) + 3\sigma^2 +
\frac{4\rho^2}{\sigma^2}(\mu^2+\mu+1) \rcb.
\end{align}

\begin{figure}[!p]
\begin{center}
\subfloat[Pearson]{\includegraphics[width=\myw, height=\myh]{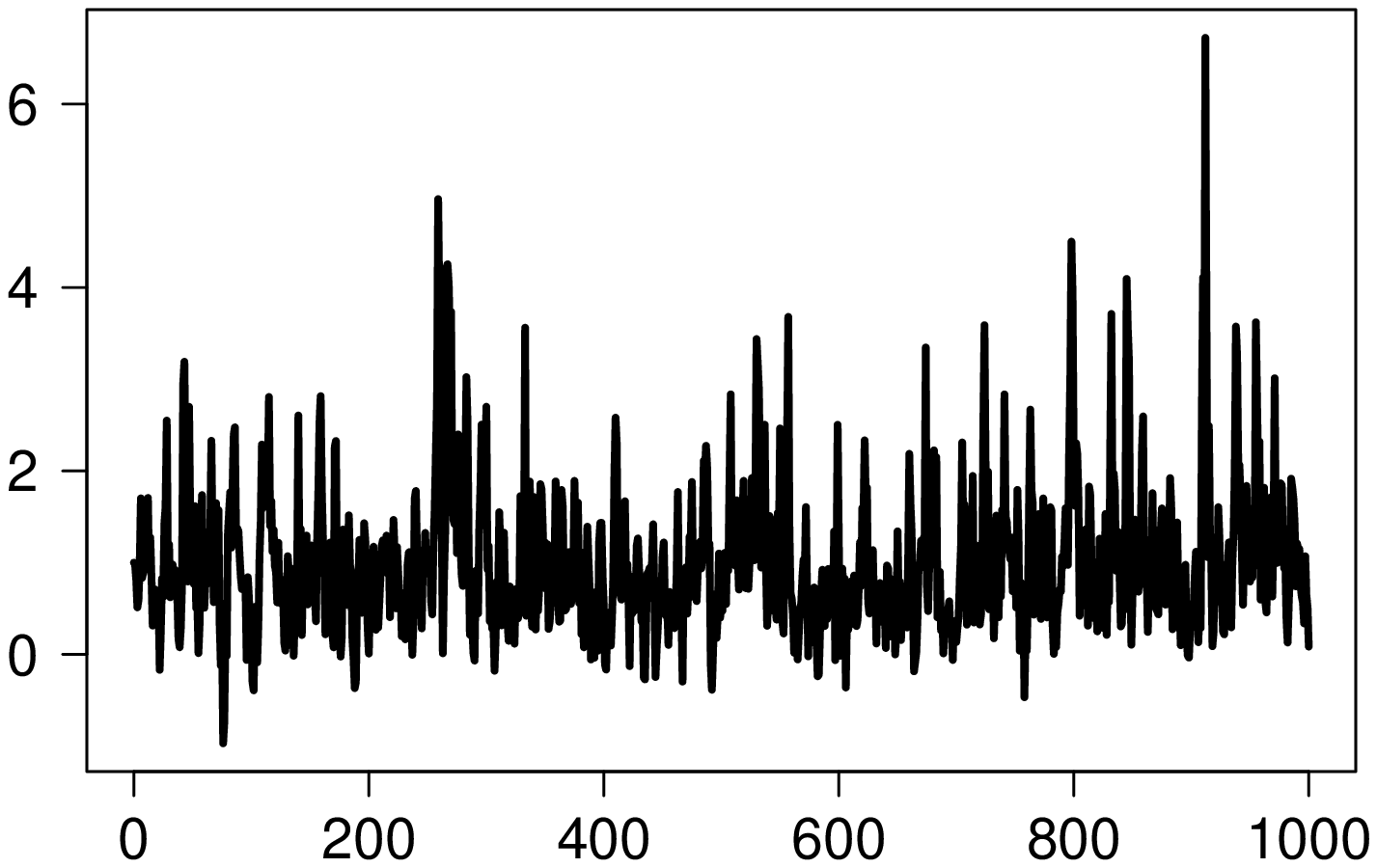}
\label{fig:pearson_data}}
\subfloat[DWELL, no error]{\includegraphics[width=\myw, height=\myh]{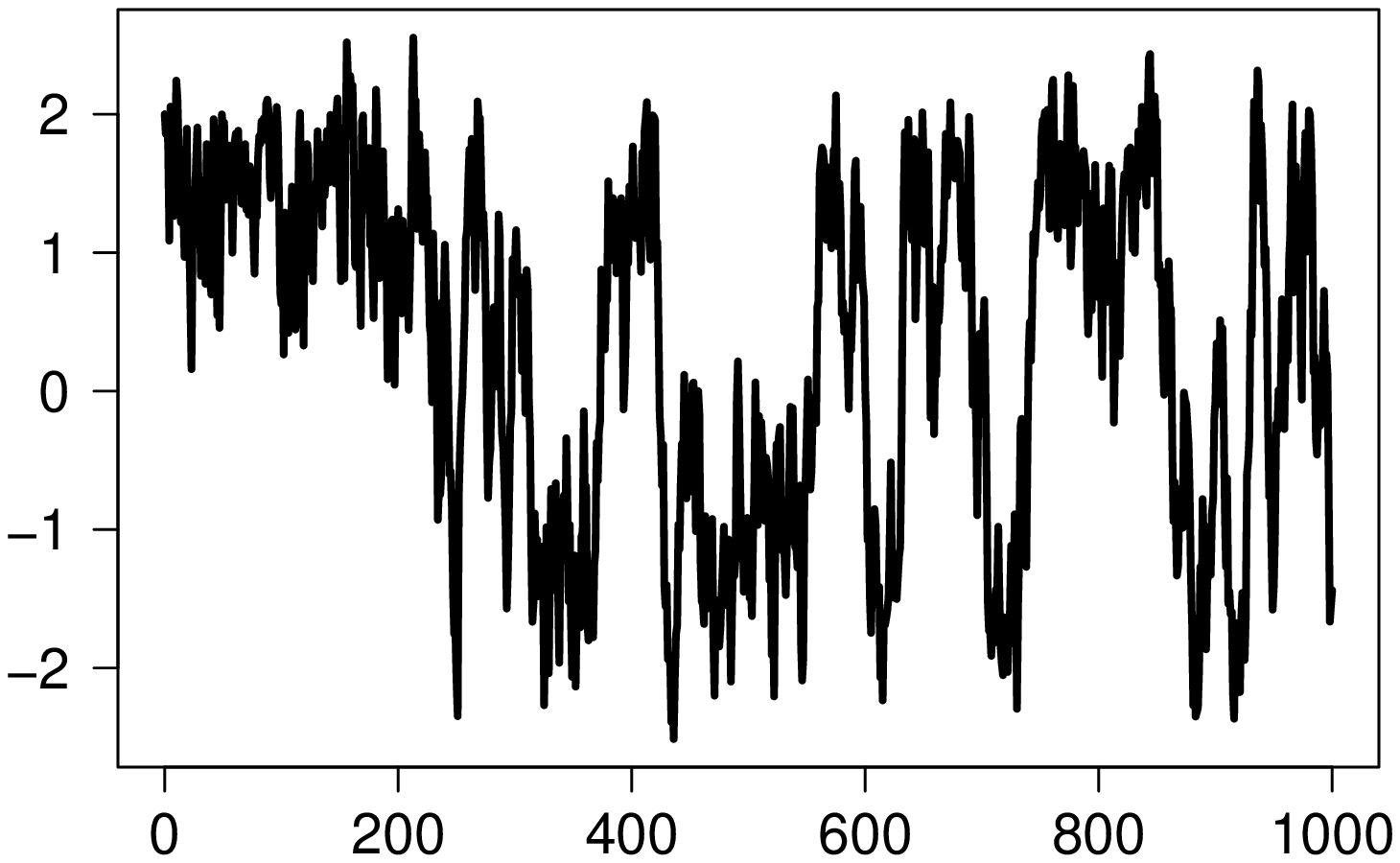}
\label{fig:dwell_data}}
\subfloat[DWELL, with error]{\includegraphics[width=\myw, height=\myh]{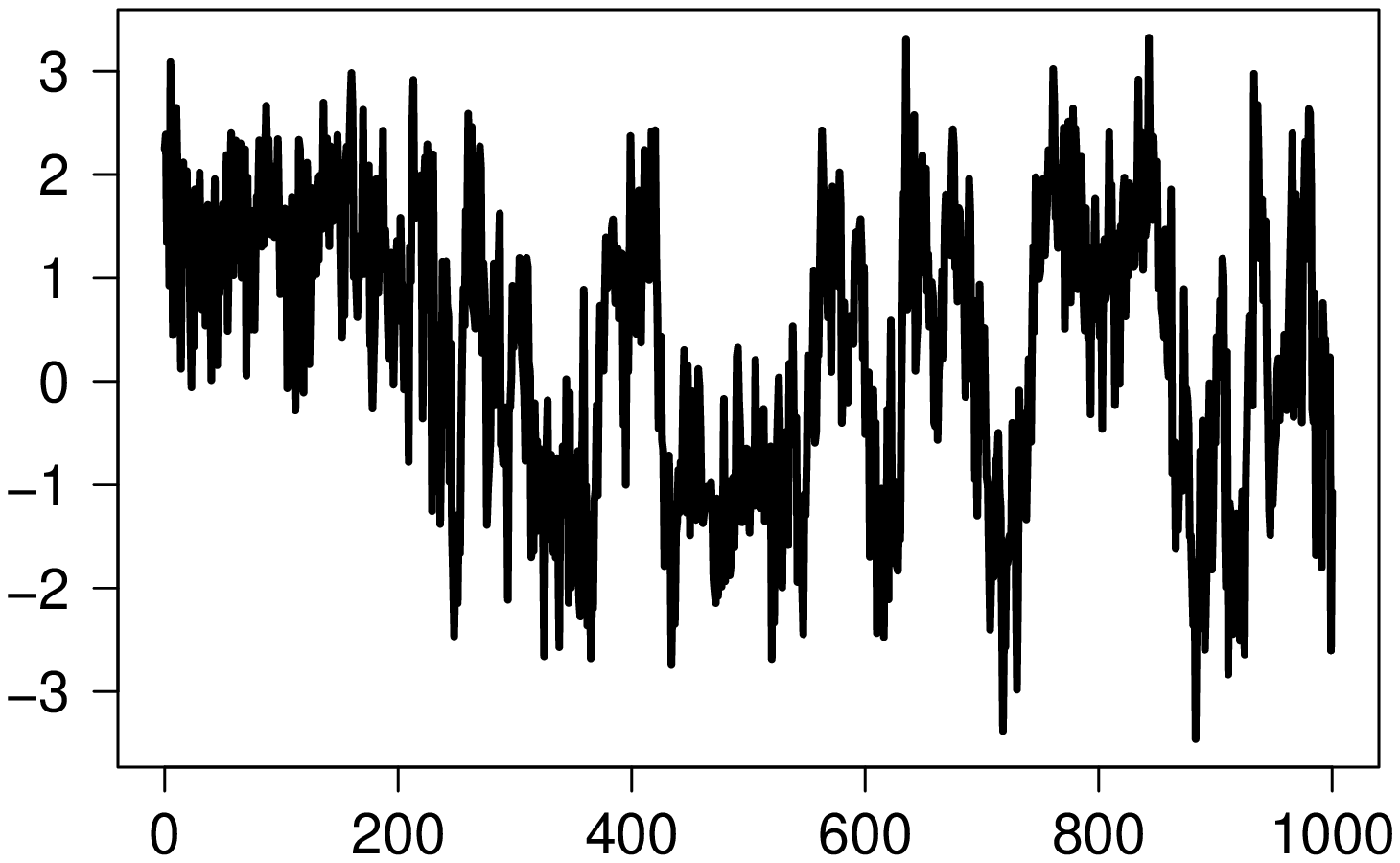}
\label{fig:dwell_er_data}}\\[-2ex]
\subfloat[MVWELL, 1st coordinate]{\includegraphics[width=\myw, height=\myh]{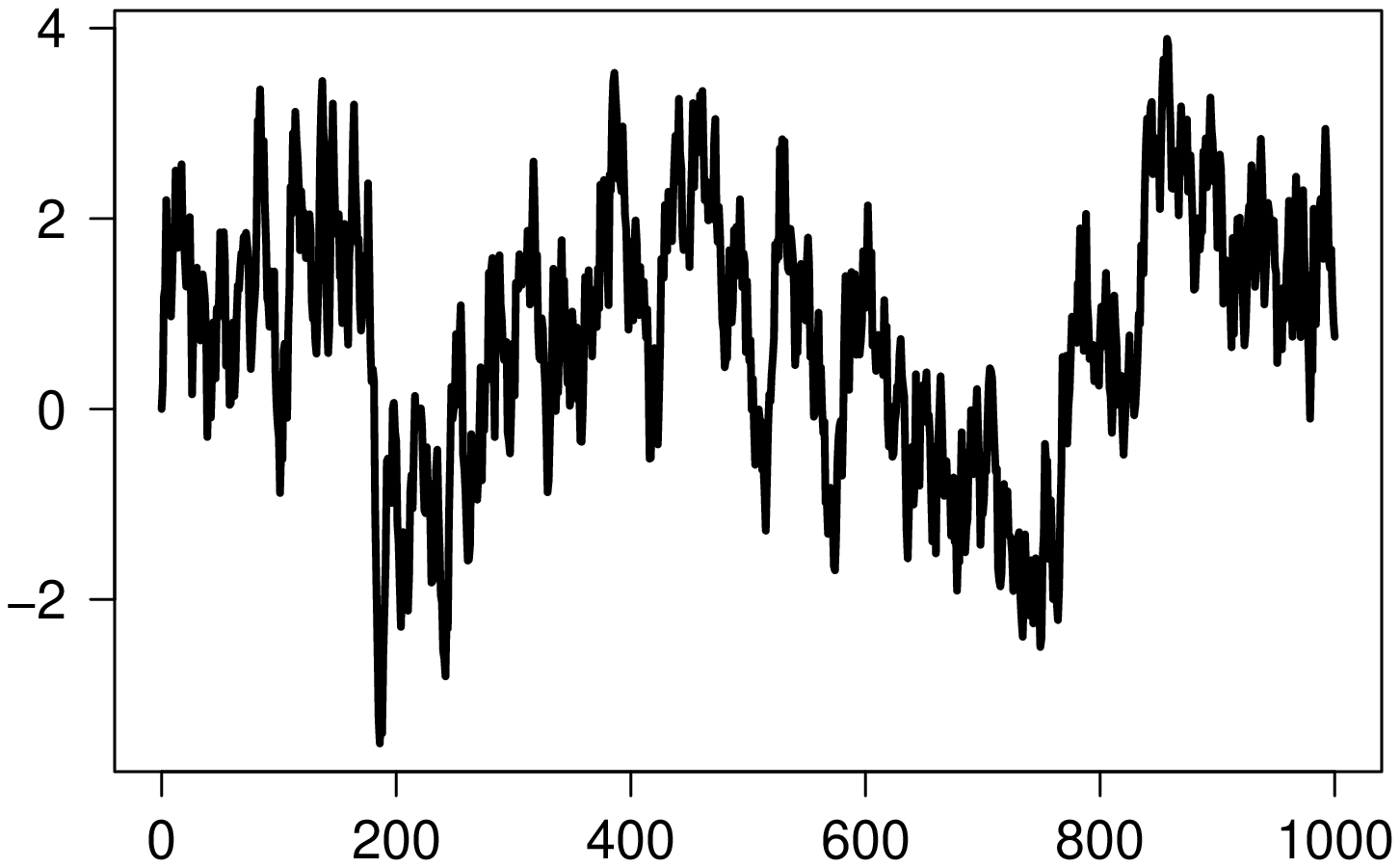}
\label{fig:mvwell_data1}}
\subfloat[MVWELL, 2nd coordinate]{\includegraphics[width=\myw, height=\myh]{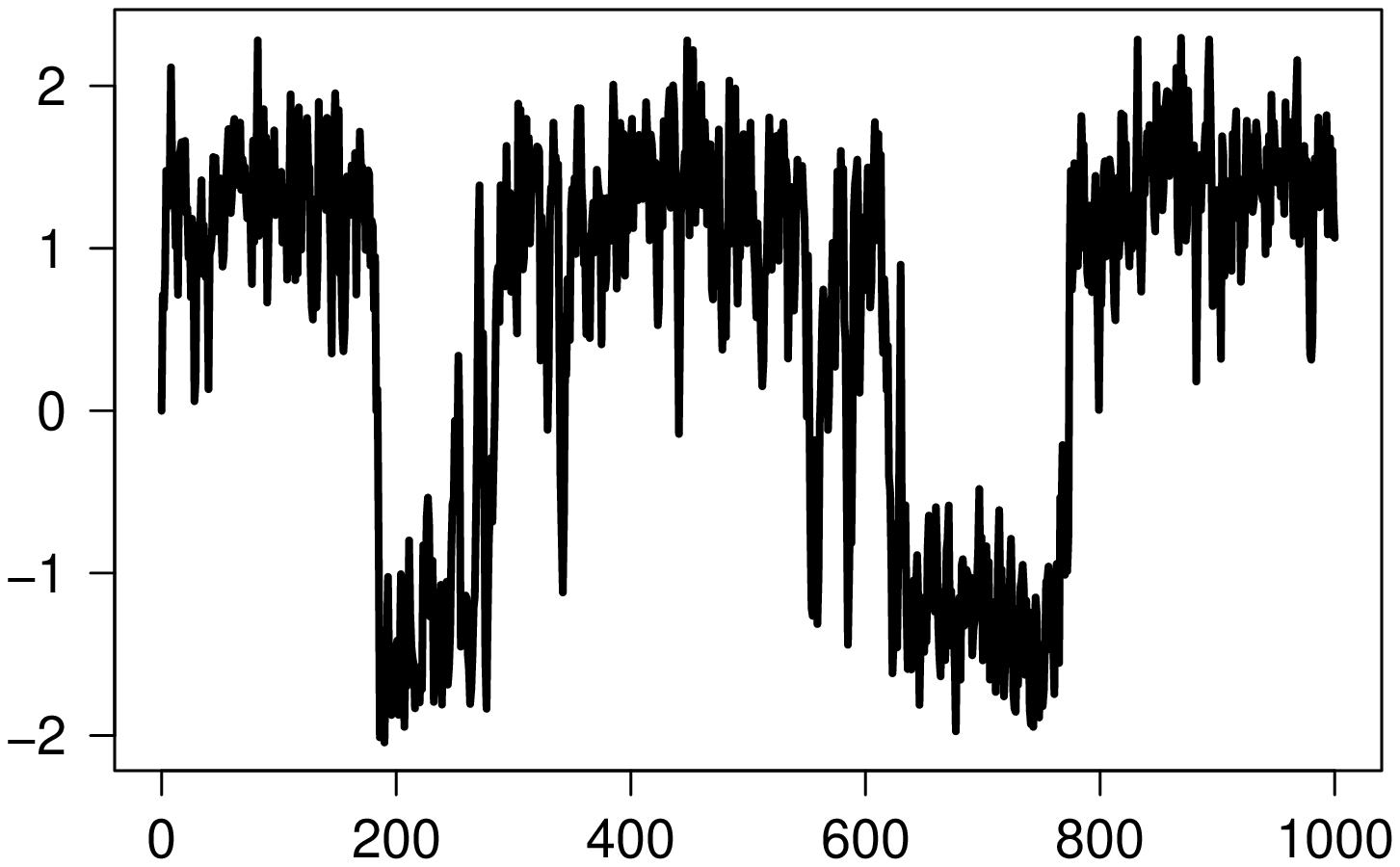}
\label{fig:mvwell_data2}}
\caption[]{Simulated datasets from \subref{fig:pearson_data} the Pearson
model, \subref{fig:dwell_data} the double well potential model,
\subref{fig:dwell_er_data} the double well observed with error, and
\subref{fig:mvwell_data1},\subref{fig:mvwell_data2} the
two-dimensional double well model.\\[1ex]
}\label{fig:obs_data}
%
%
%
\psfrag{th1}{\footnotesize $\rho$}
\psfrag{th2}{\footnotesize $\mu$}
\psfrag{th3}{\footnotesize $\sigma$}
\subfloat[{\emcmc}1, $\lambda=0$]{\includegraphics[width=\myw, height=\myh]{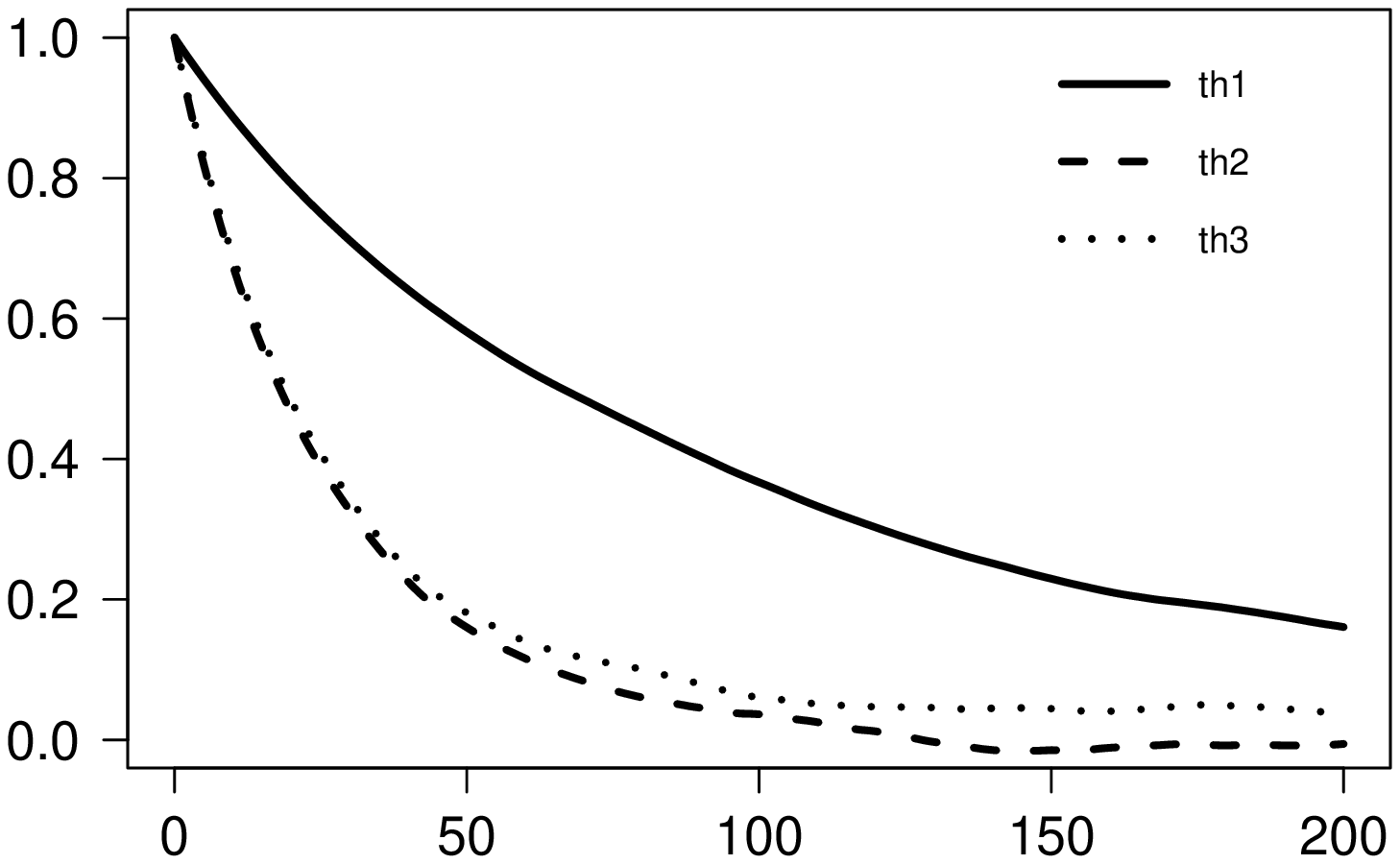}}
\subfloat[{\emcmc}1, $\lambda=5$]{\includegraphics[width=\myw, height=\myh]{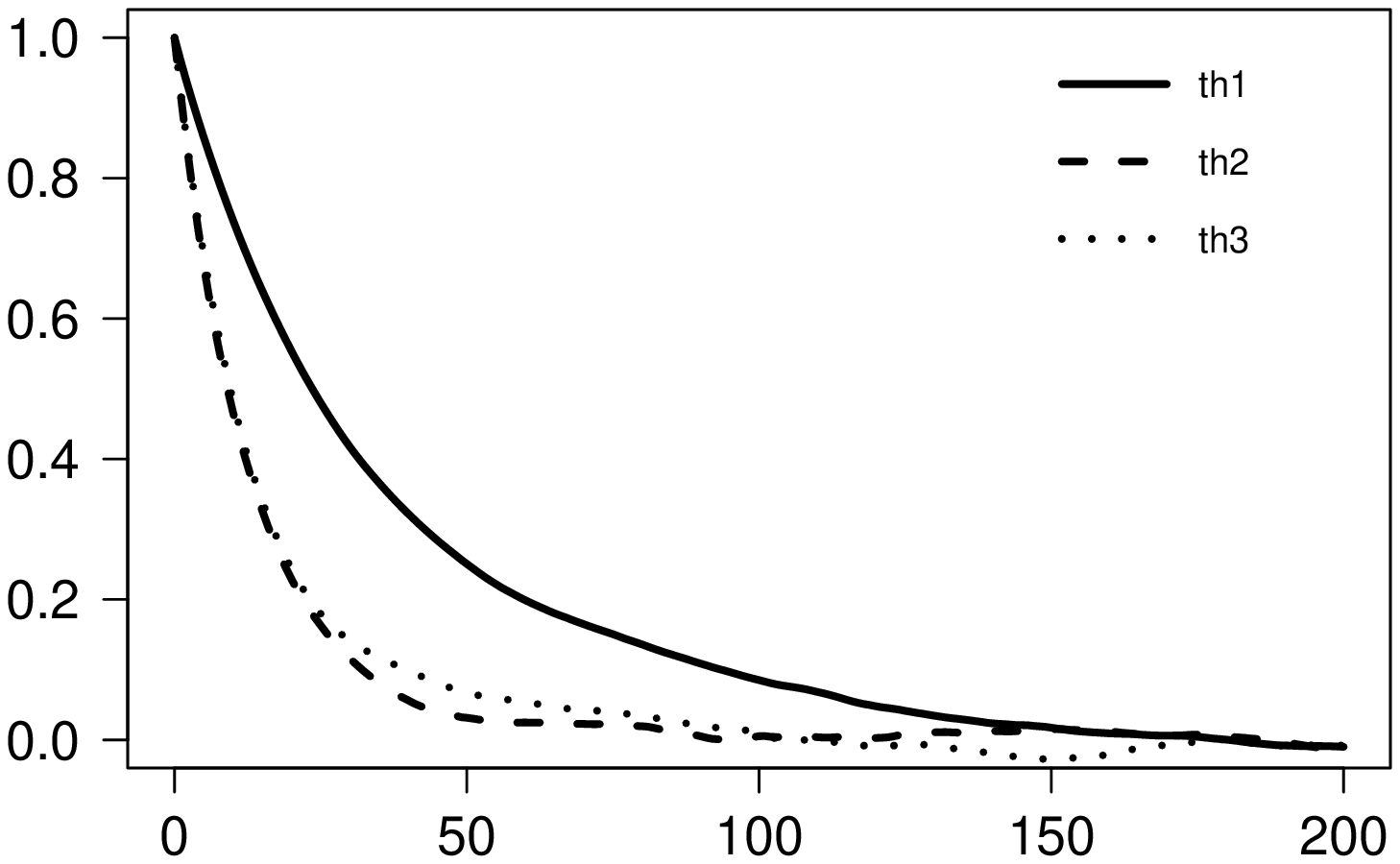}}
\subfloat[{\emcmc}1, $\lambda=10$]{\includegraphics[width=\myw, height=\myh]{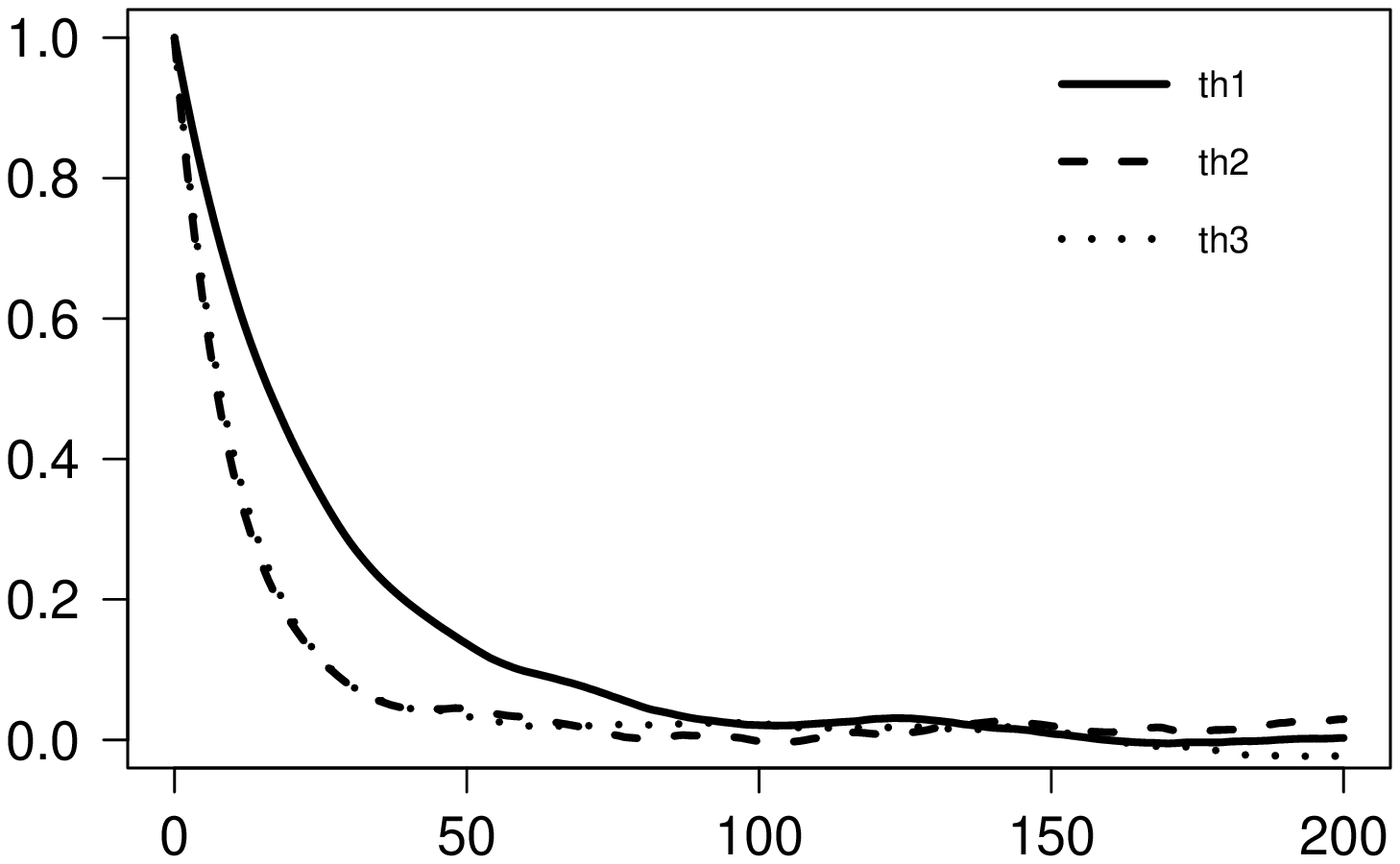}}\\[-2ex]
\subfloat[{\emcmc}1 (noncentred), $\lambda=0$]{\includegraphics[width=\myw, height=\myh]{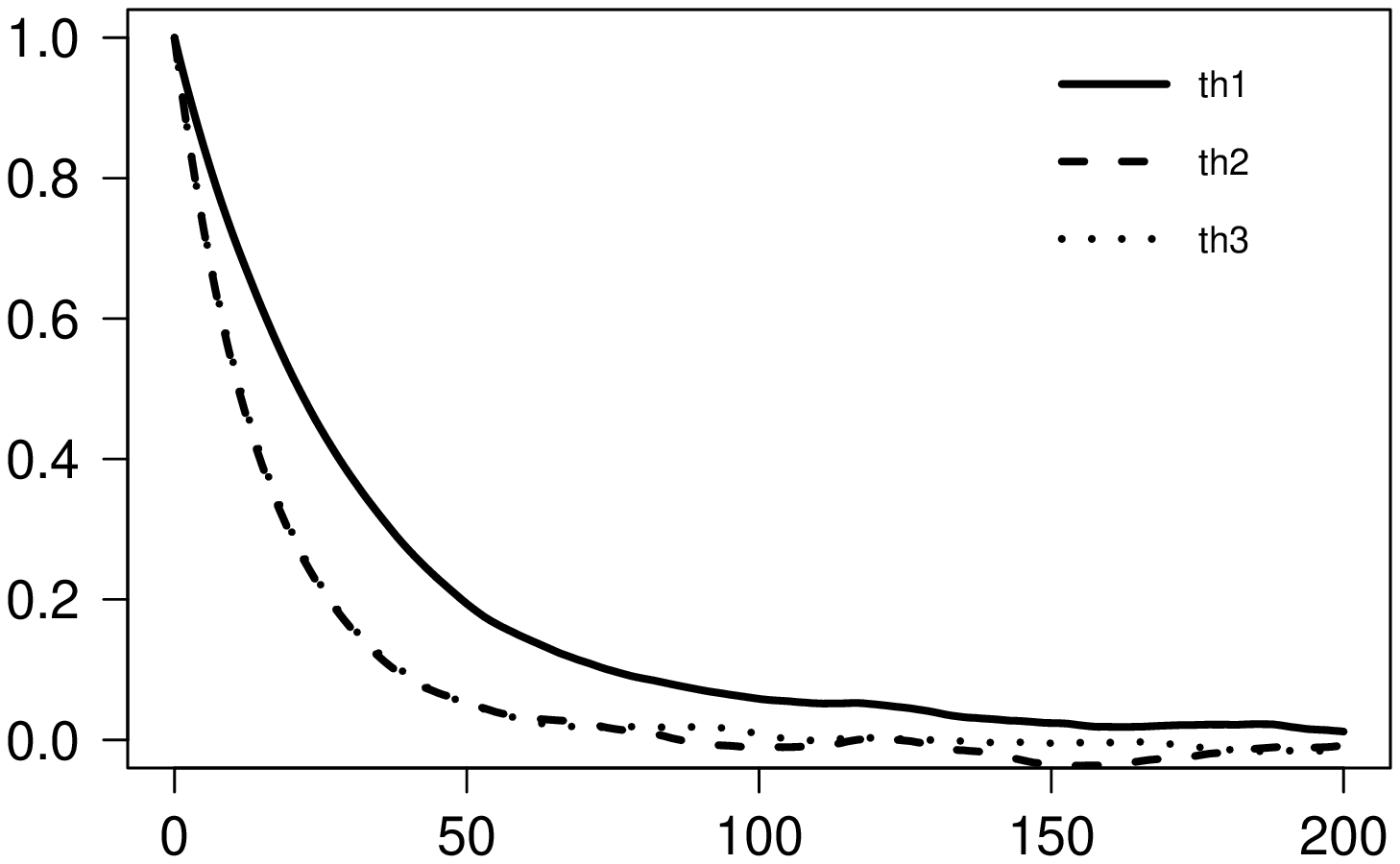}
\label{fig:tanh_acf_nca}}
 \subfloat[{\emcmc}1 (interweaved),
 $\lambda=2$]{\includegraphics[width=\myw, height=\myh]{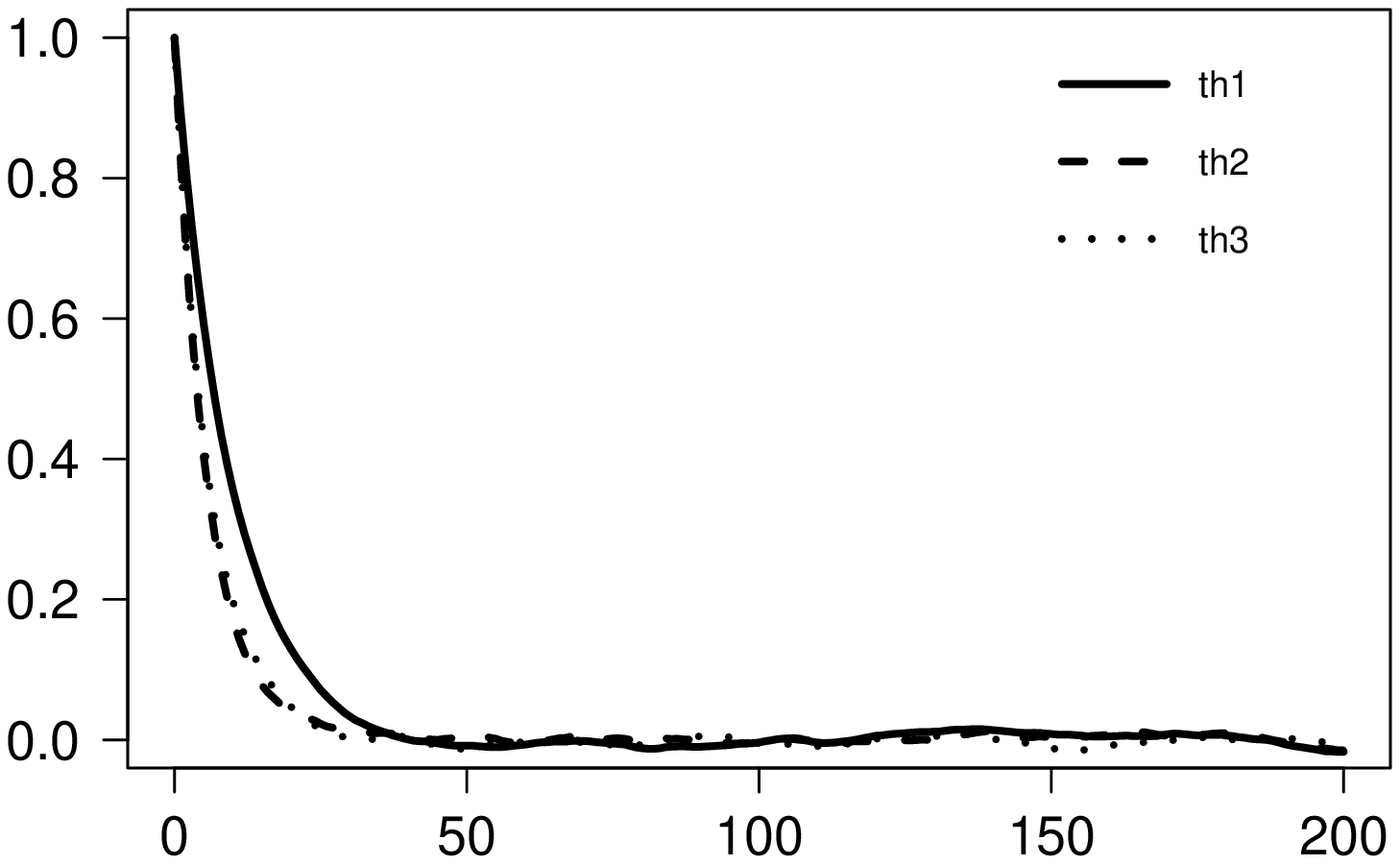}}
\subfloat[{\amcmc}$-10$]{\includegraphics[width=\myw, height=\myh]{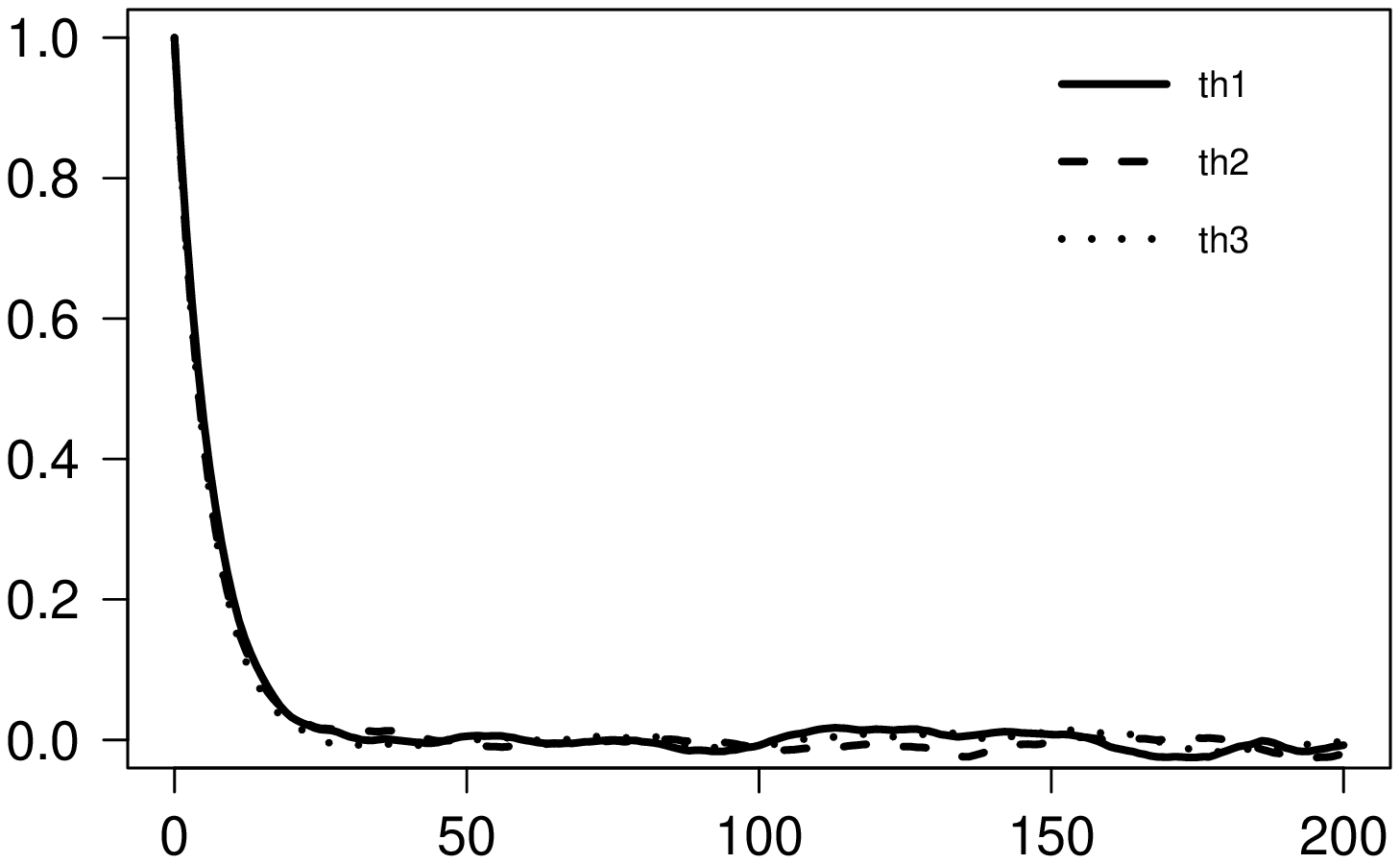}}\\[-2ex]
\scriptsize
\psfrag{ea}{{\emcmc}1}
\psfrag{hfi1}{{\amcmc}-$1$}
\psfrag{hfi5}{{\amcmc}-$5$}
\psfrag{hfi10}{{\amcmc}-$10$}
\psfrag{hfi20}{{\amcmc}-$20$}
\psfrag{hfi30}{{\amcmc}-$30$}
\subfloat[$\rho$]{\includegraphics[width=\myw, height=\myh]{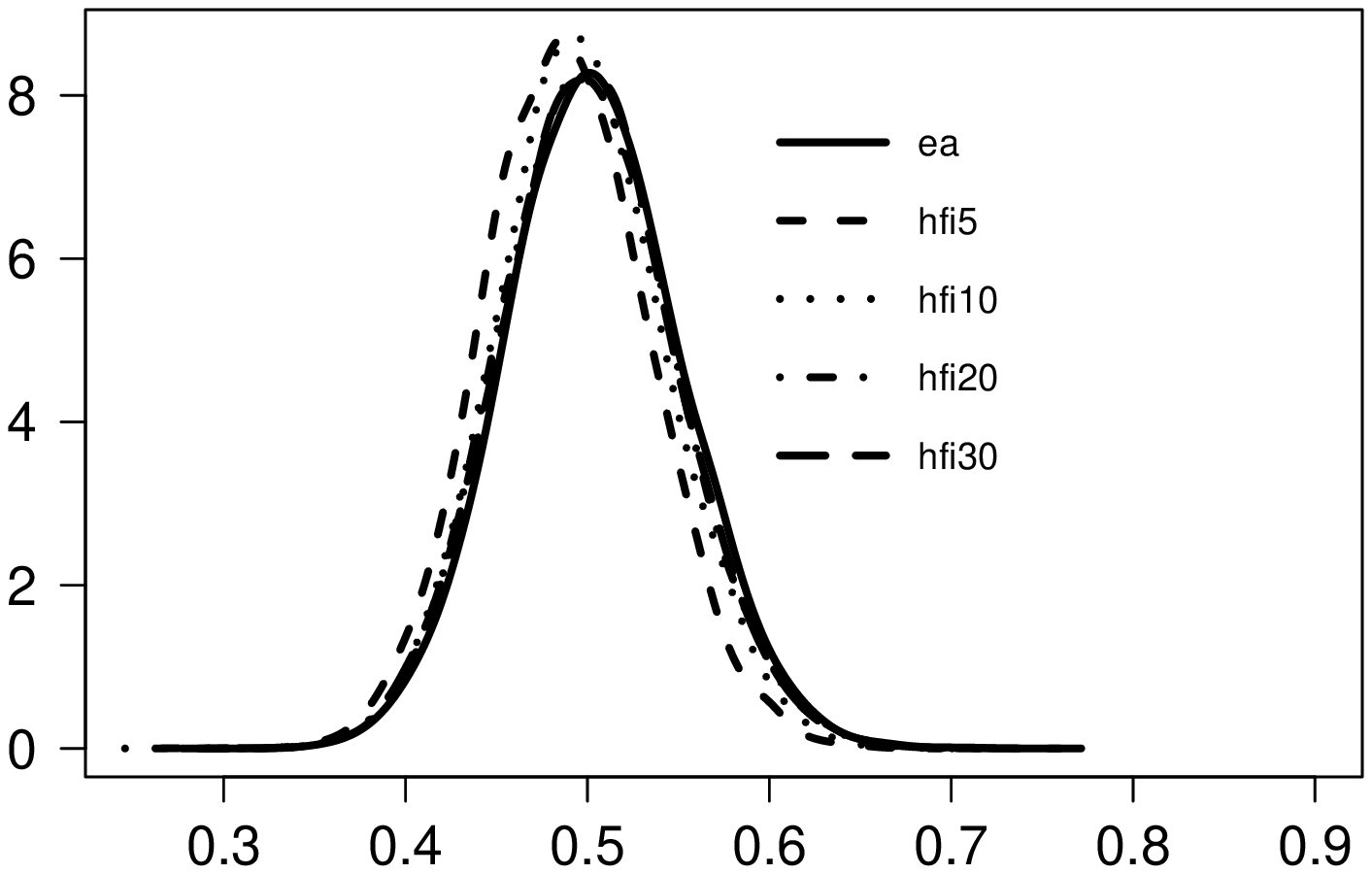}
\label{fig:tanh_density_a}}
\subfloat[$\mu$]{\includegraphics[width=\myw, height=\myh]{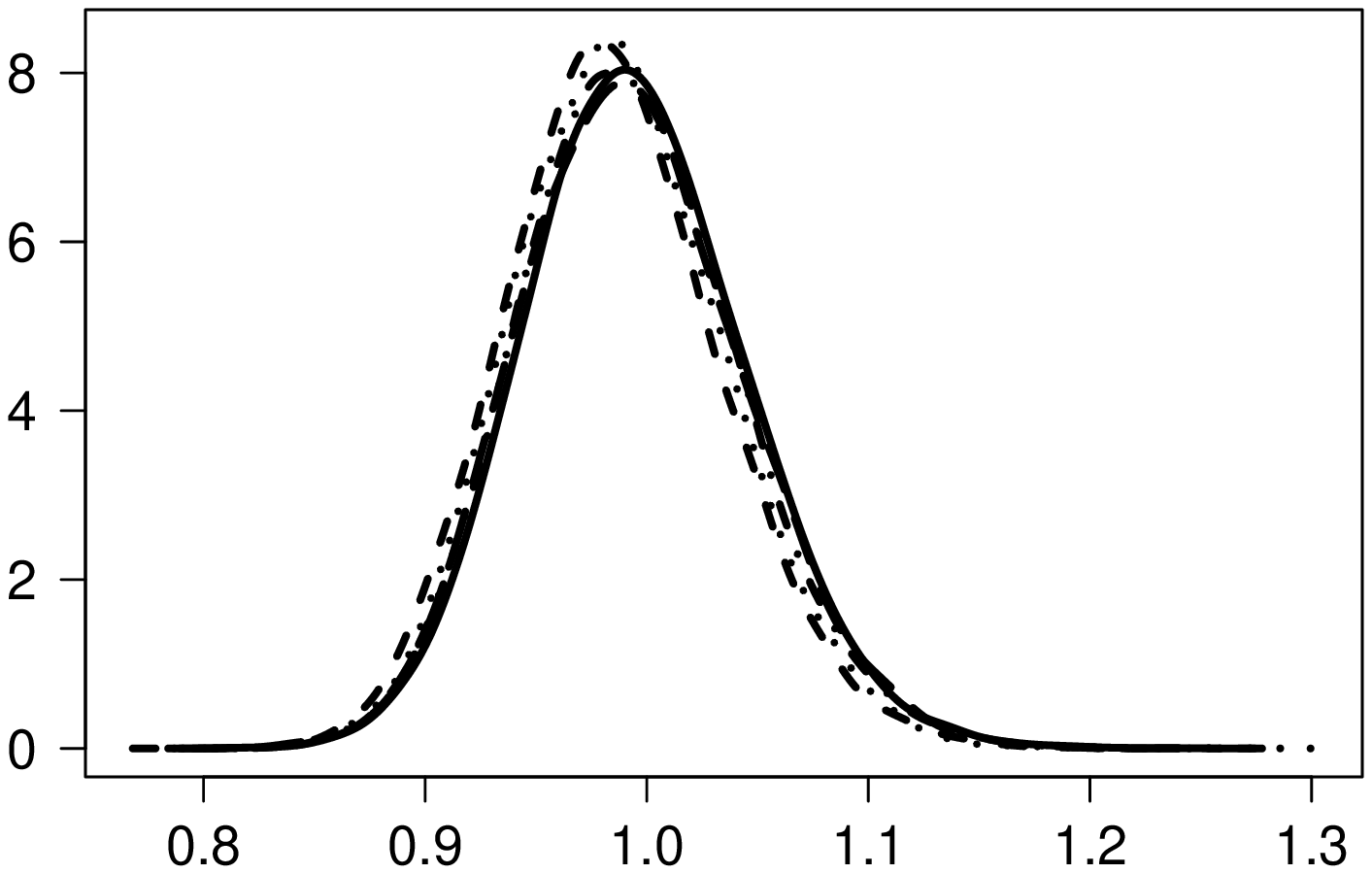}
\label{fig:tanh_density_b}}
\subfloat[$\sigma$]{\includegraphics[width=\myw, height=\myh]{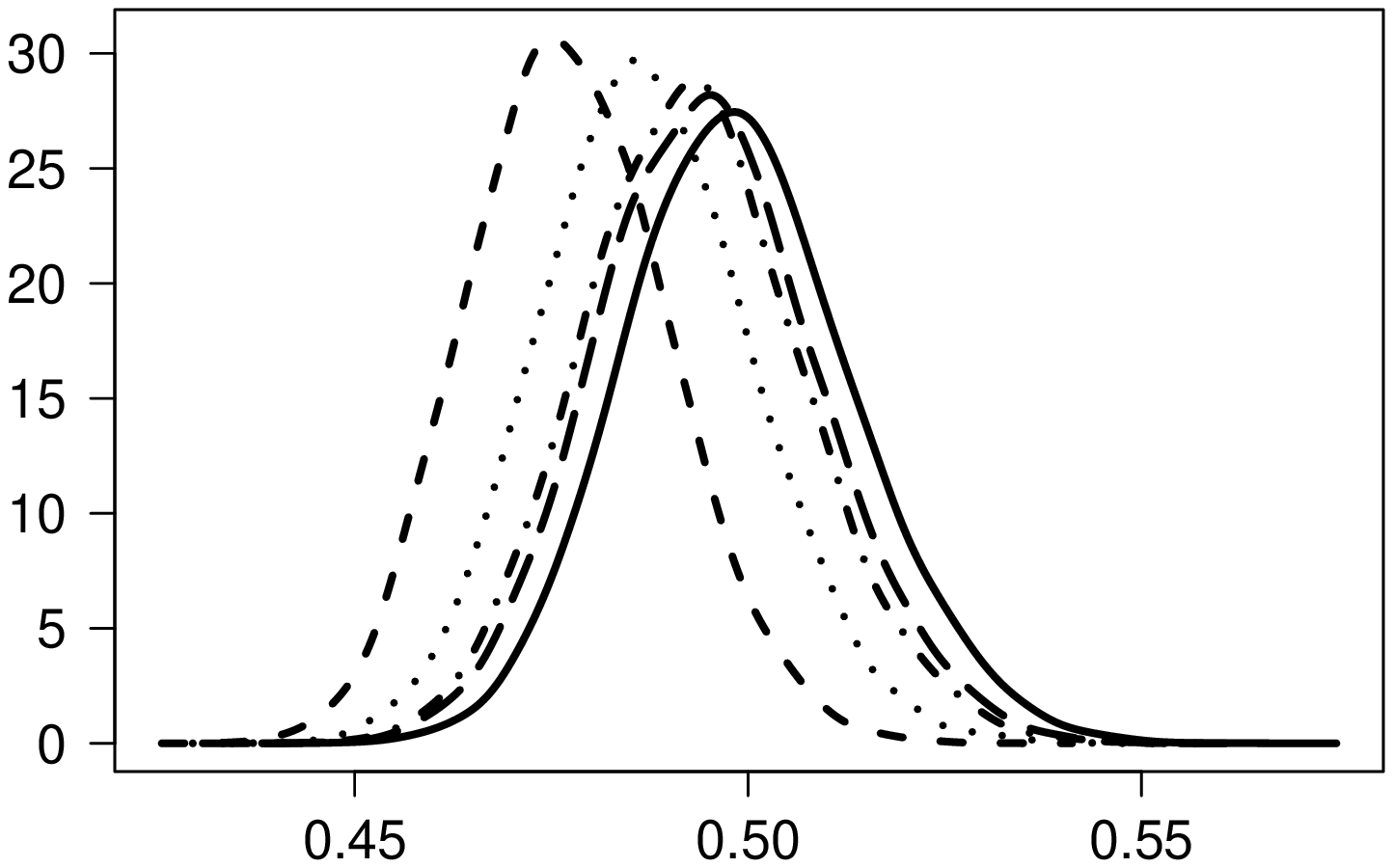}
\label{fig:tanh_density_c}}
\caption[]{The Pearson diffusion model with $n=1000$ simulated data points. True
values are
$(\rho,\mu,\sigma)=(1/2, 1, 1/2)$. Autocorrelations are reported after a
burn in of $5000$ iterations. Posterior density
estimates using EMCMC1 (interweaved) and {\amcmc} algorithms
for \subref{fig:tanh_density_a} $\rho$,
\subref{fig:tanh_density_b} $\mu$ and \subref{fig:tanh_density_c} $\sigma$.
}\label{fig:pearson_figs}
\end{center}
\end{figure}

We test the methods on a simulated data set from this process, based
on $n=1000$ (excluding the initial point) equidistant points with
$\tinc{i}=1$, $V_0=1$ and parameter values
$(\rho,\mu,\sigma)=(1/2,1,1/2)$ (Figure \ref{fig:pearson_data}). We
have used improper prior densities for the parameters, $\pi(\rho)
\propto 1$, $\pi(\mu) \propto 1$, and $\pi(\sigma) \propto
1/\sigma$. For all algorithms, sampling from the conditional density
of the parameters was performed using a block Metropolis-Hastings
step. The chains were run for $10^5$ iterations.



Figure \ref{fig:pearson_figs} shows the autocorrelation plots along
with posterior density estimates derived from the Markov
chains. Starting from EMCMC1 under the original parametrisation,
notice that for $\lambda=0$ the chain exhibits strong serial
dependence even at large lags, particularly for $\rho$. This is due to
strong a priori dependence between the parameter and the number of
Poisson points, as shown in (\ref{eq:tanh_lr}), which remains
significant in the posterior distribution. In particular, the
posterior correlation between $\sum_{i}^n{\kappa_i}$ and $\rho$ was
estimated equal to $0.93$, thus suggesting that noncentring the
Poisson process can result in better mixing rates, as Figure
\ref{fig:tanh_acf_nca} confirms. To improve the performance of the
exact methods, we consider various values for $\lambda=\{2, 5, 10\}$,
thus revealing the path at additionally $2, 5$ and $10$ points between
consecutive observations respectively. The increase in performance is
reflected in the autocorrelation function, which now decays to $0$
more quickly. As expected, the chains of the {\amcmc} algorithms mix
more rapidly than the exact ones due to the less amount of
augmentation. The posterior distributions estimated from the {\isi}
algorithms provide evidence of bias even for $M=30$ (Figures
\ref{fig:tanh_density_a} to \ref{fig:tanh_density_c}).

Table \ref{table:tanh_table} presents posterior summary statistics.
Notice that even for $M=30$, the {\amcmc} algorithm fails to pass the
Kolmogorov-Smirnov tests at a $5\%$ significance level, whereas less
amount of imputation ($M=10$) combined with the integration by parts
yields less biased approximations, clearly illustrating the importance
of eliminating the stochastic integrals as in \eqref{eq:cd_lik_ibp}.
In terms of computational performance, the interweaved algorithm with
$\lambda=2$ outperforms the rest and exhibits adjusted ESS for $\mu$
and $\sigma$ which are respectively $36\%$ and $23\%$ larger than that
of the sufficiently accurate {\amcmc} methods.

The algorithms were also run using proper priors, an exponential
distribution for $\rho$, a Gaussian for $\mu$ and an inverse Gamma for
$\sigma^2$, yielding no significant differences from the results
presented above.

\setlength{\belowcaptionskip}{0pt}
\begin{table} \footnotesize
\renewcommand{\tabcolsep}{5.5pt}
\begin{center}
  \begin{tabular}{lcrrrrrrrrrrr} \hline\hline Method &
\multicolumn{1}{c}{Par.} &\multicolumn{1}{r}{$\lambda$} &
\multicolumn{1}{c}{$M$} & \multicolumn{1}{c}{Mean}
&\multicolumn{1}{c}{SD} & \multicolumn{1}{c}{ESS$_{adj}$} &
\multicolumn{1}{c}{ESS} & \multicolumn{1}{c}{KS} &
\multicolumn{1}{c}{Time} & \multicolumn{3}{c}{Correlation
matrix}\\[0.2cm]
EMCMC1 (noncentred) &  $\rho$ &  0 &  2.316 &  0.505 &  0.048 &  6.599 &  16.789 &  &  2.544 &  1.000 &  -0.447 &  0.539 \\
 &  $\mu$ &  &  &  0.995 &  0.050 &  12.432 &  31.629 &  &  &   &  1.000 &  0.008 \\
 &  $\sigma$ &  &  &  0.499 &  0.015 &  12.117 &  30.829 &  &  &   &   &  1.000 \\[0.1cm]
EMCMC1 (interweaved) &  $\rho$ &  0 &  2.314 &  0.506 &  0.049 &  6.686 &  22.170 &  &  3.316 &  1.000 &  -0.438 &  0.540 \\
 &  $\mu$ &  &  &  0.994 &  0.050 &  15.035 &  49.859 &  &  &   &  1.000 &  0.021 \\
 &  $\sigma$ &  &  &  0.499 &  0.015 &  14.408 &  47.779 &  &  &   &   &  1.000 \\[0.1cm]
EMCMC1 (interweaved) &  $\rho$ &  2 &  4.346 &  0.504 &  0.048 &  11.094 &  53.153 &  &  4.791 &  1.000 &  -0.416 &  0.537 \\
 &  $\mu$ &  &  &  0.995 &  0.050 &  18.951 &  90.792 &  &  &   &  1.000 &  0.026 \\
 &  $\sigma$ &  &  &  0.499 &  0.015 &  17.022 &  81.554 &  &  &   &   &  1.000 \\[0.1cm]
{\amcmc}-30 &  $\rho$ &   &  30.000 &  0.501 &  0.048 &  4.916 &  80.518 &  0.005 &  16.380 &  1.000 &  -0.439 &  0.519 \\
 &  $\mu$ &  &  &  0.993 &  0.050 &  5.160 &  84.516 &  0.017 &  &   &  1.000 &  0.021 \\
 &  $\sigma$ &  &  &  0.495 &  0.014 &  5.329 &  87.293 & $< 0.001$ &  &   &   &  1.000 \\[0.1cm]
{\amcmc}-10 (int-by-parts) &  $\rho$ &   &  10.000 &  0.504 &  0.049 &  12.898 &  81.865 &  0.286 &  6.347 &  1.000 &  -0.422 &  0.537 \\
 &  $\mu$ &  &  &  0.995 &  0.050 &  13.878 &  88.087 &  0.749 &  &   &  1.000 &  0.020 \\
 &  $\sigma$ &  &  &  0.499 &  0.015 &  13.793 &  87.545 &  0.583 &  &   &   &  1.000 \\[0.1cm]
 \hline\hline
 \end{tabular}
\end{center}
\caption{The Pearson diffusion model with $n=1000$ simulated data
points. True values are $(\rho,\mu,\sigma)=(1/2, 1, 1/2)$. Statistics
are reported after a burn-in period of $5000$ iterations. The $M$
column shows the average number of imputed points between consecutive
observations. The SD column shows the standard deviation. The ESS
column shows the effective sample size per $1000$ iterations. The Time
column shows the time in seconds required for $1000$ iterations of
each chain. The adjusted effective sample size is shown in column
ESS$_{adj}$. The KS column shows the $p$-value for the
Kolmogorov-Smirnov test with null hypothesis that draws from the exact and
approximate marginals come from the same distribution.}
\label{table:tanh_table}
\end{table}

\subsection{A double well potential model}

\begin{figure}[!t]
\begin{center} 
\psfrag{th1}{\footnotesize $\rho$}
\psfrag{th2}{\footnotesize $\mu$}
\psfrag{th3}{\footnotesize $\sigma$}
\subfloat[{\emcmc}3, $\lambda=10$]{\includegraphics[width=\myw, height=\myh]{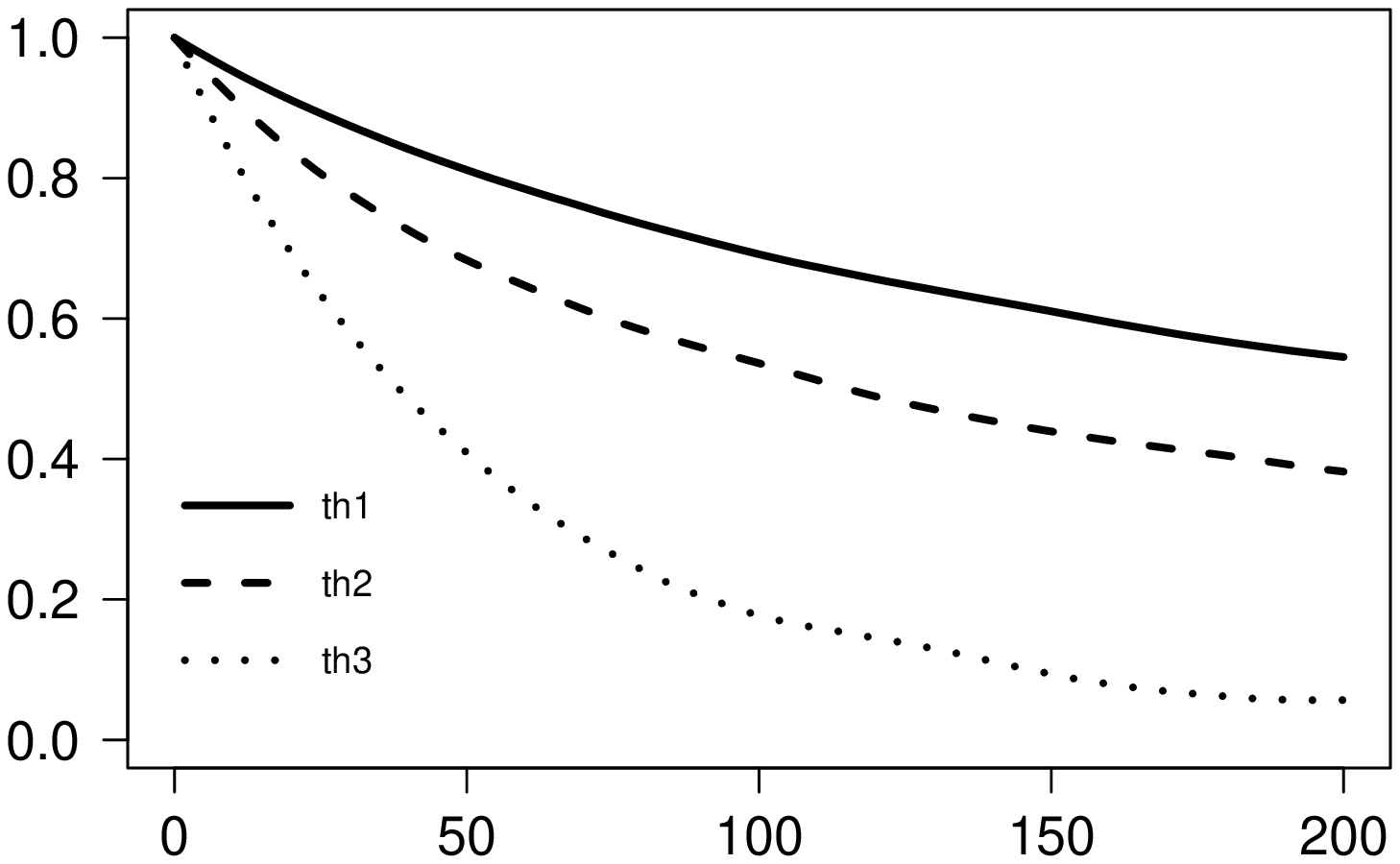}}
\subfloat[{\emcmc}3 (noncentred), $\lambda=2$]{\includegraphics[width=\myw, height=\myh]{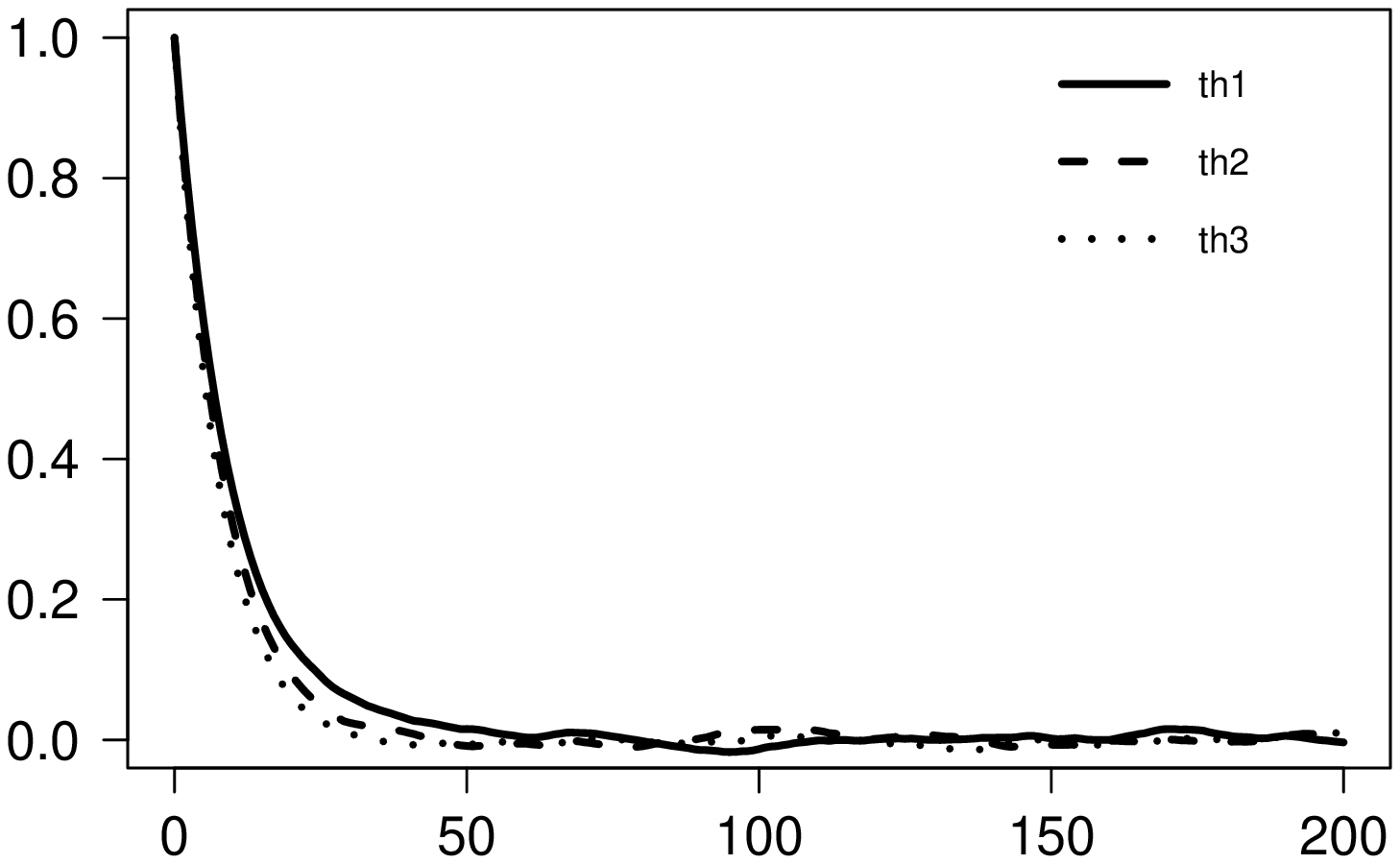}
\label{fig:dwell_acf_nca}} 
\subfloat[{\amcmc}$-10$]{\includegraphics[width=\myw, height=\myh]{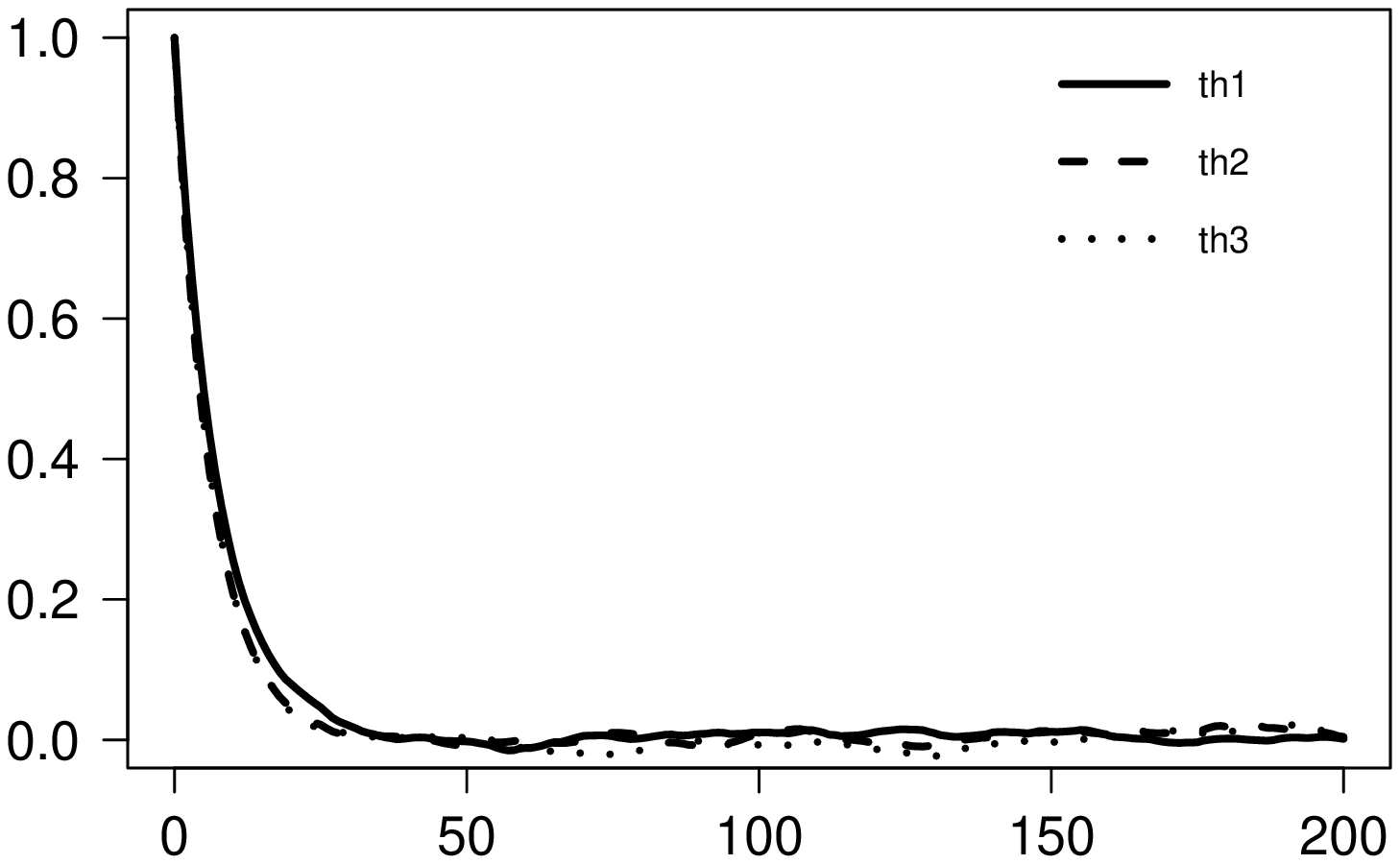}}\\[-2ex]
\scriptsize 
\psfrag{ea}{EMCMC3} 
\psfrag{hfi5}{{\amcmc}-$5$} 
\psfrag{hfi10}{{\amcmc}-$10$}
\psfrag{hfi20}{{\amcmc}-$20$}
\psfrag{hfi30}{{\amcmc}-$30$}
\subfloat[$\rho$]{\includegraphics[width=\myw, height=\myh]{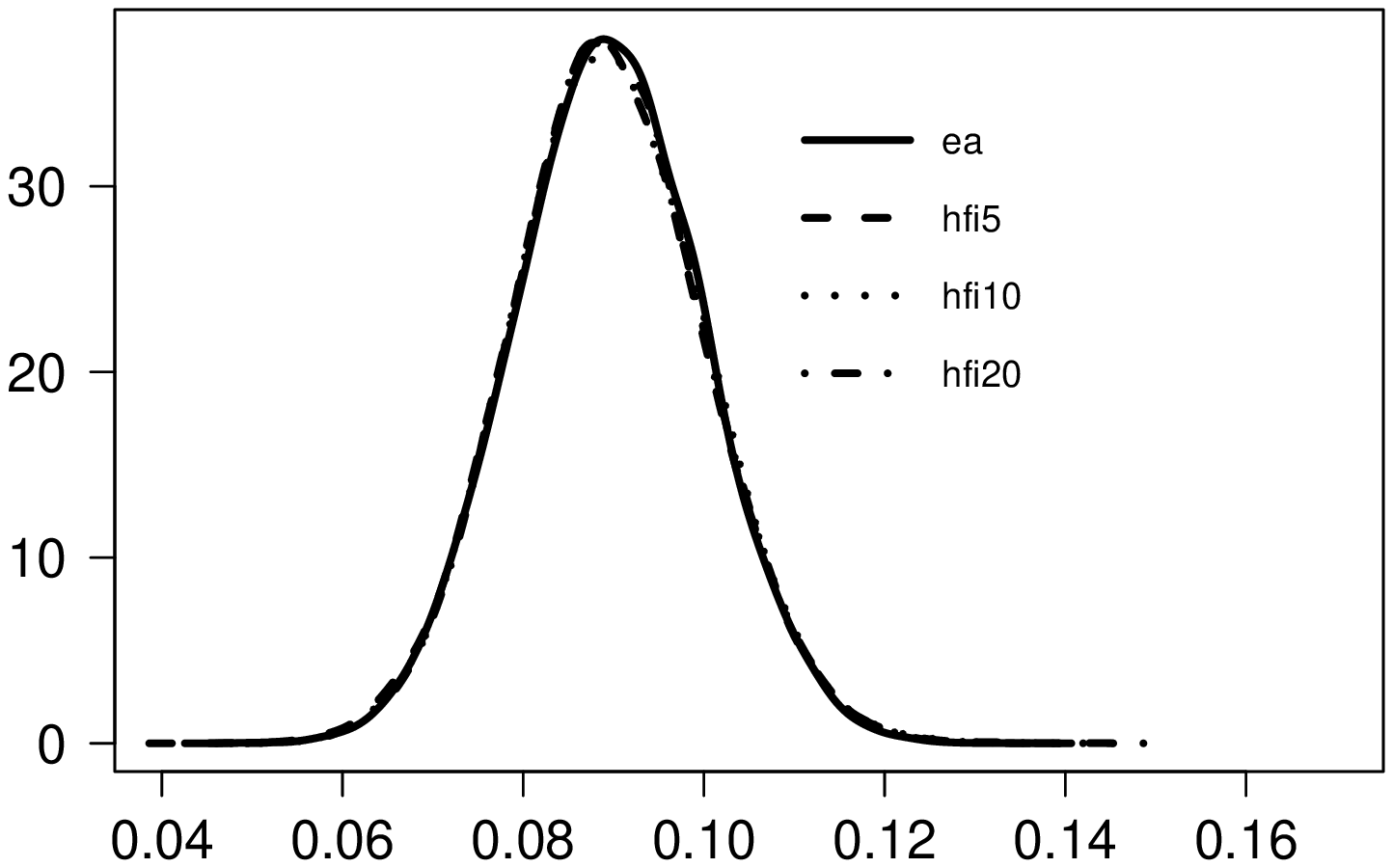}
\label{fig:dwell_density_a}}
\subfloat[$\mu$]{\includegraphics[width=\myw, height=\myh]{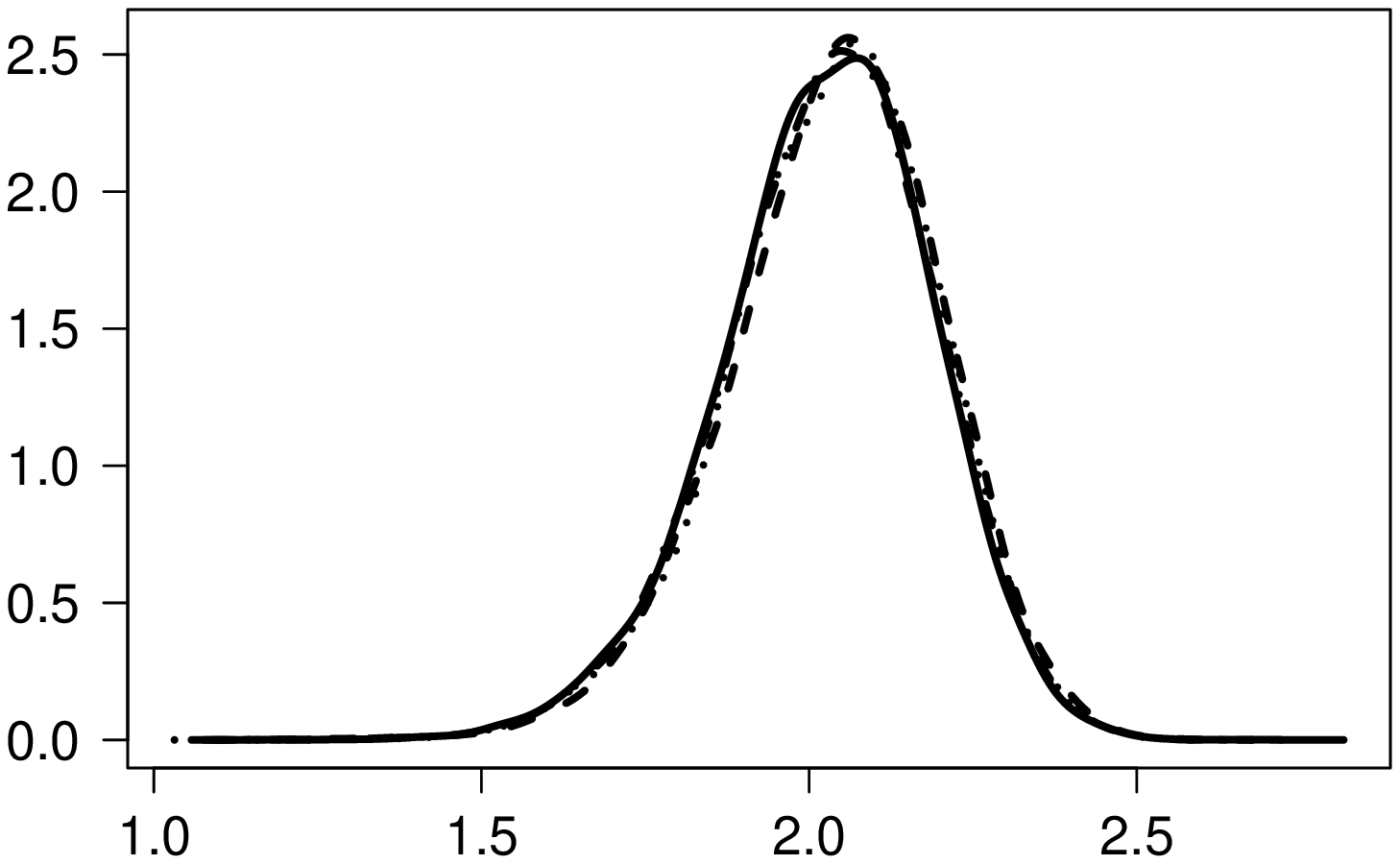}
\label{fig:dwell_density_b}}
\subfloat[$\sigma$]{\includegraphics[width=\myw, height=\myh]{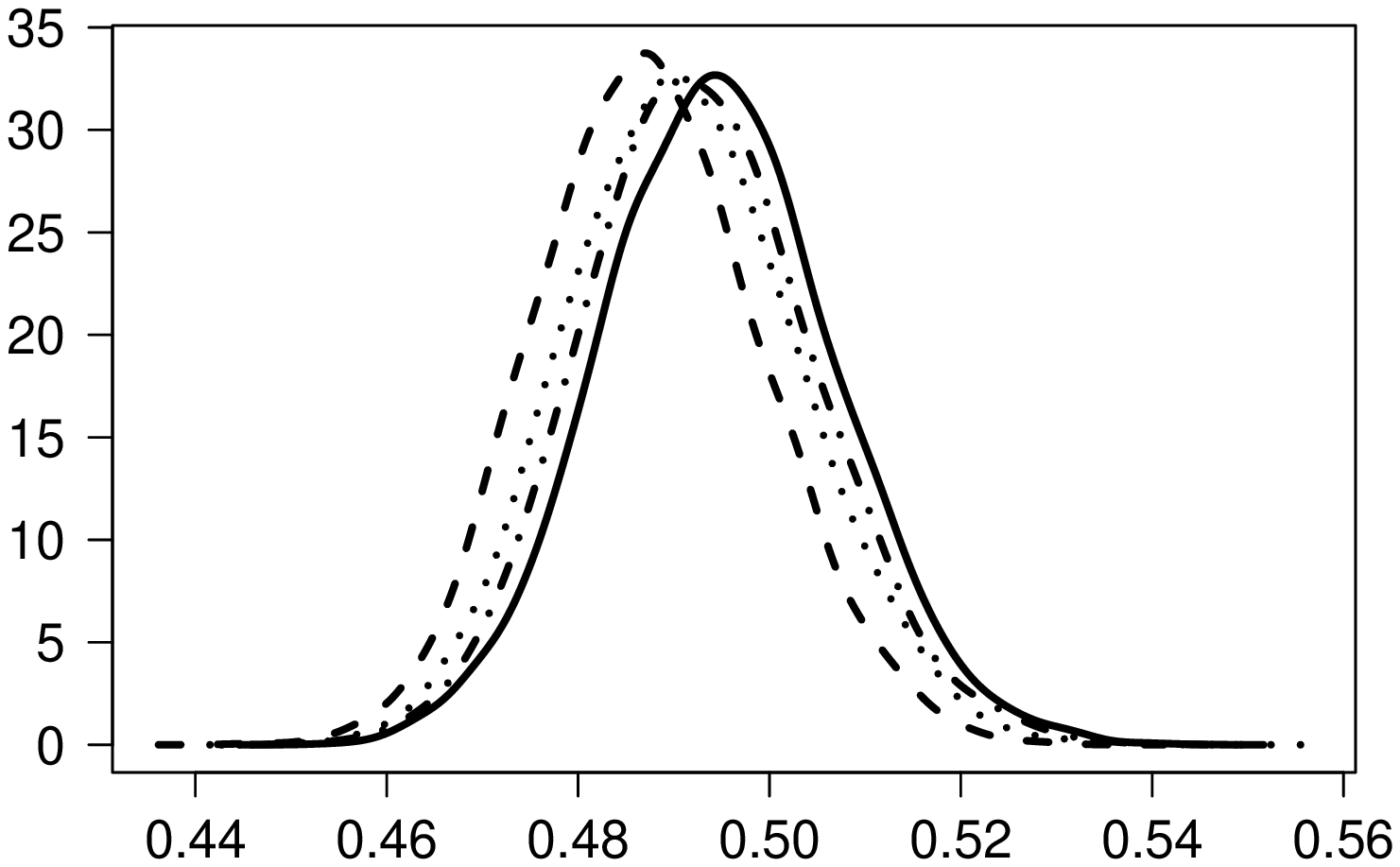}
\label{fig:dwell_density_c}}
\caption[]{The DWELL diffusion model with $n=1000$ simulated data
points. True values are $(\rho,\mu,\sigma)=(0.1, 2,
1/2)$. Autocorrelations are reported after a burn in of $5000$
iterations. Posterior density estimates using EMCMC3 (noncentred) and
{\amcmc} algorithms for \subref{fig:dwell_density_a} $\rho$,
\subref{fig:dwell_density_b} $\mu$ and \subref{fig:dwell_density_c}
$\sigma$.\\[1ex]}\label{fig:dwell_figs}
%
\psfrag{th1}{\footnotesize $\rho$}
\psfrag{th2}{\footnotesize $\mu$}
\psfrag{th3}{\footnotesize $\sigma$} 
\psfrag{th4}{\footnotesize$\tau$}
\psfrag{noer}{\scriptsize No error}
\psfrag{ea}{\scriptsize {\emcmc}3} 
\psfrag{hfi5}{\scriptsize {\amcmc}-$5$}
\psfrag{hfi10}{\scriptsize {\amcmc}-$10$}
\psfrag{hfi20}{\scriptsize {\amcmc}-$20$}
\psfrag{hfi40}{\scriptsize {\amcmc}-$40$} %
\subfloat[{\emcmc}3 (noncentred), $\lambda=0$]{\includegraphics[width=\myw, height=\myh]{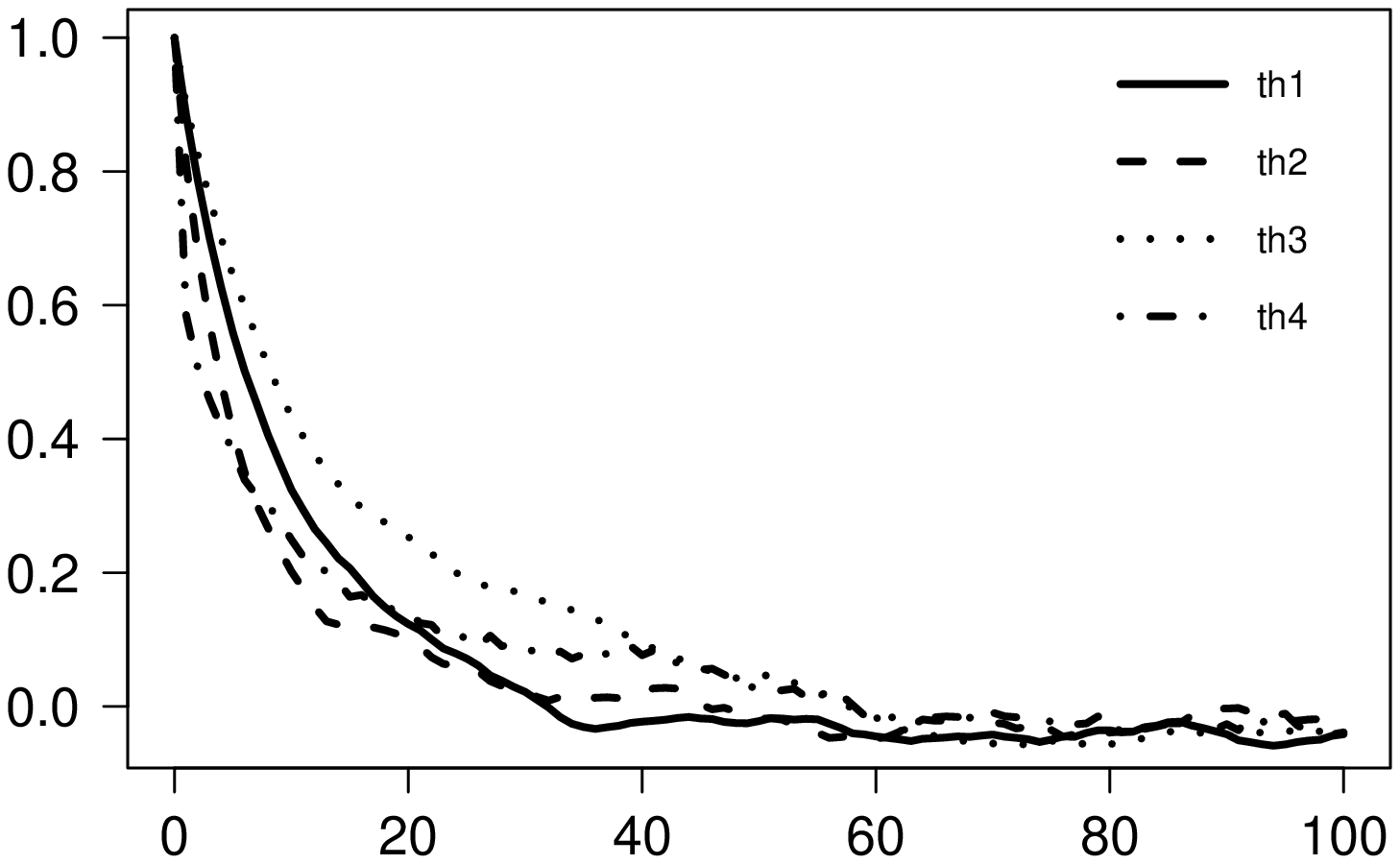}}
\subfloat[{\amcmc}$-20$]{\includegraphics[width=\myw, height=\myh]{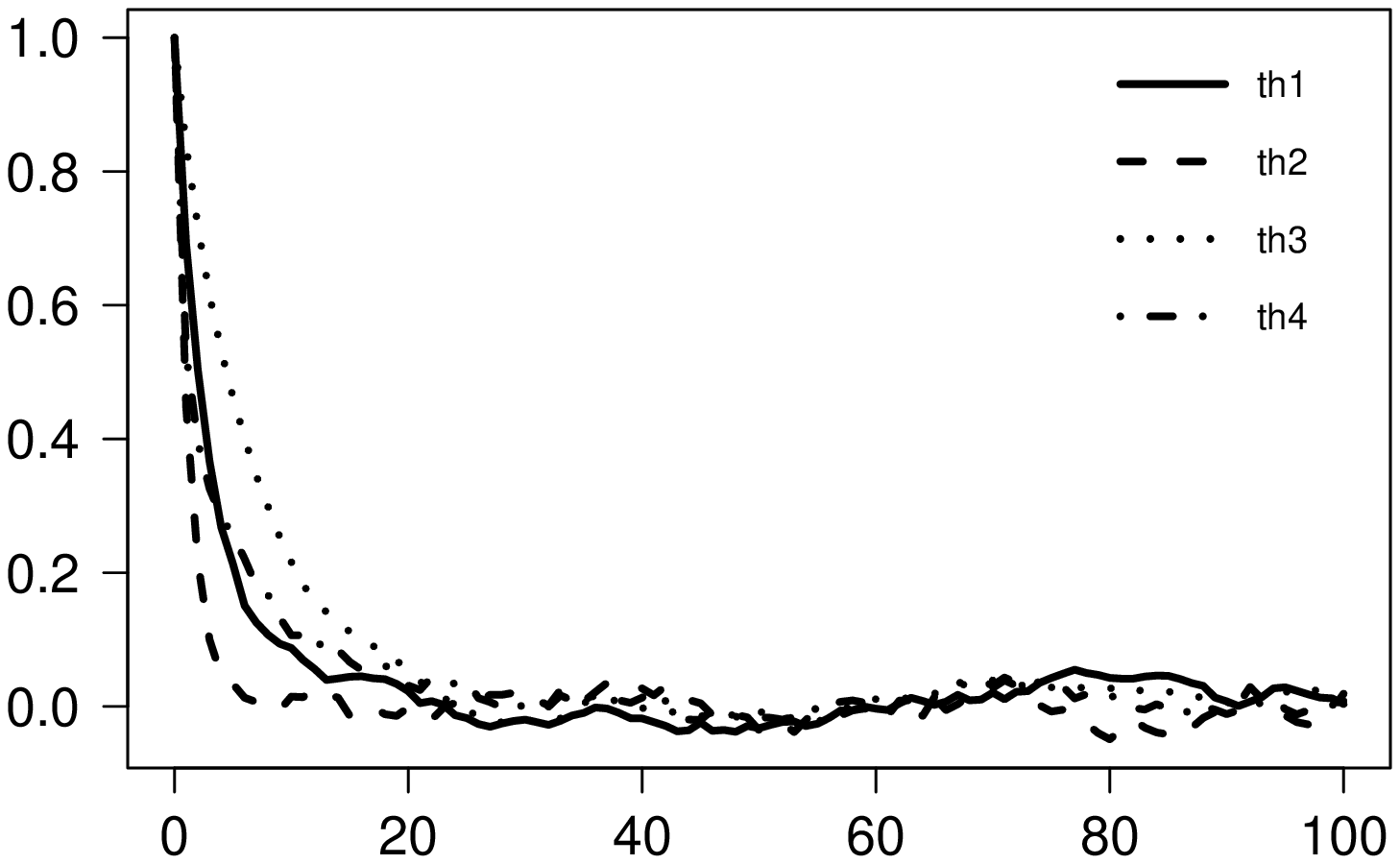}}
\subfloat[$\rho$]{\includegraphics[width=\myw, height=\myh]{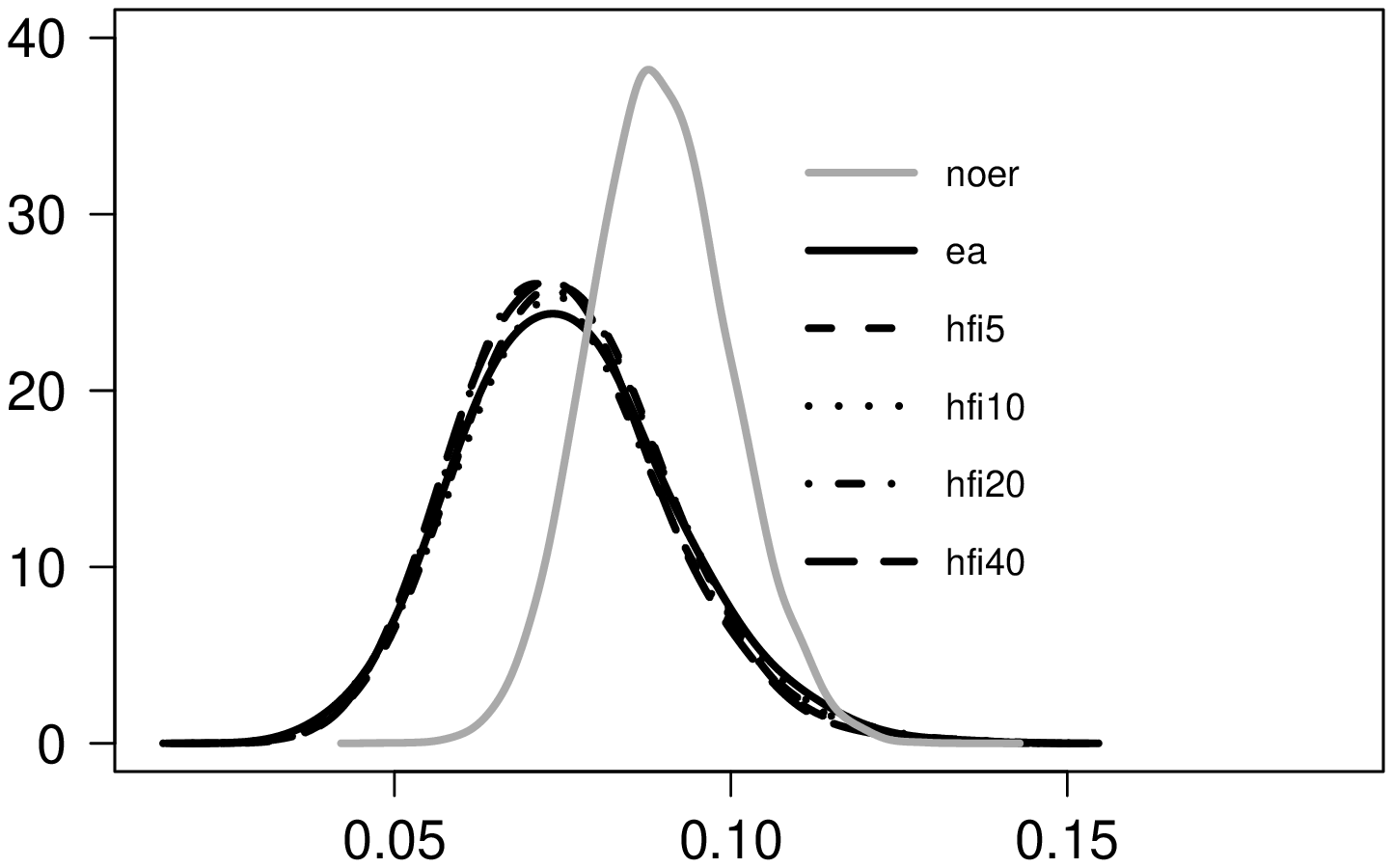}
\label{fig:dweller_density_a}}\\[-2ex]
\subfloat[$\mu$]{\includegraphics[width=\myw, height=\myh]{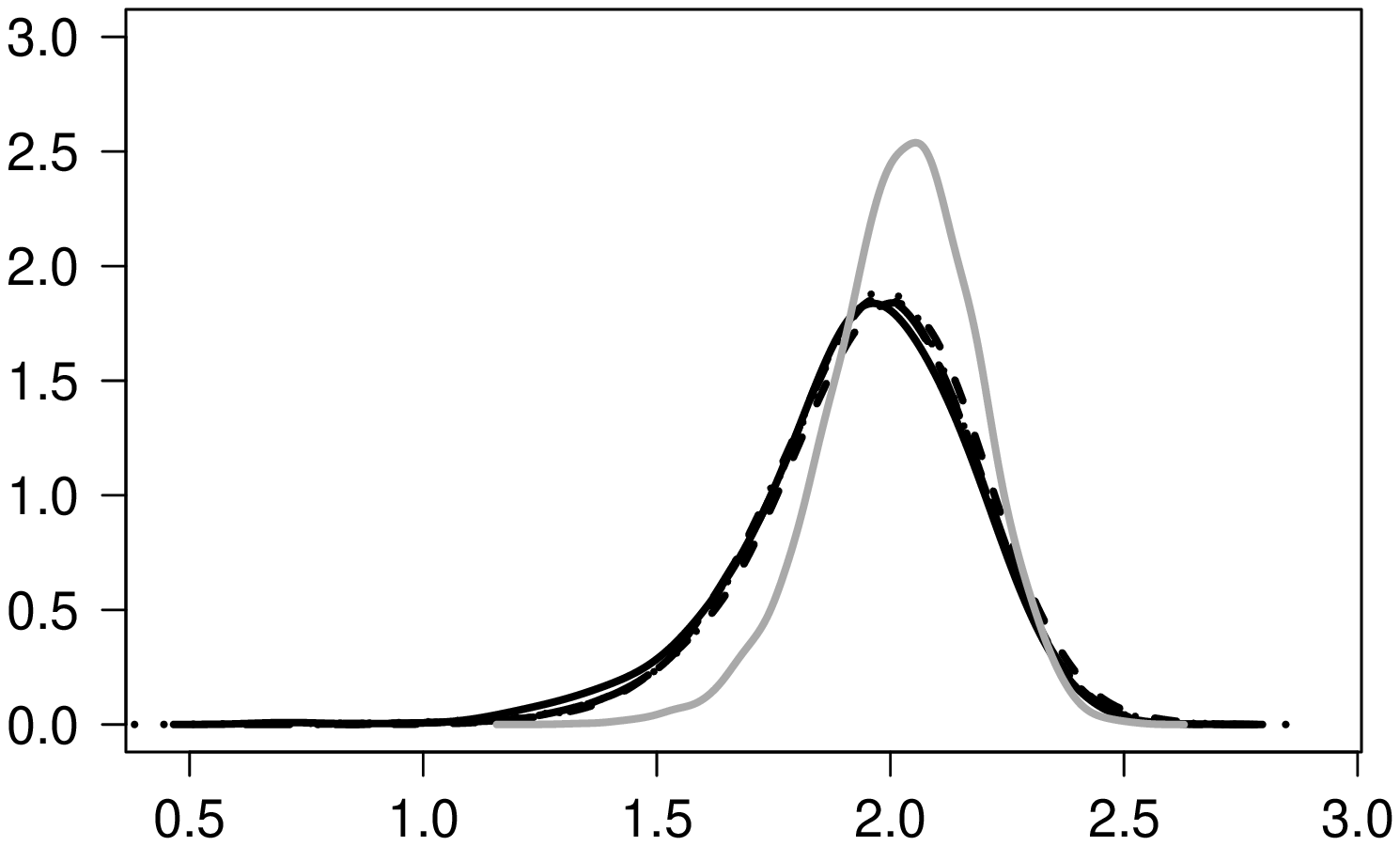}
\label{fig:dweller_density_b}}
\subfloat[$\sigma$]{\includegraphics[width=\myw, height=\myh]{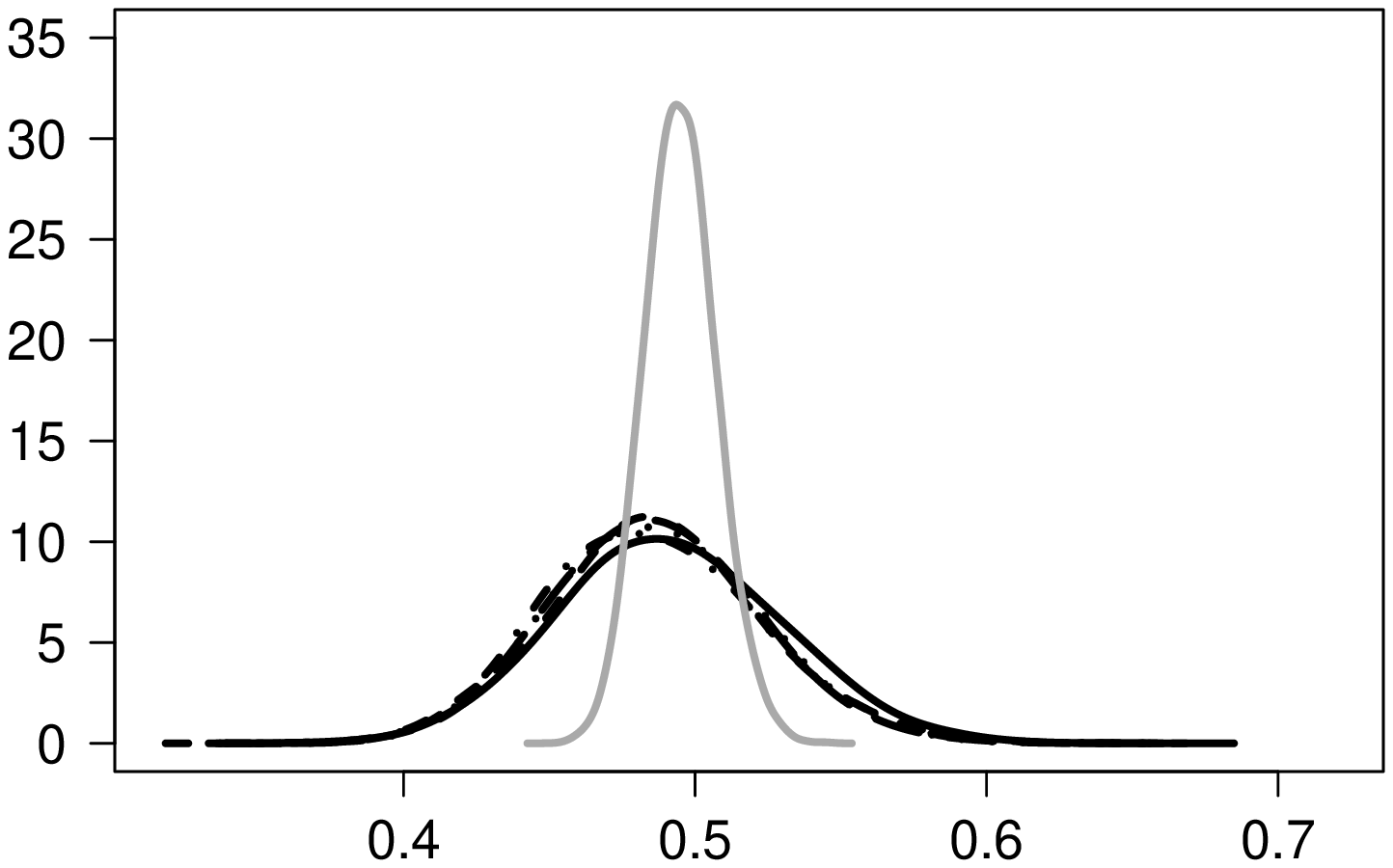}
\label{fig:dweller_density_c}}
\subfloat[$\tau$]{\includegraphics[width=\myw, height=\myh]{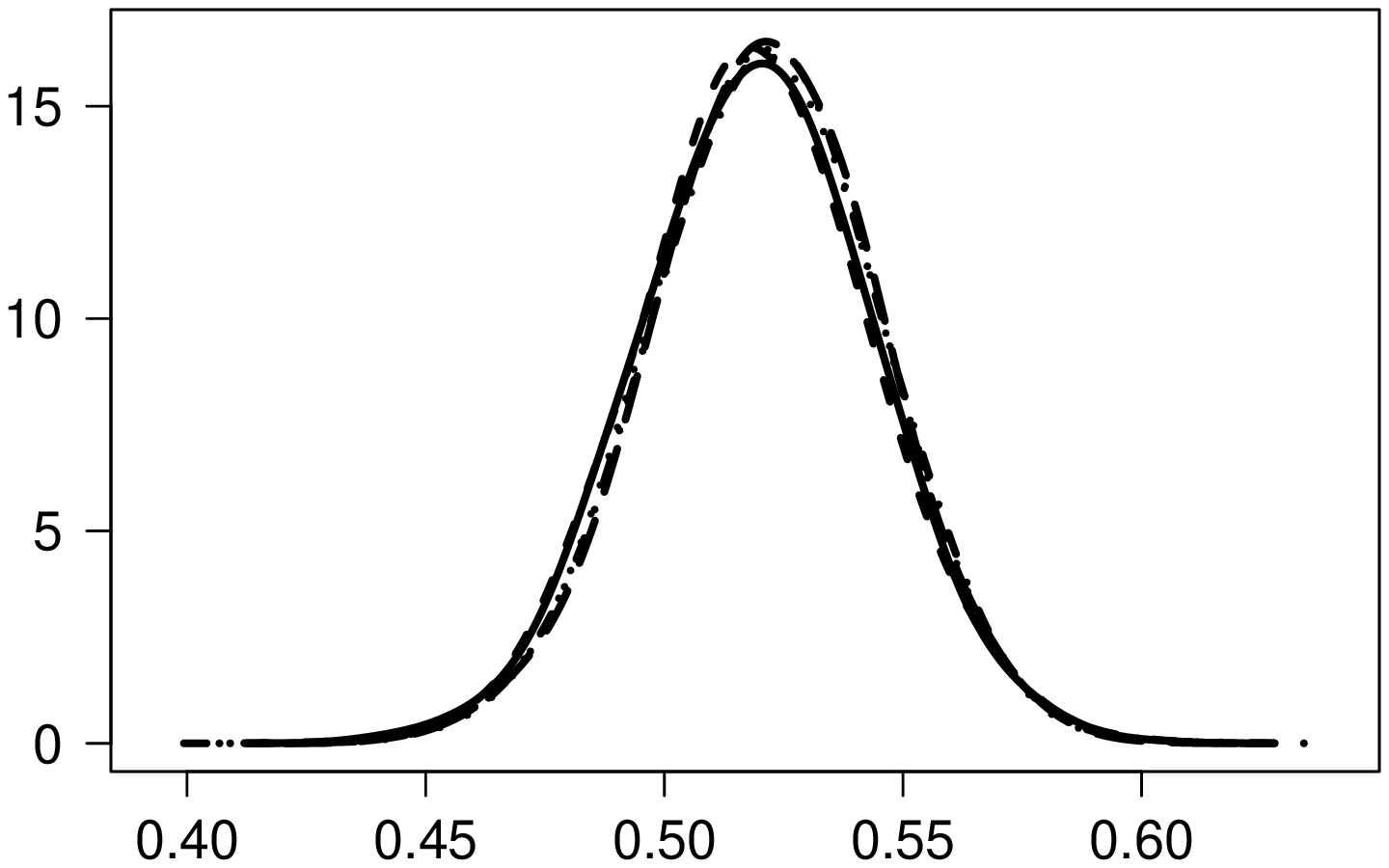}
\label{fig:dweller_density_d}}
\caption[]{The DWELL diffusion model observed with error with $n=1000$
simulated data points. True values are $(\rho,\mu,\sigma,\tau)=(1/2,
1, 1/2, 1/2)$. The outputs of the MCMC chains are subsampled every
$10$ iterations. Autocorrelations are reported after a burn in
of $5000$ iterations. Posterior density estimates using EMCMC3
(noncentred) and {\amcmc} algorithms for \subref{fig:dweller_density_a}
$\rho$, \subref{fig:dweller_density_b} $\mu$,
\subref{fig:dweller_density_c} $\sigma$, and
\subref{fig:dweller_density_d} $\tau$. In
\subref{fig:dweller_density_a}-\subref{fig:dweller_density_c} we
superimpose the posterior density obtained when the same process is
directly observed.}
\label{fig:dweller_figs}
\end{center}
\end{figure}

We consider the solution process to
\begin{align*} \d V_s = - \rho V_s \lp V_s^2 - \mu\rp\d s + \sigma \d
W_s,
\end{align*}
where $\rho>0, \mu>0, \sigma>0$. The parameter vectors are identified
as $\dpar=(\rho,\mu)^{T}$ and $\spar=\sigma$. The process is known as
the double well potential process (denoted by DWELL hereafter). We
simulated $n=1000$ (excluding the initial point) equidistant
observations with $\tinc{i}=1$ for parameter setting $(\rho, \mu,
\sigma)=(0.1, 2, 1/2)$ and $V_0\sim N(2,1/4)$ (Figure
\ref{fig:dwell_data}). Reduction to a unit volatility process is
easily achieved using $X_s:=V_s/\sigma$, which solves the SDE
\begin{align*} \d X_s = -\rho X_s\lp \sigma^2 X_s^2 - \mu \rp\d s + \d
W_s.
\end{align*}
Simple calculations reveal that the function $\asqapf{u} :=
\driftsqd{u}/2$ is given by
\begin{align*} \asqapf{u} = \frac{\rho}{2}\lcb \rho\sigma^4 u^6 -
2\rho\mu\sigma^2u^4 + \lp\rho\mu^2 - 3\sigma^2\rp u^2 + \mu\rcb,
\end{align*}
and that the algorithms are applicable with $l(\theta) =
\asqapf{u_l}$, where 
\begin{align*} 
u^2_l = \frac{2\rho\mu + \sqrt{\rho(\rho\mu^2 +
9\sigma^2)}}{3\rho\sigma^2}.
\end{align*}
Finally, for a given realisation of the layer $\player$ the upper
bound (\ref{eq:poisson_rate_ea3}) is easy found by noticing
that
\begin{align*} \asqapf{u} \leq \frac{\rho}{2}\lp \rho\sigma^4 u^6 +
\rho\mu^2 u^2 + \mu\rp =: g(u;\theta),
\end{align*}
which is convex and has a minimum at $u=0$, implying that
\begin{align}\label{eq:dwell_rate} \maxint{\player} = \lsb
g\lcb\xb(\spar) -\player\delta ;\theta\rcb\vee g\lcb
\yb(\spar)+\player\delta;\theta\rcb\rsb - l(\theta).
\end{align}
We assign improper prior densities to the parameters with $\pi(\rho)
\propto 1$, $\pi(\mu) \propto 1$, and $\pi(\sigma) \propto 1/\sigma$,
and run the chains for $10^5$ iterations. The performance of the
algorithms and posterior density estimates are shown in Figure
\ref{fig:dwell_figs}. Notice that the performance of {\emcmc}3 under the
original parametrisation is very poor, due to strong posterior
correlation between Poisson points and parameters; a fact attributed
to the sensitivity of $\maxint{\alayer}$ to the parameters (see
expression (\ref{eq:dwell_rate})). On the other hand, the noncentred
algorithm exhibits a much stronger performance with low serial
correlation for each parameter even after $50$ lags.

Posterior summaries from the output of the chains and computational
performance are gathered in Table \ref{table:dwell_table}. In contrast
to the EA1 example presented earlier, the interweaved strategy, after
accounting for the additional computational cost, does not offer any
significant improvement over the noncentred algorithm. Finally, we
found that {\amcmc} with $M=40$ paired with integration by
parts provides a reasonable approximation to the posterior marginal
densities and exhibits slightly larger adjusted ESS than the
noncentred algorithm with $\lambda=2$. The results were robust in
changes in parameter prior distributions.

\setlength{\belowcaptionskip}{-10pt}

\setlength{\belowcaptionskip}{0pt}
\begin{table}[!t]
\renewcommand{\tabcolsep}{5.5pt}
\footnotesize
\begin{center}
  \begin{tabular}{lcrrrrrrrrrrr}
\hline\hline
Method & \multicolumn{1}{c}{Par.} &\multicolumn{1}{r}{$\lambda$} &
\multicolumn{1}{c}{$M$} & \multicolumn{1}{c}{Mean}
&\multicolumn{1}{c}{SD}
& \multicolumn{1}{c}{ESS$_{adj}$} & \multicolumn{1}{c}{ESS} &
\multicolumn{1}{c}{KS}
& \multicolumn{1}{c}{Time} &
\multicolumn{3}{c}{Correlation matrix}\\[0.2cm]
EMCMC3 (noncentred) &  $\rho$ &  2 &  8.466 &  0.089 &  0.010 &  2.820 &  52.718 &  &  18.695 &  1.000 &  0.464 &  0.370 \\
 &  $\mu$ &  &  &  2.026 &  0.161 &  3.130 &  58.521 &  &  &   &  1.000 &  -0.004 \\
 &  $\sigma$ &  &  &  0.495 &  0.012 &  3.510 &  65.616 &  &  &   &   &  1.000 \\[0.1cm]
EMCMC3 (interweaved) &  $\rho$ &  2 &  8.313 &  0.089 &  0.010 &  2.769 &  54.197 &  &  19.574 &  1.000 &  0.465 &  0.384 \\
 &  $\mu$ &  &  &  2.027 &  0.160 &  3.189 &  62.422 &  &  &   &  1.000 &  0.012 \\
 &  $\sigma$ &  &  &  0.495 &  0.012 &  3.655 &  71.534 &  &  &   &   &  1.000 \\[0.1cm]
{\amcmc}-40 &  $\rho$ &   &  40.000 &  0.090 &  0.011 &  3.337 &  73.816 &  0.059 &  22.123 &  1.000 &  0.472 &  0.386 \\
 &  $\mu$ &  &  &  2.029 &  0.161 &  3.631 &  80.322 &  0.162 &  &   &  1.000 &  0.026 \\
 &  $\sigma$ &  &  &  0.494 &  0.012 &  3.594 &  79.502 & $< 0.001$ &  &   &   &  1.000 \\[0.1cm]
{\amcmc}-40 (int-by-parts) &  $\rho$ &   &  40.000 &  0.089 &  0.010 &  3.409 &  74.472 &  0.082 &  21.848 &  1.000 &  0.467 &  0.376 \\
 &  $\mu$ &  &  &  2.023 &  0.160 &  3.916 &  85.557 &  0.704 &  &   &  1.000 &  0.006 \\
 &  $\sigma$ &  &  &  0.495 &  0.012 &  3.796 &  82.930 &  0.293 &  &   &   &  1.000 \\[0.1cm]

\hline\hline
 \end{tabular}
\end{center}
\caption{The DWELL diffusion model with $n=1000$ simulated data
points. True values are $(\rho,\mu,\sigma)=(0.1, 2, 1/2)$. Statistics are
reported after a burn-in period of $5000$ iterations. Column details
as in Table \ref{table:tanh_table}.}
\label{table:dwell_table}
\end{table}
\setlength{\abovecaptionskip}{-5pt}

\subsection{Noisy observations}

We now illustrate the performance of {\emcmc} by adding to the
observations of the previous example a Gaussian error with mean $0$
and variance $\tau^2=1/4$ (Figure \ref{fig:dwell_er_data}). We have
assigned the same priors for the diffusion parameters as before, and
an improper prior proportional to $1/\tau$ for $\tau$. We employ
{\emcmc}3 under the noncentred parametrisation and the {\isi} scheme with
increasing values of $M=\{5,10,20,40\}$. The algorithms are run for
$10^5$ iterations and the MCMC outputs are thinned every $10$
iterations. Figure \ref{fig:dweller_figs} presents autocorrelation
plots for the parameters along with the marginal posterior density
estimates. It is interesting to notice that the presence of noise in
the observations seems to aid the approximation of the {\amcmc} methods,
since the posterior densities do not change significantly as $M$
increases, and seem to provide a reasonable approximation to the exact
ones.

\subsection{A bivariate double well potential}
\label{sec:mvwell}

We consider a double well potential process in two dimensions (denote
by MVWELL hereafter), solution to
\begin{align}
\label{eq:mvwell_model} \d V_s = -\frac{\sigma^2}{2}\nabla_v G(V_s)\d
s + \sigma\d W_s,~~\mbox{where }G(v) = \rho_1\lsb \lp
v^{\{2\}}\rp^2-\mu_1\rsb^2 + \rho_2\lp v^{\{2\}} - \mu_2
v^{\{1\}}\rp^2,
\end{align}
%
%
%
and $\rho_1, \rho_2, \mu_1, \mu_2, \sigma>0$. The parameter vectors
are identified as $\dpar=(\rho_1,\mu_1, \rho_2, \mu_2)^{T}$ and
$\spar=\sigma$. The invariant density of the process is proportional
to $\exp\lcb - G(v)\rcb$ and has two modes, at $\lp\sqrt{\mu_1}/\mu_2,
\sqrt{\mu_1}\rp$ and $\lp-\sqrt{\mu_1}/\mu_2, -\sqrt{\mu_1}\rp$. This
model belongs in the EA3 class and reduction to a unit volatility
process is achieved as $X_s:=V_s/\sigma$. The lower bound $l(\theta)$
and Poisson intensity $\maxint{\player}$ are given in the Appendix.

\begin{figure}[!t]
  \scriptsize
\begin{center}
\psfrag{th1}{$\rho_1$}
\psfrag{th2}{$\mu_1$}
\psfrag{th3}{$\rho_2$}
\psfrag{th4}{$\mu_2$}
\psfrag{th5}{$\sigma$}
\subfloat[{\emcmc}3, $\lambda=5$]{\includegraphics[width=\myw, height=\myh]{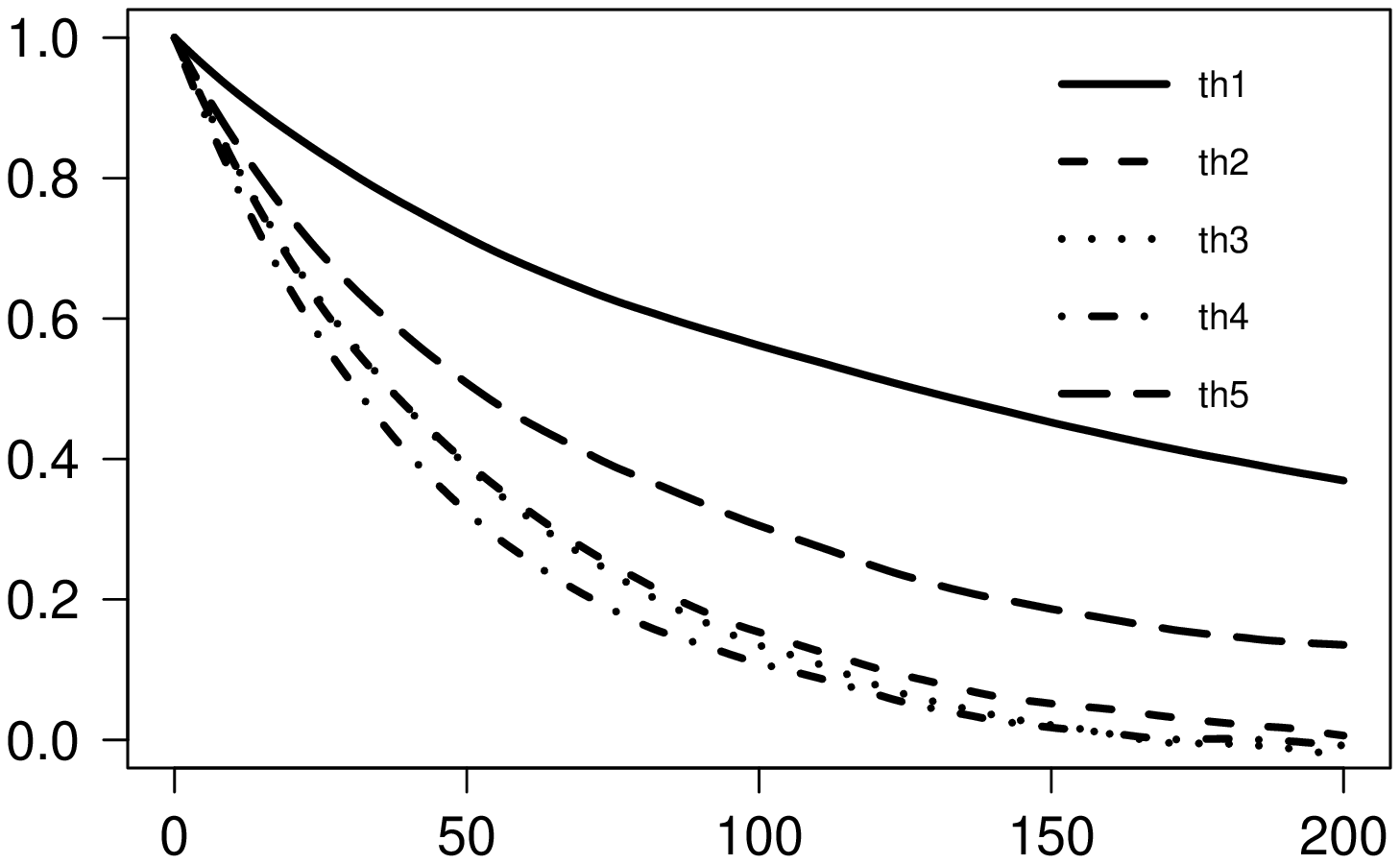}}
\subfloat[{\emcmc}3 (noncentred), $\lambda=5$]{\includegraphics[width=\myw, height=\myh]{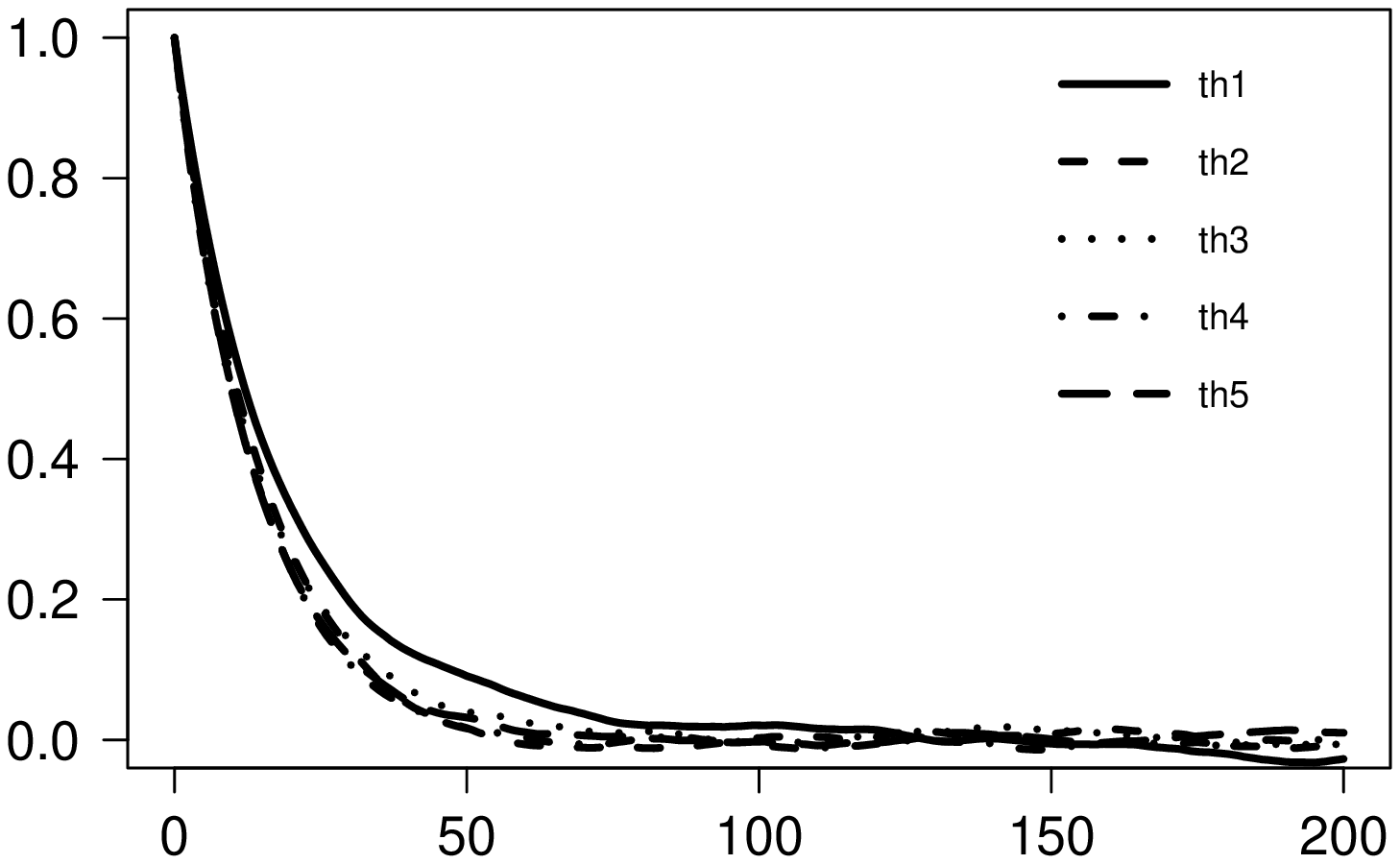}}
\subfloat[{\emcmc}3 (interweaved), $\lambda=2$]{\includegraphics[width=\myw, height=\myh]{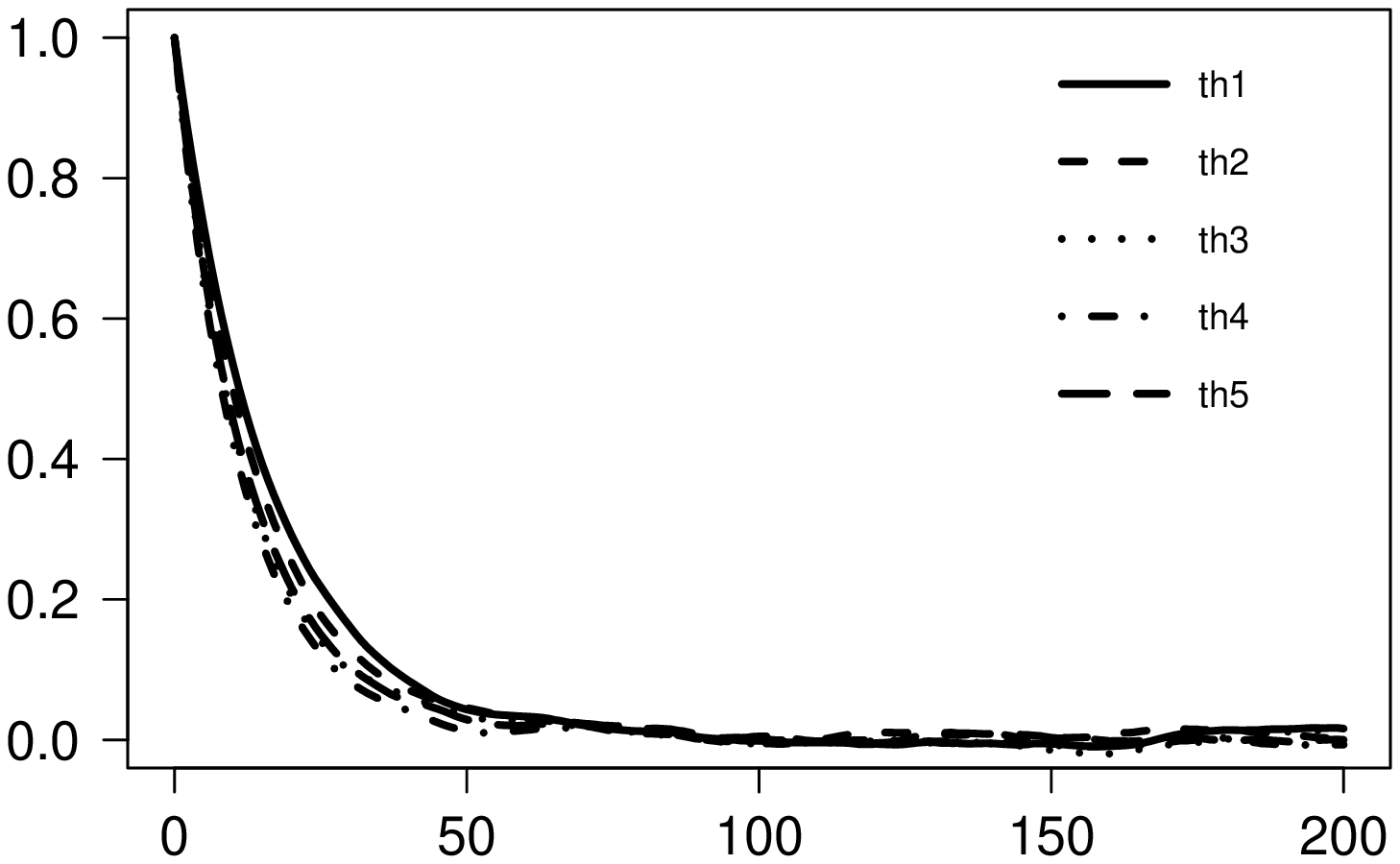}}\\
\subfloat[{\amcmc}-$20$]{\includegraphics[width=\myw, height=\myh]{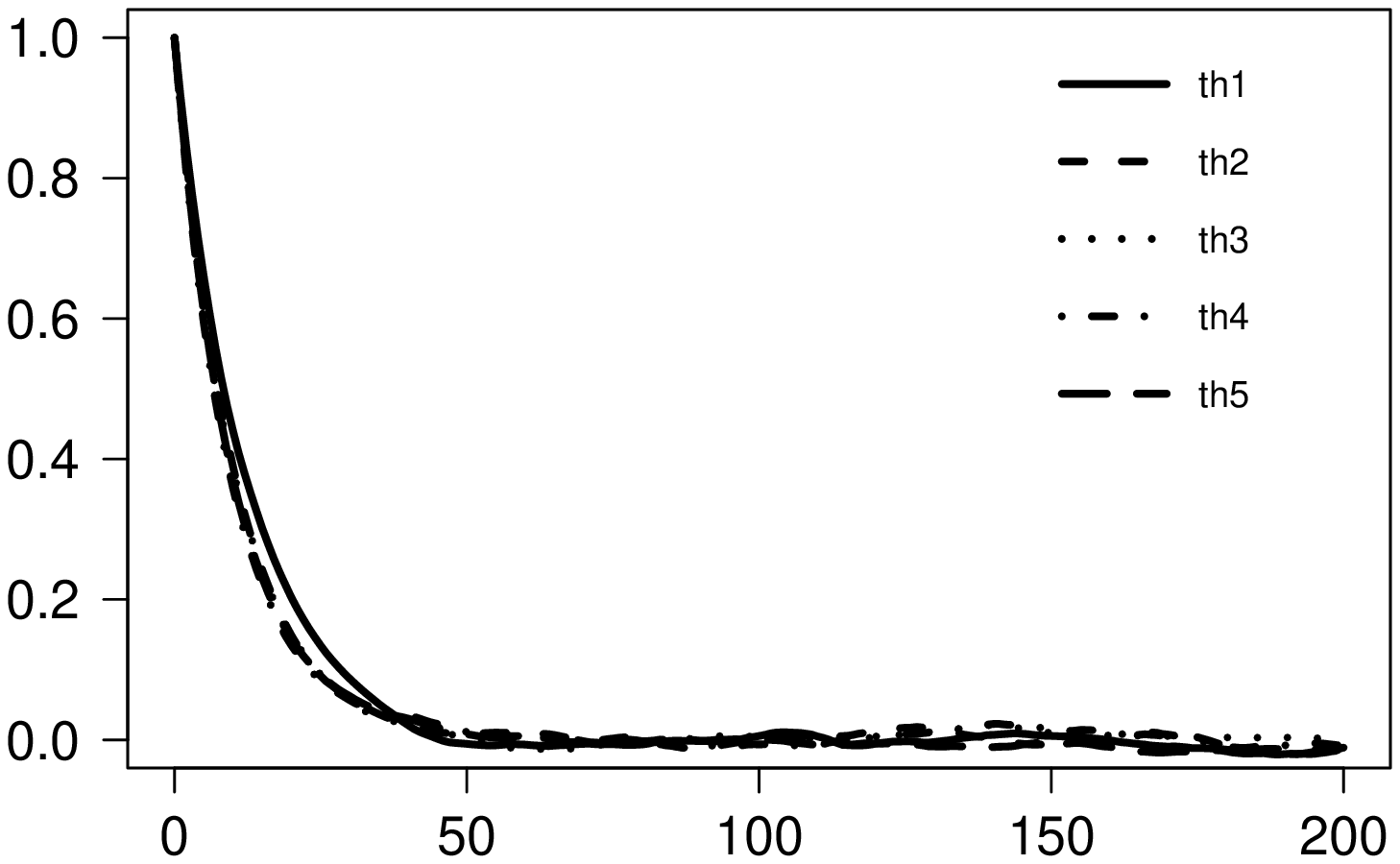}}
\scriptsize
\psfrag{ea}{{\emcmc}3}
\psfrag{hfi5}{{\amcmc}-$5$}
\psfrag{hfi10}{{\amcmc}-$10$}
\psfrag{hfi20}{{\amcmc}-$20$}
\subfloat[$\rho_1$]{\includegraphics[width=\myw, height=\myh]{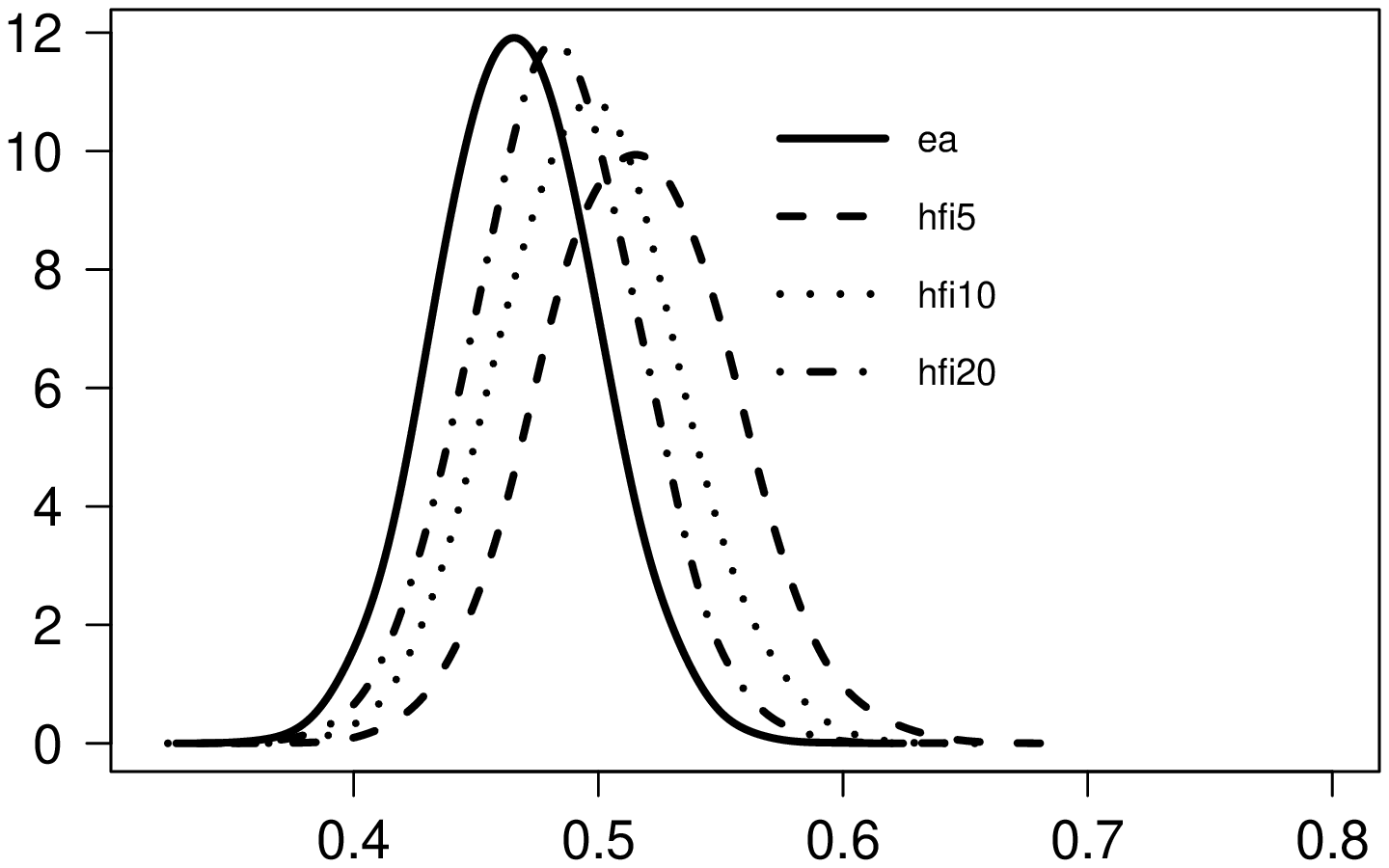}\label{fig:mv1}}
\subfloat[$\mu_1$]{\includegraphics[width=\myw, height=\myh]{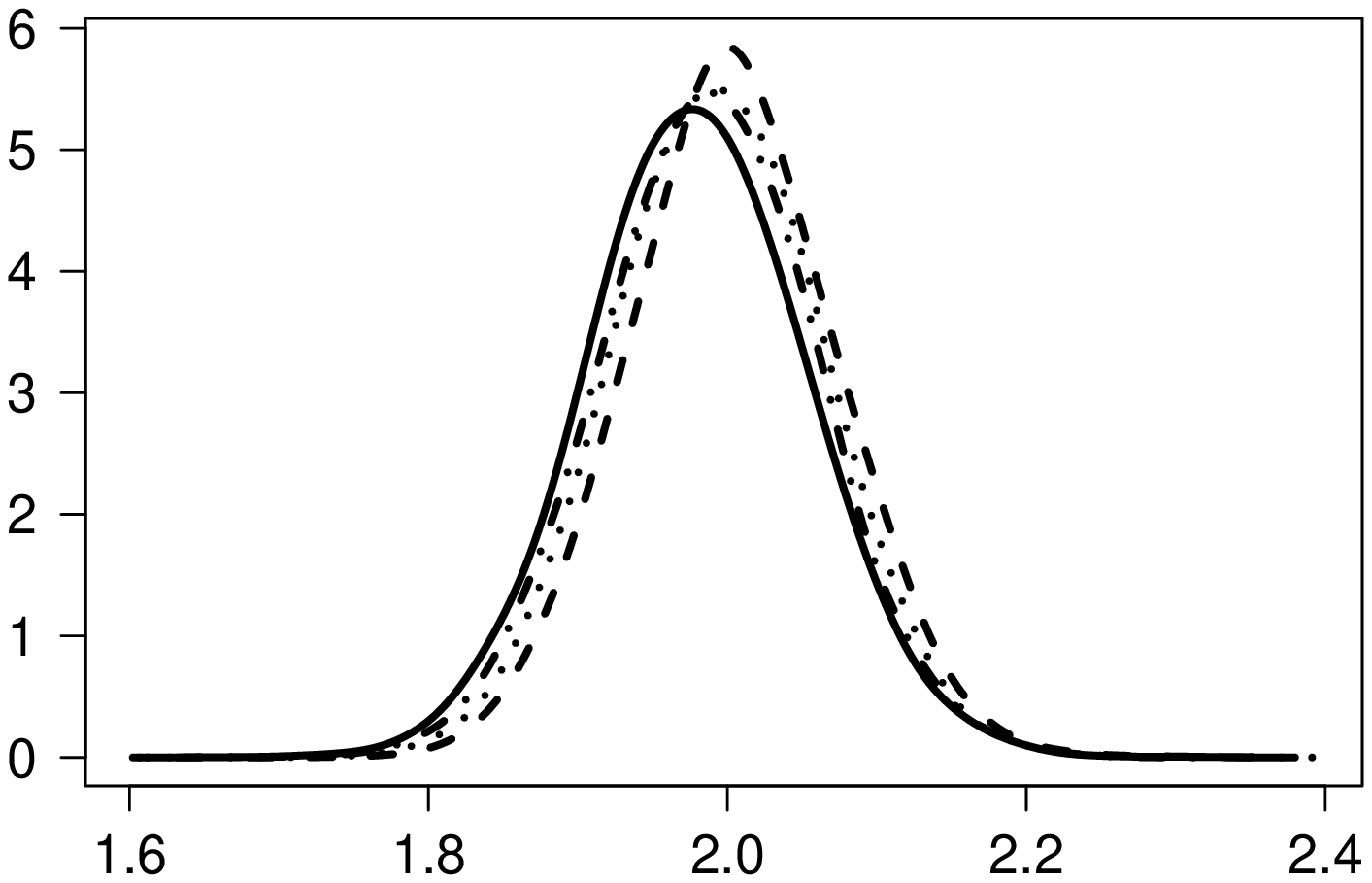}\label{fig:mv2}}\\
\subfloat[$\rho_2$]{\includegraphics[width=\myw, height=\myh]{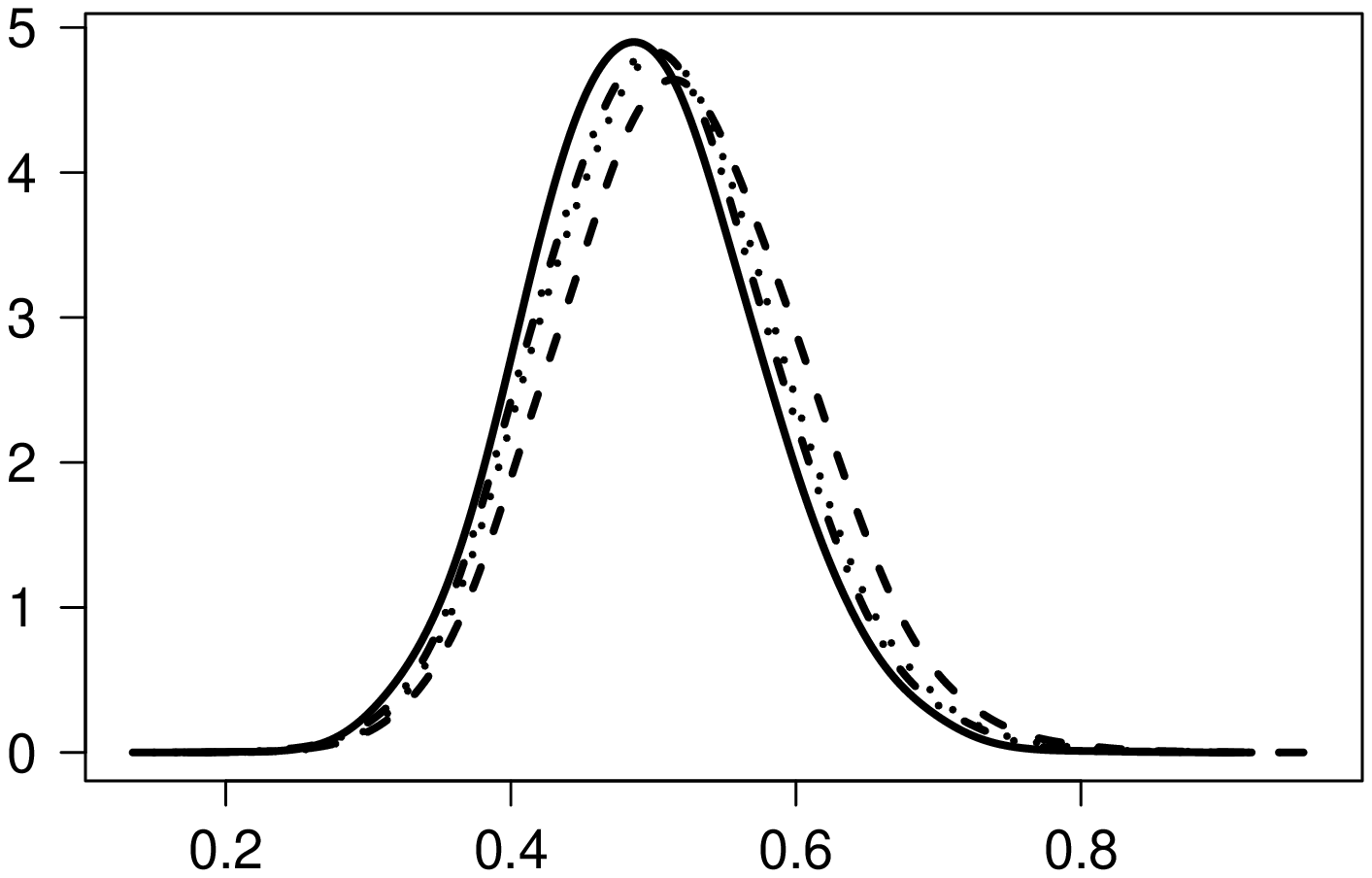}\label{fig:mv3}}
\subfloat[$\mu_2$]{\includegraphics[width=\myw, height=\myh]{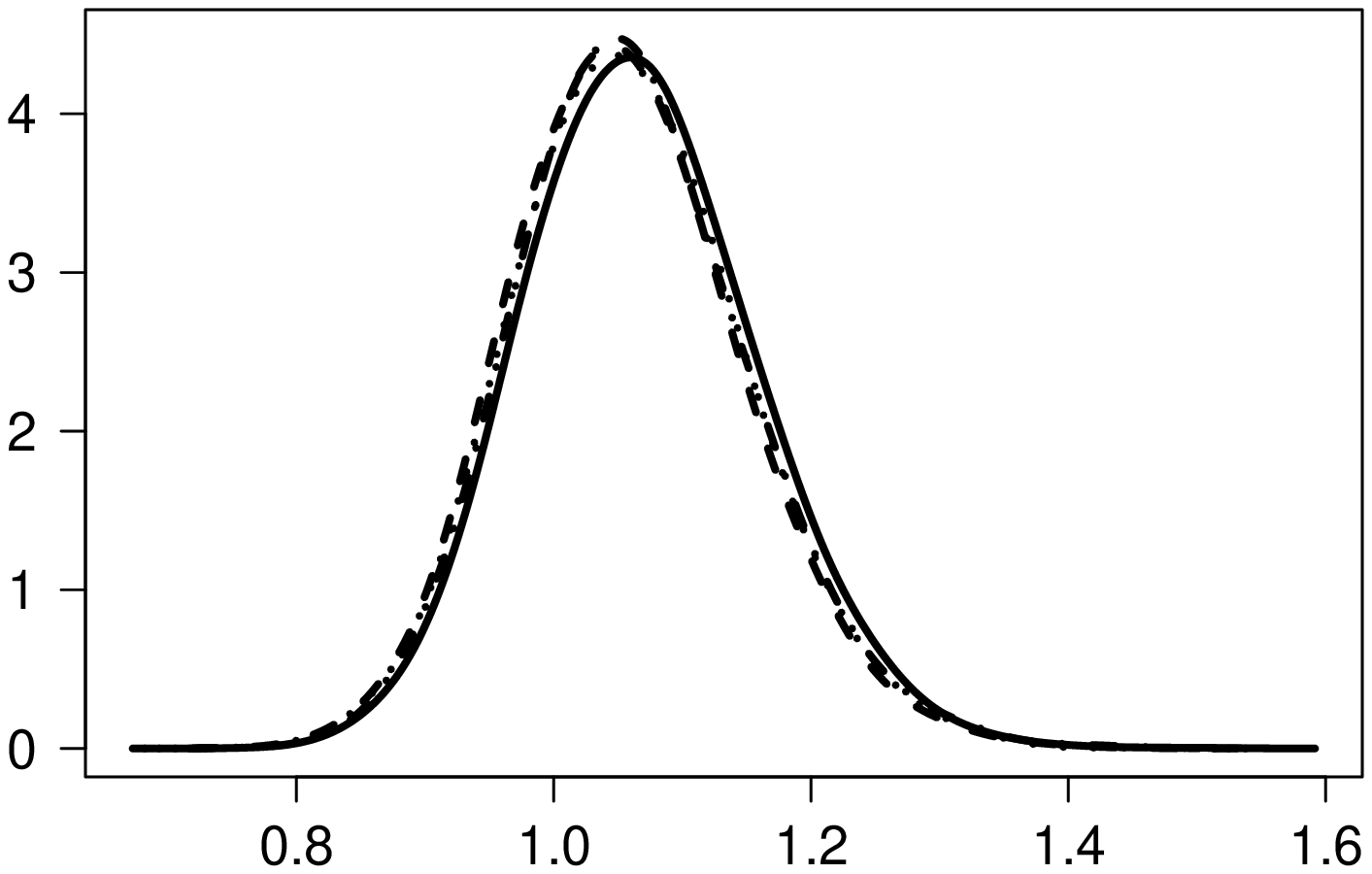}\label{fig:mv4}}
\subfloat[$\sigma$]{\includegraphics[width=\myw, height=\myh]{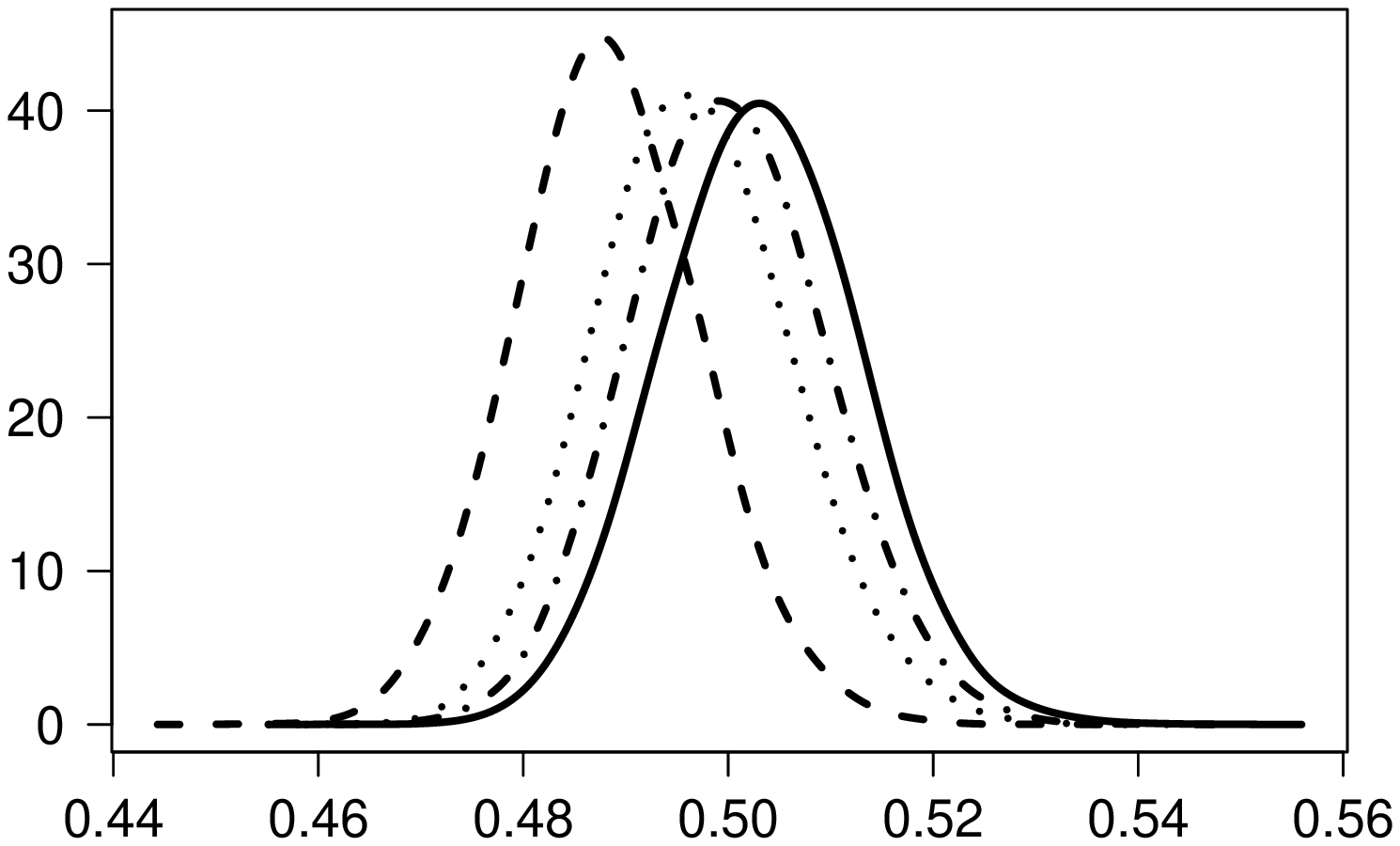}\label{fig:mv5}}
\caption[]{The MVWELL diffusion model with $n=1000$ simulated data points. True
values are
$(\rho_1,\mu_1, \rho_2,\mu_2,\sigma)=(1/2, 2, 1/2, 1, 1/2)$. Autocorrelations are reported after a
burn in of $5000$ iterations. Posterior density
estimates using EMCMC3 (interweaved) and {\amcmc} algorithms
for \subref{fig:mv1} $\rho_1$, \subref{fig:mv2} $\mu_1$,
\subref{fig:mv3} $\rho_2$, \subref{fig:mv4} $\mu_2$ and \subref{fig:mv5} $\sigma$. 
}\label{fig:mvwell_figs}
\end{center}
\end{figure}

We simulate $n=1000$ equidistant observations with $\tinc{i}=1$,
$V_0=(0, 0)^{T}$ and parameters $(\rho_1, \mu_1, \rho_2, \mu_2,
\sigma) = (1/2, 2, 1/2, 1, 1/2)$. The simulated dataset is shown in
Figures \ref{fig:mvwell_data1} and \ref{fig:mvwell_data2}. We assign
improper priors to the parameters with $\pi(\rho_i)\propto 1,
\pi(\mu_i)\propto 1, \pi(\sigma)\propto 1/\sigma$, $i=1,2$. All MCMC
chains were run for $10^5$ iterations and the performance of the
algorithms is shown in Figure \ref{fig:mvwell_figs}. For $\lambda=5$,
the {\emcmc}3 algorithm under the original parametrisation performs
poorly, whereas the noncentred exhibits a much more rapid mixing. On
the other hand, the interweaved strategy with $\lambda=2$ shows a
comparable performance to that of the noncentred. Posterior summary
statistics from the algorithms are shown in Table
\ref{table:mvwell_table}. From the approximate methods, we found that
{\amcmc}-$60$ paired with integration by parts was the most efficient
algorithm which provided a sufficiently accurate approximation to the
posterior marginal distributions.

Finally, as we pointed out in Section \ref{sec:ea_cost}, the EA3 is
inflicted by an additional computational cost due to the rejection
sampler for the layered Brownian bridge, which is clearly
reflected in the CPU time of the interweaved algorithm. In particular,
a computational profiling of the algorithm showed that approximately
$91\%$ of the total time was used by EA3, out of which $85\%$ was due
to the simulation of layered Brownian bridges. This suggests that an
alternative more efficient design of the layered Brownian bridge
simulation would boost substantially the performance of {\emcmc}3.

\setlength{\belowcaptionskip}{-10pt}

\setlength{\belowcaptionskip}{0pt}
\begin{table}[!t]
\renewcommand{\tabcolsep}{3.6pt}
\footnotesize
\begin{center}
  \begin{tabular}{lcrrrrrrrrrrrrr}
\hline\hline
Method & \multicolumn{1}{c}{Par.} &\multicolumn{1}{r}{$\lambda$} &
\multicolumn{1}{c}{$M$} & \multicolumn{1}{c}{Mean}
&\multicolumn{1}{c}{SD}
& \multicolumn{1}{c}{ESS$_{adj}$} & \multicolumn{1}{c}{ESS} &
\multicolumn{1}{c}{KS}
& \multicolumn{1}{c}{Time} &
\multicolumn{5}{c}{Correlation matrix}\\[0.2cm]
EMCMC3 (noncentred) &  $\rho_1$ &  5 &  10.455 &  0.468 &  0.032 &  0.482 &  29.326 &  &  60.819 &  1.000 &  0.208 &  0.056 &  -0.032 &  -0.069 \\
 &  $\mu_1$ &  &  &  1.980 &  0.072 &  0.547 &  33.281 &  &  &   &  1.000 &  0.492 &  -0.591 &  0.016 \\
 &  $\rho_2$ &  &  &  0.494 &  0.079 &  0.558 &  33.947 &  &  &   &   &  1.000 &  -0.814 &  -0.032 \\
 &  $\mu_2$ &  &  &  1.065 &  0.090 &  0.584 &  35.517 &  &  &   &   &   &  1.000 &  0.019 \\
 &  $\sigma$ &  &  &  0.503 &  0.009 &  0.579 &  35.224 &  &  &   &   &   &   &  1.000 \\[0.1cm]
EMCMC3 (interweaved) &  $\rho_1$ &  2 &  7.233 &  0.467 &  0.032 &  0.674 &  31.721 &  &  47.057 &  1.000 &  0.192 &  0.018 &  -0.019 &  -0.062 \\
 &  $\mu_1$ &  &  &  1.980 &  0.073 &  0.751 &  35.345 &  &  &   &  1.000 &  0.482 &  -0.582 &  0.020 \\
 &  $\rho_2$ &  &  &  0.492 &  0.078 &  0.872 &  41.032 &  &  &   &   &  1.000 &  -0.816 &  -0.018 \\
 &  $\mu_2$ &  &  &  1.068 &  0.089 &  0.936 &  44.029 &  &  &   &   &   &  1.000 &  -0.001 \\
 &  $\sigma$ &  &  &  0.503 &  0.009 &  0.843 &  39.673 &  &  &   &   &   &   &  1.000 \\[0.1cm]
{\amcmc}-60 (int-by-parts) &  $\rho_1$ &   &  60.000 &  0.467 &  0.032 &  1.447 &  44.949 &  0.100 &  31.065 &  1.000 &  0.184 &  0.022 &  -0.008 &  -0.053 \\
 &  $\mu_1$ &  &  &  1.979 &  0.072 &  1.518 &  47.154 &  0.831 &  &   &  1.000 &  0.490 &  -0.594 &  -0.008 \\
 &  $\rho_2$ &  &  &  0.494 &  0.079 &  1.594 &  49.515 &  0.361 &  &   &   &  1.000 &  -0.819 &  -0.031 \\
 &  $\mu_2$ &  &  &  1.066 &  0.090 &  1.734 &  53.862 &  0.268 &  &   &   &   &  1.000 &  0.027 \\
 &  $\sigma$ &  &  &  0.503 &  0.009 &  1.621 &  50.371 &  0.932 &  &   &   &   &   &  1.000 \\[0.1cm]

\hline\hline
 \end{tabular}
\end{center}
\caption{The MVWELL diffusion model with $n=1000$ simulated data
points. True values are $(\rho_1,\mu_1, \rho_2,\mu_2,\sigma)=(1/2, 2,
1/2, 1, 1/2)$. Statistics are reported after a burn-in period of
$5000$ iterations. Column details as in Table \ref{table:tanh_table}.}
\label{table:mvwell_table}
\end{table}


\section{Discussion}
\label{sec:discuss}

We have developed exact data augmentation methods for discretely
directly and indirectly observed diffusions. We have established the
precise connection between this paradigm and the best existing
alternative method when the variance-stabilising transformation can be
performed, the {\isi}. The empirical comparison of the two methods
showed that in univariate processes {\emcmc} can perform at least as
well as a sufficiently accurate {\amcmc}, even when ignoring the
additional computational cost needed by the latter to determine a good
value of $M$ through experimentation. For the considered bivariate
example, {\emcmc} is outperformed by {\amcmc} since the cost of the
former is dominated by the simulation of the layered Brownian bridges,
and thus could be improved by considering alternative designs for this
simulation.

We have also pointed out an intriguing connection
between exact and approximate methods: the degree of freedom rendered
by the Poisson sampling rate. On going work involves the rigorous
proof of the effect of the auxiliary Poisson sampling. The auxiliary
sampling can be seen as a variance reduction scheme. In general, there
is a large scope for investigating other such schemes both for {\eda}
and {\isi}. In this article we have already demonstrated the effect of
performing integration by parts where possible to the efficiency of
the algorithms.

The extension of these methods outside the class of processes
prescribed by the current version of EA3 is definitely an exciting
direction. Another direction of interest for future research is to
explore further the connection between unbiased estimation of
transition density and MCMC. There is a large and growing literature
which develops MCMC algorithms for models with intractable likelihoods
using unbiased estimators thereof; see for example
\cite{andri:robe:2009,pmcmc}. The class of diffusions where such
estimators can be obtained is much larger than that simulated by EA3,
see for example Section 4.6 of \cite{pap}. 

There exists available software for implementing all the
methods in this paper, which is available on request by the authors. 


\section*{Acknowledgements}

O. Papaspiliopoulos would like to acknowledge financial support by the Spanish government
through a ``Ramon y Cajal'' fellowship and grant MTM2009-09063. G. Sermaidis
was funded by the Greek State Scholarships Foundation. G. Roberts
acknowledges CRiSM and EPSRC.



\bibliographystyle{Styles/rss}
\bibliography{main}

\begin{thebibliography}{35}
\expandafter\ifx\csname natexlab\endcsname\relax\def\natexlab#1{#1}\fi
\expandafter\ifx\csname url\endcsname\relax
  \def\url#1{\texttt{#1}}\fi
\expandafter\ifx\csname urlprefix\endcsname\relax\def\urlprefix{URL }\fi

\bibitem[{A{\"i}t-Sahalia({2008})}]{sah:2008}
A{\"i}t-Sahalia, Y. ({2008}) Closed-form likelihood expansions for multivariate
  diffusions.
\newblock \emph{Ann. Statist.}, \textbf{{36}}, {906--937}.

\bibitem[{A{\"i}t-Sahalia and Kimmel(2007)}]{saha:real}
A{\"i}t-Sahalia, Y. and Kimmel, R. (2007) Maximum likelihood estimation of
  stochastic volatility models.
\newblock \emph{Journal of Financial Economics}, \textbf{83}, 413--452.

\bibitem[{Andrieu \emph{et~al.}(2010)Andrieu, Doucet and Holenstein}]{pmcmc}
Andrieu, C., Doucet, A. and Holenstein, R. (2010) Particle {M}arkov chain
  {M}onte {C}arlo.
\newblock \emph{J. R. Stat. Soc. Ser. B Stat. Methodol.}, \textbf{3}, 269--342.

\bibitem[{Andrieu and Roberts(2009)}]{andri:robe:2009}
Andrieu, C. and Roberts, G.~O. (2009) The pseudo-marginal approach for
  efficient {M}onte {C}arlo computations.
\newblock \emph{Ann. Statist.}, \textbf{37}, 697--725.

\bibitem[{Beskos \emph{et~al.}(2006{\natexlab{a}})Beskos, Papaspiliopoulos and
  Roberts}]{besk:papa:robe:2004}
Beskos, A., Papaspiliopoulos, O. and Roberts, G.~O. (2006{\natexlab{a}})
  Retrospective exact simulation of diffusion sample paths with applications.
\newblock \emph{Bernoulli}, \textbf{12}, 1077--1098.

\bibitem[{Beskos \emph{et~al.}(2008)Beskos, Papaspiliopoulos and
  Roberts}]{besk:papa:robe:2008}
Beskos, A., Papaspiliopoulos, O. and Roberts, G.~O. (2008) A factorisation of
  diffusion measure and finite sample path constructions.
\newblock \emph{Methodol. Comput. Appl. Probab.}, \textbf{10}, 85--104.

\bibitem[{Beskos \emph{et~al.}(2006{\natexlab{b}})Beskos, Papaspiliopoulos,
  Roberts and Fearnhead}]{besk:papa:robe:fear:2006}
Beskos, A., Papaspiliopoulos, O., Roberts, G.~O. and Fearnhead, P.
  (2006{\natexlab{b}}) Exact and computationally efficient likelihood-based
  estimation for discretely observed diffusion processes.
\newblock \emph{J. R. Stat. Soc. Ser. B Stat. Methodol.}, \textbf{68},
  333--382.
\newblock With discussion and a reply by the authors.

\bibitem[{Brown \emph{et~al.}(2000)Brown, K\aa{}resen, Roberts and
  Tonellato}]{tone}
Brown, P.~E., K\aa{}resen, K.~F., Roberts, G.~O. and Tonellato, S. (2000)
  Blur-generated non-separable space-time models.
\newblock \emph{J. R. Stat. Soc. Ser. B Stat. Methodol.}, \textbf{62},
  847--860.

\bibitem[{Cotter \emph{et~al.}(2012)Cotter, Roberts, Stuart and
  White}]{robe:stu:2012}
Cotter, S.~L., Roberts, G.~O., Stuart, A.~M. and White, D. (2012) {MCMC}
  methods for functions: modifying old algorithms to make them faster.
\newblock Submitted.

\bibitem[{Elerian \emph{et~al.}(2001)Elerian, Chib and
  Shephard}]{eler:chib:shep:2001}
Elerian, O., Chib, S. and Shephard, N. (2001) Likelihood inference for
  discretely observed nonlinear diffusions.
\newblock \emph{Econometrica}, \textbf{69}, 959--993.

\bibitem[{Eraker(2001)}]{eraker:2001}
Eraker, B. (2001) M{CMC} analysis of diffusion models with application to
  finance.
\newblock \emph{J. Bus. Econom. Statist.}, \textbf{19}, 177--191.

\bibitem[{\'Etor\'e and Martinez(2011)}]{ea:local}
\'Etor\'e, P. and Martinez, M. (2011) Exact simulation of one-dimensional
  stochastic differential equations involving the local time at zero of the
  unknown process.
\newblock Tech. rep.
\newblock Available online from \texttt{http://arxiv.org/abs/1102.2565}.

\bibitem[{Forman and S{\o}rensen(2008)}]{forman:sorensen}
Forman, J.~L. and S{\o}rensen, M. (2008) The {P}earson diffusions: a class of
  statistically tractable diffusion processes.
\newblock \emph{Scand. J. Statist.}, \textbf{35}, 438--465.

\bibitem[{Girolami and Calderhead(2011)}]{giro:riem}
Girolami, M. and Calderhead, B. (2011) Riemann manifold {L}angevin and
  {H}amiltonian {M}onte {C}arlo methods.
\newblock \emph{J. R. Stat. Soc. Ser. B Stat. Methodol.}, \textbf{73},
  123--214.
\newblock With discussion and a reply by the authors.

\bibitem[{Golightly and Wilkinson(2006)}]{gol:wil:2006}
Golightly, A. and Wilkinson, D.~J. (2006) Bayesian sequential inference for
  stochastic kinetic biochemical network models.
\newblock \emph{J. Comput. Biol.}, \textbf{13}, 838--851.

\bibitem[{Golightly and Wilkinson(2008)}]{gol:wilk:2008}
Golightly, A. and Wilkinson, D.~J. (2008) Bayesian inference for nonlinear
  multivariate diffusion models observed with error.
\newblock \emph{Comput. Statist. Data Anal.}, \textbf{52}, 1674--1693.

\bibitem[{Gon\c{c}alves and Roberts(2012)}]{ea:jump}
Gon\c{c}alves, F. and Roberts, G. (2012) Exact simulation problems for
  jump-diffusions.
\newblock Submitted.

\bibitem[{Horenko and Sch{\"u}tte(2008)}]{hore:hmmsde}
Horenko, I. and Sch{\"u}tte, C. (2008) Likelihood-based estimation of
  multidimensional {L}angevin models and its application to biomolecular
  dynamics.
\newblock \emph{Multiscale Model. Simul.}, \textbf{7}, 731--773.

\bibitem[{Kalogeropoulos \emph{et~al.}(2010)Kalogeropoulos, Roberts and
  Dellaportas}]{kalog}
Kalogeropoulos, K., Roberts, G. and Dellaportas, P. (2010) {Inference for
  stochastic volatility models using time change transformations}.
\newblock \emph{Ann. Statist.}, \textbf{38}, 784--807.

\bibitem[{Metzner \emph{et~al.}(2006)Metzner, Sch\"{u}tte and
  {Vanden-Eijnden}}]{Metzner:2006}
Metzner, P., Sch\"{u}tte, C. and {Vanden-Eijnden}, E. (2006) Illustration of
  transition path theory on a collection of simple examples.
\newblock \emph{The Journal of Chemical Physics}, \textbf{125}, 084110.

\bibitem[{{\O}ksendal(2003)}]{oksendal}
{\O}ksendal, B. (2003) \emph{Stochastic differential equations. An introduction
  with applications}.
\newblock Universitext. Berlin: Springer-Verlag, 6th edn.

\bibitem[{Papaspiliopoulos(2011)}]{pap}
Papaspiliopoulos, O. (2011) A methodological framework for {M}onte {C}arlo
  probabilistic inference for diffusion processes.
\newblock In \emph{Bayesian Time Series Models}. Cambridge University Press.

\bibitem[{Papaspiliopoulos and Roberts(2012)}]{paprobsemstat}
Papaspiliopoulos, O. and Roberts, G.~O. (2012) Importance sampling techniques
  for estimation of diffusion models.
\newblock In \emph{Statistical Methods for Stochastic Differential Equations},
  311--337. Monographs on Statistics and Applied Probability, Chapman and Hall.

\bibitem[{Papaspiliopoulos \emph{et~al.}(2007)Papaspiliopoulos, Roberts and
  Sk{\"o}ld}]{papa:robe:2007}
Papaspiliopoulos, O., Roberts, G.~O. and Sk{\"o}ld, M. (2007) A general
  framework for the parametrization of hierarchical models.
\newblock \emph{Statistical Science}, \textbf{22}, 59--73.

\bibitem[{Peluchetti and Roberts(2008)}]{pel:robe:ea}
Peluchetti, S. and Roberts, G.~O. (2008) An empirical study of the efficiency
  of the {EA} for diffusion simulation.
\newblock Tech. rep., University of Warwick.

\bibitem[{Picchini \emph{et~al.}(2010)Picchini, Gaetano and Ditlevsen}]{ditl}
Picchini, U., Gaetano, A. and Ditlevsen, S. (2010) {Stochastic Differential
  Mixed-Effects Models}.
\newblock \emph{Scandinavian Journal of Statistics}, \textbf{37}, 67--90.

\bibitem[{Pitt and Shephard(1999)}]{pitt1999}
Pitt, M. and Shephard, N. (1999) {Analytic convergence rates and
  parameterization issues for the Gibbs sampler applied to state space models}.
\newblock \emph{Journal of Time Series Analysis}, \textbf{20}, 63--85.

\bibitem[{Plummer \emph{et~al.}(2010)Plummer, Best, Cowles and Vines}]{rcoda}
Plummer, M., Best, N., Cowles, K. and Vines, K. (2010) \emph{coda: Output
  analysis and diagnostics for MCMC}.
\newblock \urlprefix\url{http://CRAN.R-project.org/package=coda}.
\newblock R package version 0.13-5.

\bibitem[{{R Development Core Team}(2010)}]{rlang}
{R Development Core Team} (2010) \emph{R: A Language and Environment for
  Statistical Computing}.
\newblock R Foundation for Statistical Computing, Vienna, Austria.
\newblock \urlprefix\url{http://www.R-project.org}.

\bibitem[{Roberts \emph{et~al.}(2004)Roberts, Papaspiliopoulos and
  Dellaportas}]{robe:papa:della:2004}
Roberts, G.~O., Papaspiliopoulos, O. and Dellaportas, P. (2004) Bayesian
  inference for non-{G}aussian {O}rnstein-{U}hlenbeck stochastic volatility
  processes.
\newblock \emph{J. R. Stat. Soc. Ser. B Stat. Methodol.}, \textbf{66},
  369--393.

\bibitem[{Roberts and Stramer(2001)}]{robe:stra:2001}
Roberts, G.~O. and Stramer, O. (2001) On inference for partially observed
  nonlinear diffusion models using the {M}etropolis-{H}astings algorithm.
\newblock \emph{Biometrika}, \textbf{88}, 603--621.

\bibitem[{Rydberg(1997)}]{Rydberg:1997}
Rydberg, T.~H. (1997) A note on the existence of unique equivalent martingale
  measures in a {M}arkovian setting.
\newblock \emph{Finance and Stochastics}, \textbf{1}, 251--257.

\bibitem[{Sekhon(2007)}]{rmatching}
Sekhon, J.~S. (2007) Multivariate and propensity score matching software with
  automated balance optimization: The matching package for {R}.
\newblock \emph{Journal of Statistical Software}.

\bibitem[{Sermaidis(2010)}]{sermaidis:thesis}
Sermaidis, G. (2010) \emph{Likelihood-based inference for discretely observed
  diffusions.}
\newblock Ph.D. thesis, Department of Statistics, University of Warwick.

\bibitem[{Yu and Meng(2011)}]{meng:yun}
Yu, Y. and Meng, X. (2011) To center or not to center, that is not the
  question: An ancillarity-sufficiency interweaving strategy ({ASIS}) for
  boosting {MCMC} efficiency.
\newblock \emph{J. Comput. Graph. Statist.}, \textbf{20}, 531--570.

\end{thebibliography}


\section{Appendix}
\label{sec:app}

\subsubsection*{Proof of Lemma \ref{lem:aux-den}}

\proof 
Expression (\ref{eq:md_density_ea3}) is derived by writing the density
of the accepted random variables $(\player, \plpath,\pPois)$ with
respect to the law of the proposed
%
\begin{align*}
 \pi(\player,\plpath,\pPois\mid
\sp,\ep,\theta) = \frac{1}{a(x,y,t,\theta)}\prod_{j=1}^{\pkappa}
\ind{\lsb\frac{1}{\maxint{\player}}\phifcb{\plpath_{\ppsi_j
}
+\lp 1-\frac{\psi_j}{t}\rp\xt{} + \frac{\psi_j}{t}\yt{}}
<\pups_j\rsb },
\end{align*}
and by invoking a change of measure from the law of a Poisson process
of intensity $\maxint{\player}$ to the law of one of unit intensity,
thus ensuring a parameter-independent dominating measure. Finally,
integrating out the marks $\pUps=\{\pups_j,1\leq j\leq \pkappa\}$, we
obtain expression (\ref{eq:md_density_ea3}). \qed

\subsubsection*{Proof of Theorem \ref{th:mcmc_ea3_ca}}

\proof The factorisation of the density in the three terms is
elementary. For $V_0=\sp$, $V_t=\ep$ and $x=\lf{}{\sp}$,
$y=\lf{}{\ep}$, taking expectations on both sides of (\ref{eq:girs})
with respect to $\cmW{t,x,y}$ we derive the fundamental identity
\begin{align*}
%
\ttrans{t}{x}{y}&=\gdens{y-x}{t}\exp\lcb \adrift{y} - \adrift{x}
-l(\theta)t\rcb a(x,y,t,\theta),
\end{align*}
which combined with \eqref{eq:Jacobian} gives $a(x,y,t,\theta)$ as a
function of $\trans{t}{\sp}{\ep}$. Combining this with
Lemma \ref{lem:aux-den} yields the expression. 

It remains to show that \eqref{eq:post} can be obtained by integrating
out the auxiliary variables and conditioning on the data. However,
this is trivial since by taking expectations with respect to
the dominating measure 
$\otimes_{i=1}^n\lp\cmMn{\tinc{i}}\times\mP{\tinc{i}}\rp$ we
obtain the marginal $\pi(\theta) \prod_{i=1}^n \trans{\Delta
  t_i}{V_{t_{i-1}}}{V_{t_i}} $ from which \eqref{eq:post} follows as a
conditional. 
 \qed

\subsubsection*{Proof of Theorem \ref{th:connection}}

\proof 
For notational simplicity, we define $K(\obs,\theta)$ to be the following
deterministic function of observations $\obs$ and $\theta$:
\begin{align*}
\exp\lcb
\adriftcb{x_n(\spar)} - \adriftcb{x_0(\spar)} -
l(\theta)(t_n-t_0)\rcb
\prod_{i=1}^n\DD{V_{t_i}}\gdenscb{x_i(\spar) - x_{i-1}(\spar)}{\tinc{i}}.
\end{align*}
The joint posterior density of $\theta$ and imputed data $\{\alayer_i,
\alpath_i,\aPsi_i,1\leq i \leq n \}$ is given (up to a constant) in
expression (\ref{eq:eda-den}). Integrating out the Poisson processes is
easily done by first integrating out the coordinates and then the number of
Poisson points. Therefore, integrating out $\aPsi_i=\{
\apsi_{i,j},1\leq j \leq\akappa_i \}$, $1\leq i \leq n$, we obtain
\begin{align*}
&\pi(\theta)K(\obs,\theta)\exp\lp -
\sum_{i=1}^n\lsb\maxint{\alayer_i}-1\rsb\tinc{i}\rp\prod_{i=1}
^n\lcb
\frac{1}{\tinc{i}}\int_0^{\tinc{i}}\lsb \maxint{\alayer_i} -
\phifcb{\alpath_{i,s} + \mbb_{i,s}(\theta)}
\rsb\d s\rcb^{\akappa_i}.
\end{align*}
Integrating out the Poisson points yields the posterior density
of $\theta$ and $\{\alayer_i, \alpath_i,1\leq i\leq n \}$ with
respect to $\Leb^p\otimes_{i=1}^n\cmMn{\tinc{i}}$, as
\begin{align*}
&\pi(\theta)K(\obs,\theta)\exp\lp -
\sum_{i=1}^n\lsb\maxint{\alayer_i}-1\rsb\tinc{i}\rp
%
\prod_{i=1}^n\exp\lcb \tinc{i}\lsb \frac{1}{\tinc{i}}
\int_0^{\tinc{i}}\lsb \maxint{\alayer_i} -
\phifcb{\alpath_{i,s}+\mbb_{i,s}(\theta)
}
\rsb\d s - 1\rsb \rcb\notag\\
&=\pi(\theta)K(\obs,\theta)\exp\lcb-\sum_{i=1}^n\int_0^{\tinc{i}}
\phifcb{\alpath_{i,s}+\mbb_{i,s}(\theta)
}
\d s\rcb.
\end{align*}
Given the construction of $\cmMn{\tinc{i}}$, by
integrating out the layers we obtain the joint 
density of $\theta$ and $\{\tX_i,1\leq i \leq n\}$ with respect
to 
$\Leb^p\otimes_{i=1}^n\cmWn{\tinc{i},0,0}$ which coincides with the posterior
density (\ref{eq:cd_lik_ibp}) targeted by {\ipi}. \qed

\subsubsection*{Proof of Theorem \ref{th:mcmc_ea3_nca}}

\proof For a pair of observations $(V_{t_{i-1}}, V_{t_i})$, the joint density
of the accepted elements of EA3 $(\alayer_i, \alpath_i,\aPoisnc_i)$
conditionally on $V_{t_{i-1}}, V_{t_i}, \theta$ is given by
\begin{align*}
 \frac{\prod_{j=1}^{\infty}\lcb
1
-\ind{\lsb\axinc_{i,j}<\maxint{\alayer_i}\rsb}
\phifcb{\alpath_{i,\apsinc_{i,j}}+\mbb_{i,\apsinc_{i,j}}(\theta)}/\maxint{\alayer_i}
\rcb}{a(x_{i-1}(\spar), x_i(\spar),\tinc{i},\theta)},
\end{align*}
with respect to the product measure $\cmMn{\tinc{i}}\times\tmL{\tinc{i}}$, where
$\tmL{t}$ is
the measure of a unit rate Poisson process on $[0,t]\times(0,\infty)$.
Using the conditional independence of the latent data given $\obs$ and $\theta$,
\begin{align*}
 \ncdensf (\{\alayer_i, \alpath_i,\aPoisnc_i,1\leq i \leq n\}\mid \obs,\theta)
=
\prod_{i=1}^n
\ncdensf
(\alayer_i, \alpath_i,\aPoisnc_i\mid V_{t_{i-1}}, V_{t_i}, \theta),
\end{align*}
and the proof follows along the same lines as Theorem
\ref{th:mcmc_ea3_ca}. \qed

\subsubsection*{Functions related to the MVWELL model}

\noindent Applying the Lamperti transformation to \eqref{eq:mvwell_model}, we
obtain the drift of the transformed process as 
\begin{align*}
  \drift{}{x} = \nabla_x \adrift{x},\,\,\mbox{where}\,\, \adrift{x} = -\frac{1}{2}G(\sigma x).
\end{align*}
Using second partial derivative tests it is straightforward to
verify that the function $\asqapf{u} := \driftsqd{u}/2$ has a global
minimum
\begin{align*}
  l(\theta) = -\frac{p_1 + p_2}{54\mu_2^2(1+\mu_2^2)},
\end{align*}
where
\begin{align*}
  p_1 := 2\mu_2\sigma^2\sqrt{2\rho_1}\lcb
  9+\mu_2^2\lp 9+ 2\rho_1\mu_1^2\rp \rcb^{3/2},~~p_2 := \mu_2^2\sigma^2\lcb 54\rho_1\mu_1(1+\mu_2^2) -
  8\rho_1^2\mu_1^3\mu_2^2 + 27\rho_2(1+\mu_2^2)^2\rcb.
\end{align*}
The function $\asqapf{u}$ has no local maxima, implying that for
a given realisation of the layer $\alayer$ the Poisson rate is
\begin{align*}
  \maxint{\alayer} = \lsb \vee_{i=1}^{4} \asqapf{u_{i}} \rsb-l(\theta),
\end{align*}
where
\begin{align*}
  u_{1} &= \lp \bl{x}{1}(\spar) -\tl{L}{1}\delta, \bl{x}{2}(\spar)
  -\tl{L}{2}\delta\rp^{T},\quad u_{2} = \lp \bl{x}{1}(\spar) -\tl{L}{1}\delta, \bl{y}{2}(\spar)
  +\tl{L}{2}\delta\rp^{T},\\
u_{3} &= \lp \bl{y}{1}(\spar) +\tl{L}{1}\delta, \bl{x}{2}(\spar)
  -\tl{L}{2}\delta\rp^{T}, \quad u_{4} = \lp \bl{y}{1}(\spar) +\tl{L}{1}\delta, \bl{y}{2}(\spar)
  +\tl{L}{2}\delta\rp^{T}.
\end{align*}

\end{document}